\documentclass{jpp}
\usepackage{graphicx}
\usepackage[labelformat=empty, position=top]{subcaption}
\usepackage[export]{adjustbox}
\usepackage{xcolor}

\usepackage[utf8]{inputenc}
\usepackage[T1]{fontenc}
\usepackage{amsmath}

\usepackage[version=4]{mhchem}

\newcommand{\rlight}{r_{\rm L}}

\newcommand{\ex}{\mathbf{e}_{\rm x}}
\newcommand{\ey}{\mathbf{e}_{\rm y}}
\newcommand{\ez}{\mathbf{e}_{\rm z}}

\shorttitle{Particle acceleration in pulsar ultra-strong fields}
\shortauthor{I. Tomczak and J. Pétri}

\title{Particle acceleration in neutron star ultra-strong electromagnetic fields}

\author{Ivan Tomczak\aff{1}
  \corresp{\email{ivan.tomczak@astro.unistra.fr}}
 \and Jérôme Pétri\aff{1}  
  \corresp{\email{jerome.petri@astro.unistra.fr}}}

\affiliation{\aff{1}Université de Strasbourg, CNRS, Observatoire astronomique de Strasbourg, UMR 7550, F-67000, France.}

\begin{document}

\maketitle

\begin{abstract}

%\revone{Referee: 1. First, it should be emphasized in the abstract and the introduction, that these results concern non-self-consistent simulations, i.e. without feedback of particles on the electromagnetic fields.}

In this paper, we discuss the results of a new particle pusher in realistic ultra-strong electromagnetic fields as those encountered around rotating neutron stars. After presenting results of this algorithm in simple fields and comparing them to expected exact analytical solutions, we present new simulations for a rotating magnetic dipole in vacuum for a millisecond pulsar by using Deutsch solution. Particles are injected within the magnetosphere, neglecting radiation reaction, interaction among them and their feedback on the fields. Our simulations are therefore not yet fully self-consistent because Maxwell equations are not solved according to the current produced by these particles. The code highlights the symmetrical behaviour of particles of opposite charge to mass ratio~$q/m$ with respect to the north and south hemispheres. The relativistic Lorentz factor of the accelerated particles is proportional to this ratio~$q/m$: protons reach up to $\gamma_p \simeq 10^{10.7}$, whereas electrons reach up to $\gamma_e \simeq 10^{14}$.

Our simulations show that particles could be either captured by the neutron star, trapped around it, or ejected far from it, well outside the light-cylinder. Actually, for a given charge to mass ratio, particles follow similar trajectories. These particle orbits show some depleted directions, especially at high magnetic inclination with respect to the rotation axis for positive charges and at low inclination for negative charges because of symmetry. Other directions are preferred and loaded with a high density of particles, some directions concentrating the highest or lowest acceleration efficiencies.
\end{abstract}

\section{Introduction}
\label{sec:Intro}

For massive enough stars, nuclear fusion is able to produce elements heavier than carbon and oxygen up to the production of the most stable nucleus, iron~\ce{^56Fe}. At this point the core of the star collapses and the outer layers fall onto the core, triggering a supernova explosion leaving behind a compact remnant which can either be a black hole or a neutron star. 
In this paper, we focus on strongly magnetized and fast rotating neutron stars. These compact objects are believed to be among the most efficient accelerators in the universe, producing ultra-relativistic particles (electrons, positrons, protons and maybe ions). In its simplest description, a neutron star is described thanks to only a few parameters, namely
\begin{itemize}
\item the inclination angle or obliquity~$\chi$ which represent the angle between the rotation axis and magnetic axis.
\item the angular speed~$\Omega$ which defines the light cylinder radius $\rlight=c/\Omega$, the distance at which an object in co-rotation with the star reaches the speed of light~$c$.
\item the radius of the neutron star $R$.
\item the magnetic field strength~$B$ at the surface on the equator.
\end{itemize}
The extreme magnetic fields of these remnant stars range from $B\simeq10^5~\text{T}$ to $B\simeq10^{10}~\text{T}$, which coupled with the angular speed between $\Omega \simeq 1000 ~\text{rad.s}^{-1}$ and $\Omega \simeq 0.1~\text{rad.s}^{-1}$ generates an intense electric field~$E$, of the order $\Omega\,B\,R \approx 10^{12}$~V/m, accelerating particles around the neutron star to ultra-relativistic Lorentz factors. 
In our applications, the neutron star mass~$M$ is irrelevant because the gravitational force exerted on charged particles $F_g$ is negligible compared to the Lorentz force $F_L$. Indeed, both forces have a typical intensity of $F_g = G \, \frac{M\, m_p}{R^2} \simeq 3.17\times10^{-15}~\text{N}$ for a proton at the surface of the neutron star compared to $F_L = q \, E = q \, \Omega R \, B \simeq 5.01\times10^{-7}~\text{N}$. The ratio of these forces is 
$$\frac{F_g}{F_L} = \frac{G \, M \, m_p}{R^2 \, q \, E} = 6.33 \times 10^{-9} \ll 1 $$
so the electromagnetic force is $\sim 10^{8}$ times stronger than the gravitational force for protons and a factor $m_p/m_e$ larger for electrons and positrons.

%\revtwo{Referee : 1) The introduction mentions the Boris (1970) and Vay (2008) schemes, but several other schemes have meanwhile been proposed, and even thoroughly compared, e.g. in Ripperda et al., ApJ Supplement 235, 21 (2018). This and other relevant papers should therefore be referenced. In the Ripperda et al. paper, a new, fully implicit, energy conserving scheme is discussed, and tested in various regimes (that admittedly do not reach the extreme Lorentz factors of the application discussed further).}

Most numerical simulations of neutron star magnetospheres use the \cite{boris_relativistic_1970} algorithm or better the \cite{vay_simulation_2008} algorithm. However these algorithms are not well suited for ultra-strong electromagnetic fields, forcing some authors to lower the true field strengths to unrealistically low values in the simulations. Although a scaling is sometimes applied to get more realistic results, such scaling cannot be straightforwardly extrapolated for instance when radiation reaction is included because of the non linearities introduced by radiative feedback \citep{vranic_classical_2016}. Moreover, Lorentz factors reached by the particles near pulsars hardly exceed $\gamma=10^4$ with those algorithms, see for instance~\citet{ElectronPositronPairFlow}, \citet{PhilippovSpitkovsky}, \citet{GuepinCeruttiKotera} and \citet{Kalapotharakos}. 

Several attempts have been proposed to faithfully follow particles in relativistic regimes, trying to conserve energy and momentum. In this spirit \cite{zenitani_boris_2018} improved the \cite{boris_relativistic_1970} particle pusher by computing the exact analytical rotation in the magnetic field and getting better accuracy. The Boris algorithm is however popular because of its simplicity and accuracy. Moreover its stability property takes its root in its phase space volume preserving properties as discussed by \cite{qin_why_2013}. \cite{umeda_three-step_2018} also improved the Boris algorithm by employing a three stage step. Relativistic simulations are even more stringent about numerical error accumulation and volume preserving schemes are highly recommended as pointed out by \cite{zhang_volume-preserving_2015}.
	
Efficient particle pushers in ultra-strong electromagnetic fields are fundamental to simulate the neutron star electrodynamics. For a comprehensive comparison of relativistic particle integrators, see the extensive work done by \cite{ripperda_comprehensive_2018}. They carefully compared the merit of the standard Boris algorithm \cite{boris_relativistic_1970}, the \cite{vay_simulation_2008} implicit scheme in space velocity, the \cite{higuera_structure-preserving_2017} second order method and the implicit midpoint method described in \cite{lapenta_particle_2011}. This leapfrog scheme already appeared in \cite{verboncoeur_particle_2005}. A fully implicit update in space and velocity parameters for relativistic particle integrators relying on \cite{vay_simulation_2008} velocity advance has been explored by \cite{petri_fully_2017-1}. There it has been shown that catching properly and accurately the simple electric drift motion in an ultra relativistic regime remains extremely difficult to achieve. Unfortunately, neutron star magnetospheres are common places for such relativistic drift velocities. It is therefore compulsory to design efficient and accurate numerical schemes to faithfully follow these trajectories. It represents a crucial step towards realistic particle acceleration and radiation in ultra-strong electromagnetic fields. Some tests of particle acceleration in a plane electromagnetic wave using standard pushers and reported by \cite{arefiev_temporal_2015} showed severe limitations in the accuracy already for modest field strengths with strength parameter $a\approx 20$ (see eq.~\ref{eq:strength_parameter}).

The limitation arises from the huge span in time scales, from the gyroperiod frequency $\omega_B=qB/m$ to the stellar rotation frequency $\Omega$ and synthesised by the strength parameter 
\begin{equation}\label{eq:strength_parameter}
a = \frac{\omega_B}{\Omega} = \frac{qB}{m\Omega}\simeq10^{10} .
\end{equation}
for a proton near a millisecond pulsar ($\Omega=10^{3}~\text{rad.s}^{-1}$ and $B=10^{5}~\text{T}$), the situation being worst for electron/positron pairs. This ratio corresponds to the number of gyrations made by a particle during the timescale of evolution of the electromagnetic field due to the stellar rotation. It shows that the difference in timescales makes it almost impossible to compute the trajectory of particles in a reasonable amount of time, since billions of time steps are needed on the pulsar period timescale. To tackle this issue, our aim is to propose a new technique based on analytical solutions of the Lorentz force equation in constant and ultra-strong electromagnetic fields meanwhile with an acceptable computational time and most importantly avoiding a necessary scaling, allowing particles to reach high Lorentz factors with realistic fields.

Neutron stars are known to act as unipolar inductors, generating huge electric potential drops between the poles and the equator, of the order
\begin{equation}\label{eq:Potential}
 \Delta \phi = \Omega \, B \, R^2 \approx 10^{16}~V.
\end{equation}
As a consequence, it expels electrons, maybe protons and ions, filling the magnetosphere with charged particles. Typical Lorentz factor for electrons in this static field are therefore 
\begin{equation}\label{eq:FacteurLorentz}
 \gamma = \frac{e \, \Delta\phi}{m_e\,c^2} \approx 10^{10} .
\end{equation}
If the particle injection rate is high enough, this plasma will screen the electric field, drastically mitigating the potential drop $\Delta \phi$ and the acceleration efficiency. Resistive \citep{li_resistive_2012} and PIC simulations \citep{cerutti_particle_2015} showed that indeed only a small fraction of the full potential is available. However, for low particle injection rates, the plasma is unable to screen the electric field and the full potential drop develops. The magnetosphere is then almost empty and known as an electrosphere \citep{krause-polstorff_electrosphere_1985, petri_global_2002}. Such electrospheres are the subject of the present paper. They represent inactive pulsars able to accelerate particles to ultra-relativistic speeds. We aim at accurately quantifying the final Lorentz factor reached by the outflowing plasma in this large amplitude low frequency electromagnetic wave. Similar studies have be performed by \cite{MichelLiElec} staying however on a more analytical side.

The outline of the paper is as follows. First in section~\ref{sec:Algo} we briefly remind the principle of the algorithm. Next in section~\ref{sec:Verif} we show some results obtained in fields where an analytical solution is known before discussing the convergence to the exact solution in section~\ref{sec:convergence}. Results of the simulations near pulsars are presented in section~\ref{sec:Deutsch}. Some conclusions are drawn in section~\ref{sec:Conclusion}.

\section{Outline of the numerical algorithm}
\label{sec:Algo}

In this section we summarize the scheme of our algorithm. A more complete and careful description can be found in \cite{petri_relativistic_2020} with a more comprehensive introduction pointing to appropriate references. The code is based on successive exact analytical solutions for the trajectory of a relativistic particle in a constant but otherwise arbitrary electromagnetic field $(\mathbf{E}, \mathbf{B})$. Switching to a frame where $\mathbf{E}$ and $\mathbf{B}$ are parallel and aligned with the $\textbf{e}_z$ axis, integration of the Lorentz force in a Cartesian coordinate system $(\textbf{e}_x, \textbf{e}_y, \textbf{e}_z)$ leads to the 4-velocity given by
\begin{subequations}
	\label{eq:4velocity}
	\begin{align}
	u^0 & = \gamma_0 c [\cosh (\omega_E \tau) + \beta_0^z \sinh (\omega_E \tau)]  \\
	u^1 & = \gamma_0 c [\beta_0^x \cos(\omega_B \tau) + \beta_0^y \sin (\omega_B \tau)]  \\
	u^2 & = \gamma_0 c [-\beta_0^x \sin(\omega_B \tau) + \beta_0^y \cos (\omega_B \tau)]  \\
	u^3 & = \gamma_0 c [\sinh (\omega_E \tau) + \beta_0^z \cosh (\omega_E \tau)]  
	\end{align}
\end{subequations}
and to the 4-position 
\begin{subequations}
	\label{eq:4position}
	\begin{align}
	c(t-t_0) & = \dfrac{\gamma_0 c}{\omega_E} [\sinh (\omega_E \tau) + \beta_0^z \cosh (\omega_E \tau) - \beta_0^z]  \\
	x-x_0 & = \dfrac{\gamma_0 c}{\omega_B} [\beta_0^x \sin(\omega_B \tau) - \beta_0^y \cos (\omega_B \tau) +\beta_0^y]  \\
	y-y_0 & = \dfrac{\gamma_0 c}{\omega_B} [\beta_0^x \cos(\omega_B \tau) -\beta_0^x + \beta_0^y \sin (\omega_B \tau)]  \\
	z-z_0 & = \dfrac{\gamma_0 c}{\omega_E} [\cosh (\omega_E \tau) -1 + \beta_0^z \sinh (\omega_E \tau)]    .
	\end{align}
\end{subequations}
$\gamma_0$ is the initial Lorentz factor, $\boldsymbol{\beta}_0$ the initial velocity normalized to the speed of light~$c$, $\omega_E = q\,E / mc$ and $\omega_B=q\,B/m$. Integration is performed according to the particle proper time~$\tau$. The initial 4-position is $(c\,t_0, x_0,y_0,z_0)$.

%\revtwo{2) The frame with parallel E and B is not possible when the fields are orthogonal. What (numerical) criterion is actually used to distinguish orthogonal versus non-orthogonal situations?}

Note that our treatment proposed above works also for orthogonal fields but not for light-like fields for which both electromagnetic invariants vanish, $I_1=\mathbf{E} \cdot \mathbf{B} = 0$ and $I_2=E^2-c^2\,B^2=0$. For non light-like fields, we can use the same algorithm, there always exist a frame where either the magnetic field or the electric field vanishes, obtained by a Lorentz boost. For instance the vanishing magnetic field case $\textbf{B}=\textbf{0}$ is obtained by taking the limiting $\lim_{\omega_B \to 0}$ of the above expression thus leading to the 4-velocity
\begin{subequations}
	\label{eq:4velocity_B=0}
	\begin{align}
	u^0 & = \gamma_0 c [\cosh (\omega_E \tau) + \beta_0^z \sinh (\omega_E \tau)]  \\
	u^1 & = \gamma_0 c \beta_0^x \\
	u^2 & = \gamma_0 c \beta_0^y \\
	u^3 & = \gamma_0 c [\sinh (\omega_E \tau) + \beta_0^z \cosh (\omega_E \tau)]  
	\end{align}
\end{subequations}
and to the 4-position:
\begin{subequations}
	\label{eq:4position_B=0}
	\begin{align}
	c(t-t_0) & = \dfrac{\gamma_0 c}{\omega_E} [\sinh (\omega_E \tau) + \beta_0^z \cosh (\omega_E \tau) - \beta_0^z]  \\
	x-x_0 & = \gamma_0 c \beta_0^x \tau  \\
	y-y_0 & = \gamma_0 c \beta_0^y \tau  \\
	z-z_0 & = \dfrac{\gamma_0 c}{\omega_E} [\cosh (\omega_E \tau) -1 + \beta_0^z \sinh (\omega_E \tau)]    .
	\end{align}
\end{subequations}
In the same way, the vanishing electric field $\textbf{E}=\textbf{0}$ is treated with the limit $\lim_{\omega_E \to 0}$, leading to the 4-velocity
\begin{subequations}
	\label{eq:4velocity_E=0}
	\begin{align}
	u^0 & = \gamma_0 c  \\
	u^1 & = \gamma_0 c [\beta_0^x \cos(\omega_B \tau) + \beta_0^y \sin (\omega_B \tau)]  \\
	u^2 & = \gamma_0 c [-\beta_0^x \sin(\omega_B \tau) + \beta_0^y \cos (\omega_B \tau)]  \\
	u^3 & = \gamma_0 c \beta_0^z  
	\end{align}
\end{subequations}
and to the 4-position 
\begin{subequations}
	\label{eq:4position_E=0}
	\begin{align}
	c(t-t_0) & = \gamma_0 c \tau  \\
	x-x_0 & = \dfrac{\gamma_0 c}{\omega_B} [\beta_0^x \sin(\omega_B \tau) - \beta_0^y \cos (\omega_B \tau) +\beta_0^y]  \\
	y-y_0 & = \dfrac{\gamma_0 c}{\omega_B} [\beta_0^x \cos(\omega_B \tau) -\beta_0^x + \beta_0^y \sin (\omega_B \tau)]  \\
	z-z_0 & = \gamma_0 c \beta_0^z \tau  .
	\end{align}
\end{subequations}
A full derivation of the equations in all field configurations is exposed in \cite{gourgoulhon_relativite_2010}.
In order to decide whether the field is null-like or not, we compare both invariant with the field strength, applying a threshold $\epsilon\ll1$ such that if $|\mathbf{E} \cdot \mathbf{B}|<\epsilon\|\mathbf{E}\|\, \|\mathbf{B}\|$ and $|E^2-c^2\,B^2|<\epsilon\,(E^2+c^2\,B^2)$ the field is said to be light like or null. As long as $\epsilon$ is chosen small enough about $10^{-8}$ or $10^{-12}$, the results are insensitive to the precise value of the threshold prescription. Moreover, in most physical applications, the light-like field is rarely met over a large volume in space-time. It happens to be true sometimes at some points but becoming very quickly again non light-like. Especially for our forthcoming applications to PIC code, such condition is highly unlikely. In some of our tests, we impose the external electromagnetic field for instance for a plane wave. In such a case the field is and remains light-like without the plasma feedback, but this situation is highly idealized.

%\revone{3. I imagine that the author considers, for a particle starting from the point $(t^n,x_i)$, that this particle is subjected to the constant electromagnetic field  (E,B)$(t^n,x_i)$ during a proper-time integration step ? May be this should be clearly written in the paper.}

The above mentioned algorithm clearly makes the assumption that the electric and the magnetic fields are constant during a time step $\Delta\tau$ and evaluated at the particle 4-position~$(t^n,x_i)$. % so as to be able to use the equations \eqref{eq:4position}, \eqref{eq:4velocity}, \eqref{eq:4position_B=0}, \eqref{eq:4velocity_B=0}, \eqref{eq:4position_E=0}, \eqref{eq:4velocity_E=0}, \eqref{eq:4position_light_like} and \eqref{eq:4velocity_light_like}.
The algorithm switches between two reference frames, the first one identified as the observer frame (R) and the second producing an electric field aligned with the magnetic field (R'). Note that a vanishing electric or magnetic field is only a special case included in the former.

At first, in the reference frame (R) of the distant observer, the 4-position $\textit{\textbf{X}}$ and 4-velocity $\textit{\textbf{U}}$ of the particle is known as well as the electric field $\textit{\textbf{E}}$ and the magnetic field $\textit{\textbf{B}}$ at the particle's position. We then search for a reference frame (R') in which $\textit{\textbf{E'}}$ and $\textit{\textbf{B'}}$ are parallel (or, if $\textit{\textbf{E}} \boldsymbol{\cdot} \textit{\textbf{B}}=0$, where $\textit{\textbf{B'}}$ or $\textit{\textbf{E'}}$ vanishes depending on the relative strength of both fields). The speed of frame (R') with respect to frame (R) is noted $\textit{\textbf{V}}$. We align the z-axis of (R) with $\textit{\textbf{V}}$ via rotation through the Euler angles, performing a rotation with help on the matrix $\mathsfbi{M_1}$. A Lorentz boost $\mathsfbi{\Lambda}(\textit{\textbf{V}})$ switches to the reference frame (R'). In this latter frame the particle's 4-position is $\textit{\textbf{X'}}$ and its 4-velocity $\textit{\textbf{U'}}$. We then rotate the axes to align the new z'-axis with $\textit{\textbf{E}}$ and $\textit{\textbf{B}}$ with the rotation matrix $\mathsfbi{M_2}$.
We next update $\textit{\textbf{X'}}$ and $\textit{\textbf{U'}}$ to the new timestep ($\tau + \Delta \tau$) thanks to the equations (\ref{eq:4velocity}) and (\ref{eq:4position}) describing the trajectory of a particle where the electric and magnetic fields are parallel to the z-axis and assumed to be constant during $\Delta \tau$, with $\tau$ the proper time of the particle, $u^i$ the i$^{th}$ component of the four-velocity, $\boldsymbol\beta$ the normalised speed of the particle. Quantities with subscript 0 are initial conditions, and superscripts x, y or z denote a projection on the x, y or z-axis respectively. Once $\textit{\textbf{X'}}$ and $\textit{\textbf{U'}}$ computed, the frames axes are set back to their initial positions with inverse rotations given by the matrix $\mathsfbi{M_2^{-1}}$, the inverse of the $\mathsfbi{M_2}$ rotation matrix. 
We boost back to the reference frame (R) with the inverse Lorentz transformation $\mathsfbi{\Lambda}(\textit{\textbf{-V}})$, since in (R') the speed of (R) is $\textit{\textbf{-V}}$.
And again, we align the axes of (R) with their initial positions thanks to the rotation matrix $\mathsfbi{M_1^{-1}}$ finding the particle's new 4-position $\textit{\textbf{X}}$ and 4-velocity $\textit{\textbf{U}}$.

The special case of the light like field, $\mathbf{E} \cdot \mathbf{B} = 0$ and $E^2 = c^2 \, B^2$, must be treated separately because there exist no frame where $\mathbf{E}$ and $\mathbf{B}$ are parallel. In the code, to consider that the field is light-like we introduce $\epsilon \ll 1$ and we verify $\vert E^2 - c^2 B^2\vert < \epsilon \vert E^2 + c^2 B^2\vert$ %(checking that the fields have the same norms, to a factor $c$) 
and $\vert \textbf{E} \cdot \textbf{B} \vert < \epsilon \vert \textbf{E} \vert \vert \textbf{B}\vert $ (verification that the cosine of the angle between $\textbf{E}$ and $\textbf{B}$ is close to 0).
Exact solutions can be found in \cite{petri_relativistic_2020} but they are reminded here for $\omega_B=\omega_E$. In \cite{petri_relativistic_2020} we allow for a more general configuration with  $\omega_B=\pm\omega_E$ taking into account possible left or right handed coordinate axes. With $\textbf{B}$ along $\textbf{e}_z$ and $\textbf{E}$ along $\textbf{e}_y$, the 4-velocity can directly be obtained by
\begin{subequations}
	\label{eq:4velocity_light_like}
	\begin{align}
	u^0 & = \gamma_0 c [1+(1-\beta_0^x)\dfrac{(\omega_B \tau)^2}{2}+\beta_0^y \omega_B \tau]  \\
	u^1 & = \gamma_0 c [\beta_0^x+(1-\beta_0^x)\dfrac{(\omega_B \tau)^2}{2}+\beta_0^y \omega_B \tau]  \\
	u^2 & = \gamma_0 c [\beta_0^y+(1-\beta_0^x)\omega_B \tau]  \\
	u^3 & = \gamma_0 c \beta_0^z 
	\end{align}
\end{subequations}
as well as the 4-position 
\begin{subequations}
	\label{eq:4position_light_like}
	\begin{align}
	c(t-t_0) & =  \gamma_0 c [\tau + (1-\beta_0^x)\dfrac{\omega_B^2 \tau^3}{6} + \beta_0^y \dfrac{\omega_B \tau^2}{2}]  \\
	x-x_0 & = \gamma_0 c [\beta_0^x \tau + (1-\beta_0^x)\dfrac{\omega_B^2 \tau^3}{6} + \beta_0^y \dfrac{\omega_B \tau^2}{2}] \\
	y-y_0 & = \gamma_0 c [\beta_0^y \tau + (1-\beta_0^x) \dfrac{\omega_B \tau^2}{2}] \\
	z-z_0 & = \gamma_0 c \beta_0^z \tau  .
	\end{align}
\end{subequations}
Here the algorithm is simpler: the 4-position $\textbf{X}$, the 4-velocity $\textbf{U}$ of the particle are known, as well as $\textbf{E}$ and $\textbf{B}$ felt by the particle. We then perform a rotation of matrix $\mathsfbi{M}$ to align $\textbf{E}$ with $\textbf{e}_z$ and $\textbf{B}$ with $\textbf{e}_y$. Next we directly update the particles 4-position $\textbf{X}$ and 4-velocity $\textbf{U}$ to the next timestep $(\tau+\Delta\tau)$ and we then rotate back to the initial set of axes with $\mathsfbi{M^{-1}}$.
%\revtwo{2') Is the result sensitive to minor changes in this switching prescription?}

It is important to note that this peculiar case is very uncommon in the simulations mentioned in \ref{sec:Deutsch}. Indeed, when choosing $\epsilon = 10^{-15}$, the algorithm never used this peculiar case. In fact, the light-like case is only found in very precise positions or far from the neutron star, at distances the particles did not reached by the end of the simulations and this part of the algorithm was added so as to be able to work in any field, especially plane waves.

%\revtwo{2'') On a similar note, the use of the analytic trajectory prescriptions works only as the fields stay truly constant over the trajectory part that is being considered, and "this assumption is verified to a certain accuracy": can the authors explain a bit more how delicate this accuracy prescription is? Does it need to be enforced at machine precision, or is a relative error of some prescribed magnitude sufficient?}

Technically, for the computation of the new velocity and position, a convergence criterion is applied on the electromagnetic field to make sure that the assumption $\textit{\textbf{E}}$ and $\textit{\textbf{B}}$ remaining almost constant is verified to a prescribed accuracy. To do so, we use a loop on the $\textbf{E}$ and $\textbf{B}$ fields: at first we move the particle at the 4-position $(c\,t, x, y, z )$ by taking $\textbf{E}_0=\textbf{E}(t,x,y,z)$ and $\textbf{B}_0=\textbf{B}(t,x,y,z)$, so the particle's 4-position becomes $(c\,(t+dt_0), x+dx_0, y+dy_0, z+dz_0 )$.
Then, we make the particle start again from $(c\,t, x, y, z )$ but we take the averaged fields: $\textbf{E}_1=\textbf{E}(t+dt_0/2,x+dx_0/2,y+dy_0/2,z+dz_0/2)$ and $\textbf{B}_1=\textbf{B}(t+dt_0/2,x+dx_0/2,y+dy_0/2,z+dz_0/2)$ for the incrementation of the time, updating the 4-position: $(c\,(t+dt_1), x+dx_1, y+dy_1, z+dz_1)$. Then again, the particle starts from $(c\,t, x, y, z )$ and we increment the time step with the fields being $\textbf{E}_2=\textbf{E}(t+dt_1/2,x+dx_1/2,y+dy_1/2,z+dz_1/2)$ and $\textbf{B}_2=\textbf{B}(t+dt_1/2,x+dx_1/2,y+dy_1/2,z+dz_1/2)$. Generally, for $k \neq0$ incrementing the time step with the field $\textbf{E}_k$ and $\textbf{B}_k$ returns the 4-position $(c\,( t+dt_k), x+dx_k, y+dy_k, z+dz_k )$, and the fields are defined as $\textbf{E}_{k+1}=\textbf{E}(t+dt_k/2,x+dx_k/2,y+dy_k/2,z+dz_k/2)$ and $\textbf{B}_{k+1}=\textbf{B}(t+dt_k/2,x+dx_k/2,y+dy_k/2,z+dz_k/2)$.
We keep the loop going while $\textbf{B}_k \neq \textbf{B}_{k-1}$ and $\textbf{E}_k \neq \textbf{E}_{k-1}$, which numerically translates to $\vert E_k^j-E_{k-1}^j \vert  > \epsilon (\vert E_{k}^j \vert + \vert E_{k-1}^j \vert) $ and $\vert B_k^j-B_{k-1}^j \vert > \epsilon (\vert B_{k}^j \vert + \vert B_{k-1}^j \vert) $, with $\epsilon \ll 1$ (we chose $\epsilon=10^{-4}$), and $ j=\{x,y,z\}$.

As a check of the implementation of our new algorithm, simulations were performed either in a constant and uniform field, or in a plane electromagnetic wave. Once passing the tests, we applied our code to realistic astrophysical cases of rotating neutron stars using the  \cite{deutsch_electromagnetic_1955} field solution, neglecting so far radiation reaction that will be included in a next step.

\section{Test cases}
\label{sec:Verif}

We carried some simulations to test the code in cases where the trajectory of the particle is known analytically. Simple checks with only an electric or magnetic field or both were made, allowing for orthogonal or parallel geometries.

When the $\textit{\textbf{E}}$ and $\textit{\textbf{B}}$ fields are parallel, in addition to the gyration around the magnetic field lines, the particle is accelerated along the electric field, so that particles describe an helix getting elongated as time goes, if not at first slowing down if its initial velocity is in a direction opposite to its acceleration.

\begin{figure}
\begin{subfigure}{.5\textwidth}
  \centering
  \includegraphics[scale=0.4]{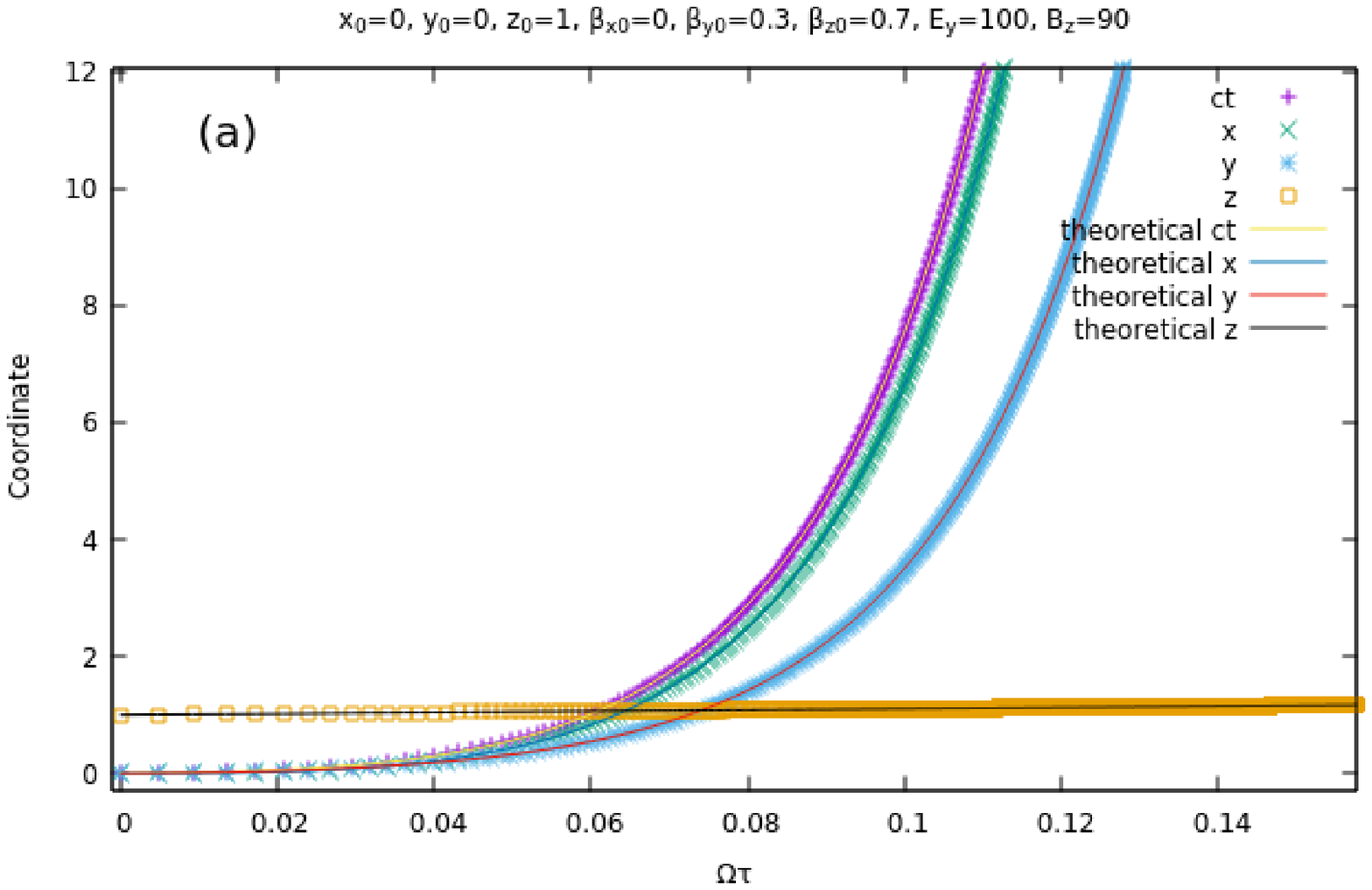}
%  \caption{Simulated and theoretical coordinates of a particle as function of its proper time $\tau$, normalised units}
\label{fig:E_dominant_xyzt}
\end{subfigure}
\begin{subfigure}{.5\textwidth}
  \centering
  \includegraphics[scale=0.4]{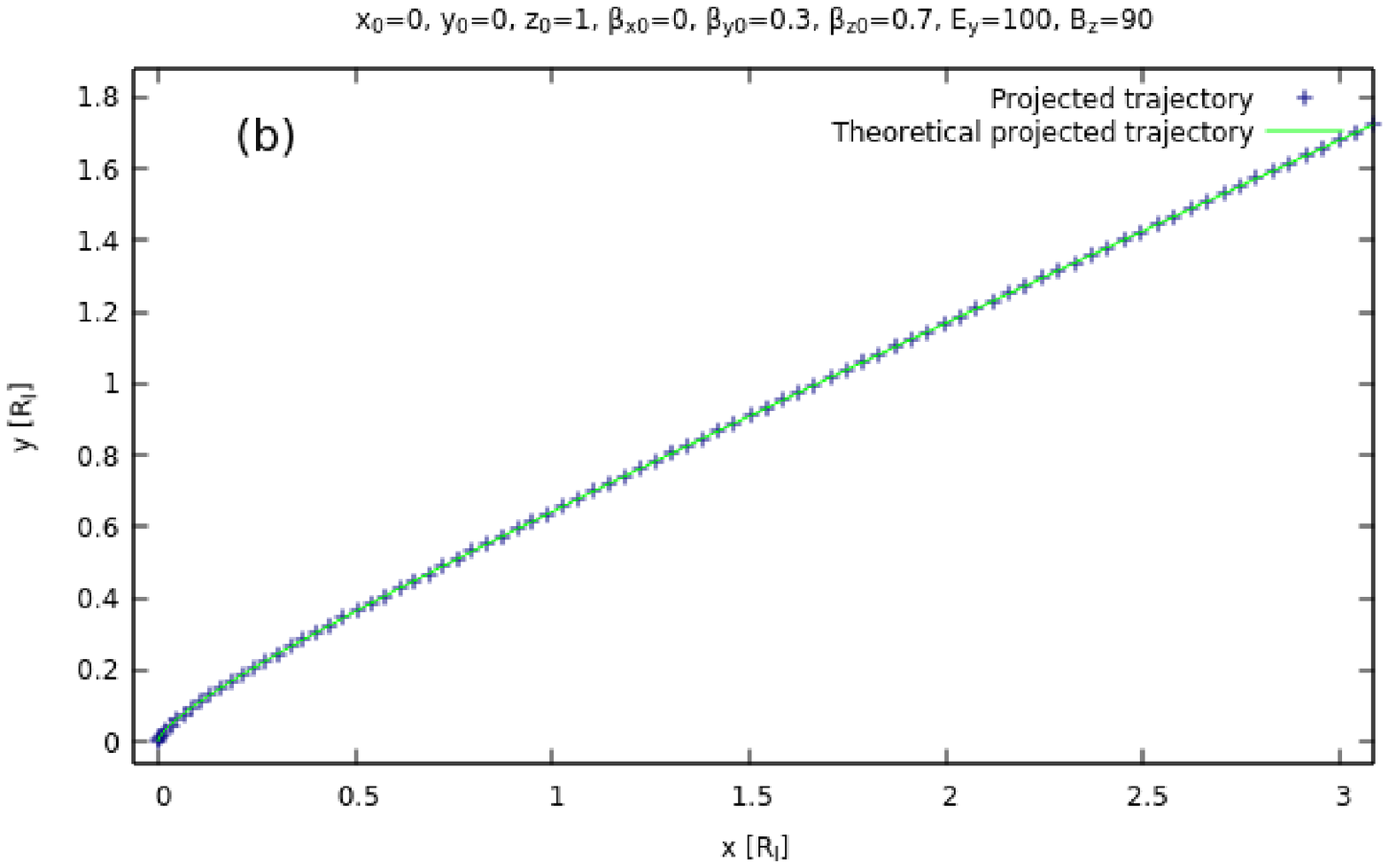}
%  \caption{Projected trajectory of a particle, normalised units}
\label{fig:E_dominant}
\end{subfigure}
\caption{Evolution of a particle in a dominant electric field, $x_0=0$, $y_0=0$, $z_0=1$, $\beta^x_{0}=0$, $\beta^y_{0}=0.3$, $\beta^z_{0}=0.7$, $\frac{qE_y}{cm}=100$, $\frac{qB_z}{m}=90$. Theoretical and simulated 4-position as function of the proper time $\tau$ (a). Projection of the trajectory of the particle in the x-y plane (b).}
\label{fig:E_dom}
\end{figure}

\begin{figure}
\begin{subfigure}{.5\textwidth}
  \centering
  \includegraphics[scale=0.4]{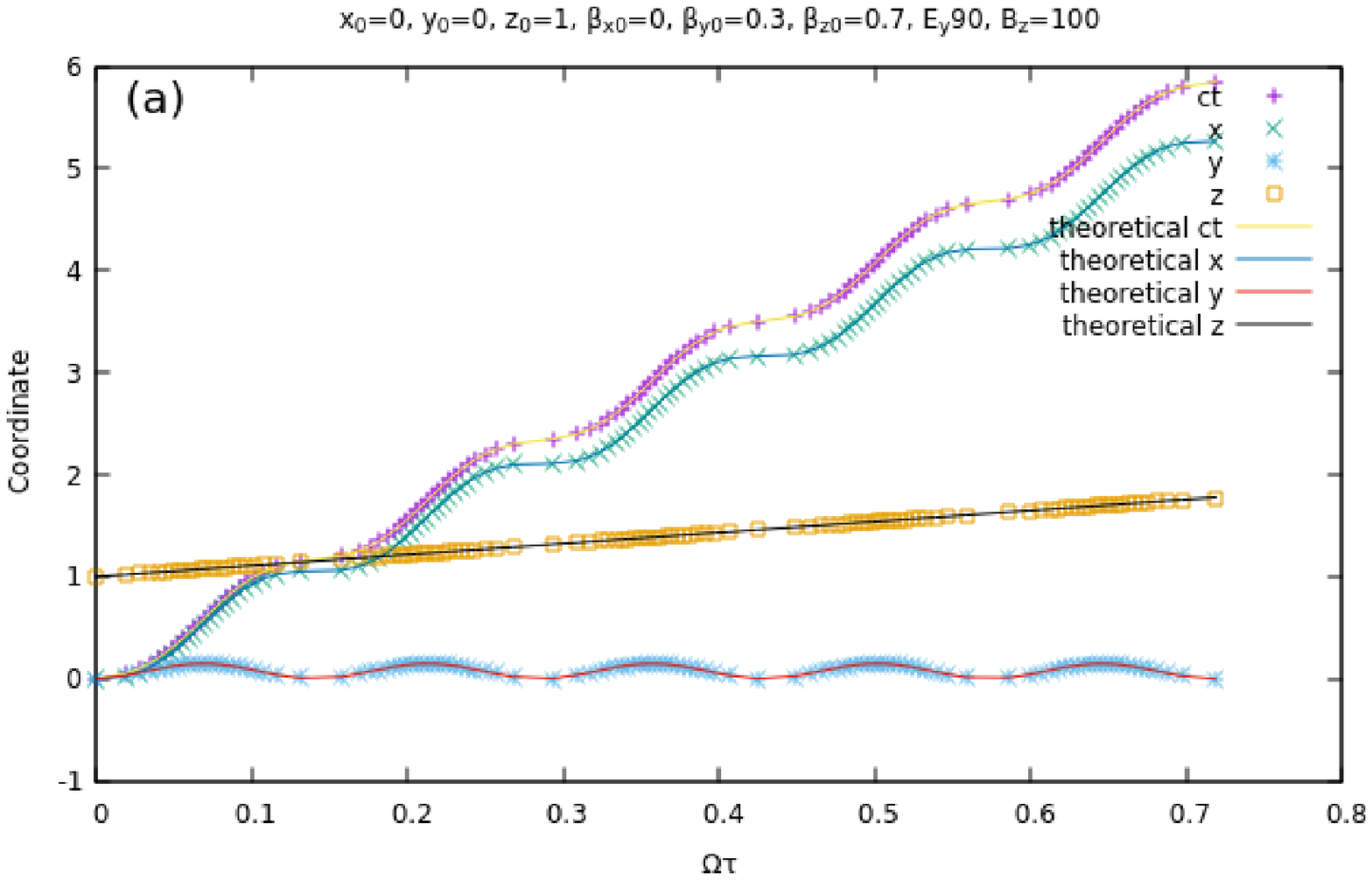}
%  \caption{Simulated and theoretical coordinates of a particle as function of its proper time $\tau$, normalised units}
\label{fig:B_dominant_xyzt}
\end{subfigure}
\begin{subfigure}{.5\textwidth}
  \centering
  \includegraphics[scale=0.4]{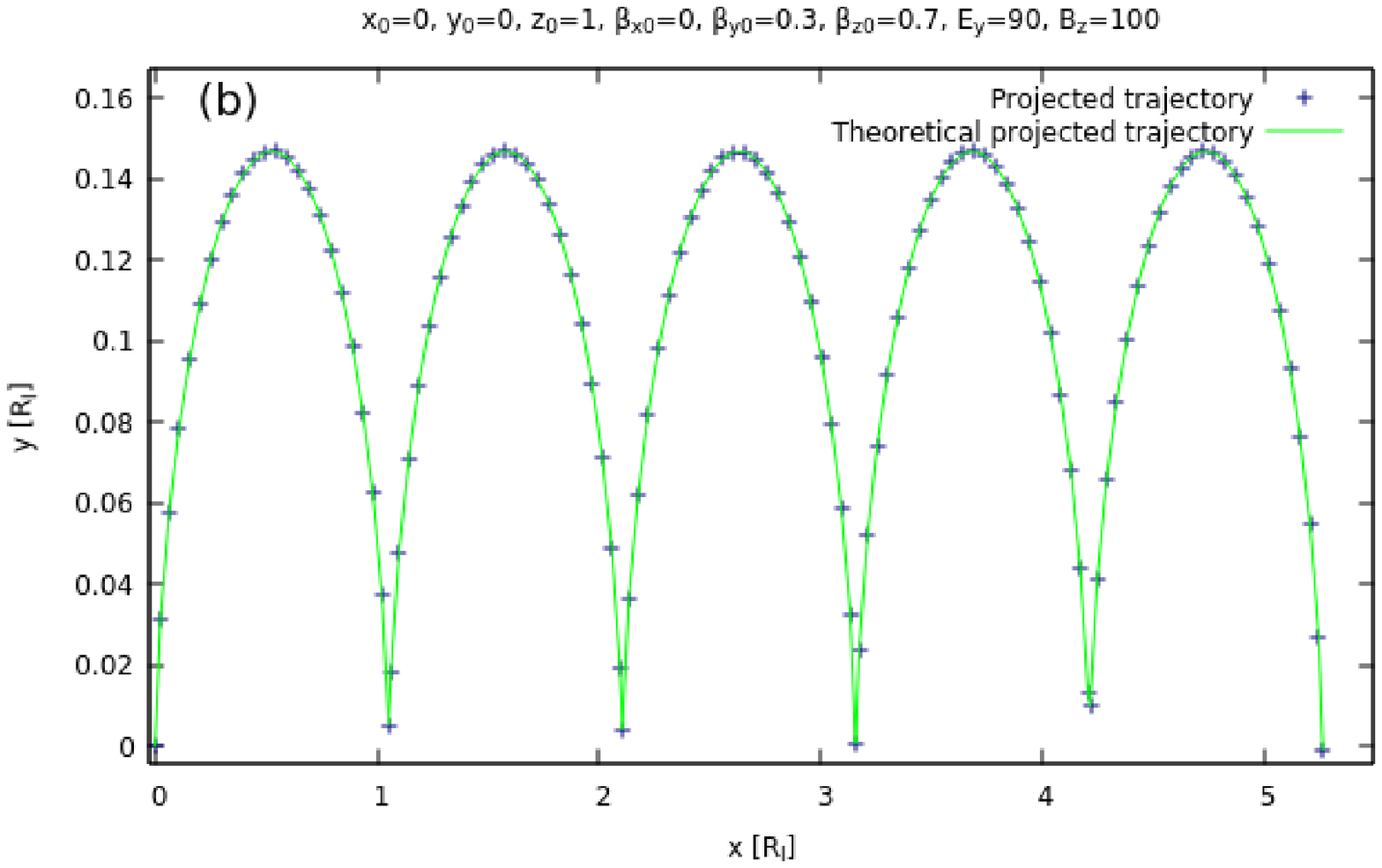}
%  \caption{Projected trajectory of a particle, normalised units}
\label{fig:B_dominant}
\end{subfigure}
\caption{Evolution of a particle in a dominant magnetic field, $x_0=0$, $y_0=0$, $z_0=1$, $\beta^x_{0}=0$, $\beta^y_{0}=0.3$, $\beta^z_{0}=0.7$, $\frac{qE_y}{mc}=90$, $\frac{qB_z}{m}=100$. Theoretical and simulated 4-position as function of the proper time $\tau$ (a). Projection of the trajectory of the particle in the x-y plane (b).}
\label{fig:B_dom}
\end{figure}

\begin{figure}
\begin{subfigure}{.5\textwidth}
  \centering
  \includegraphics[scale=0.4]{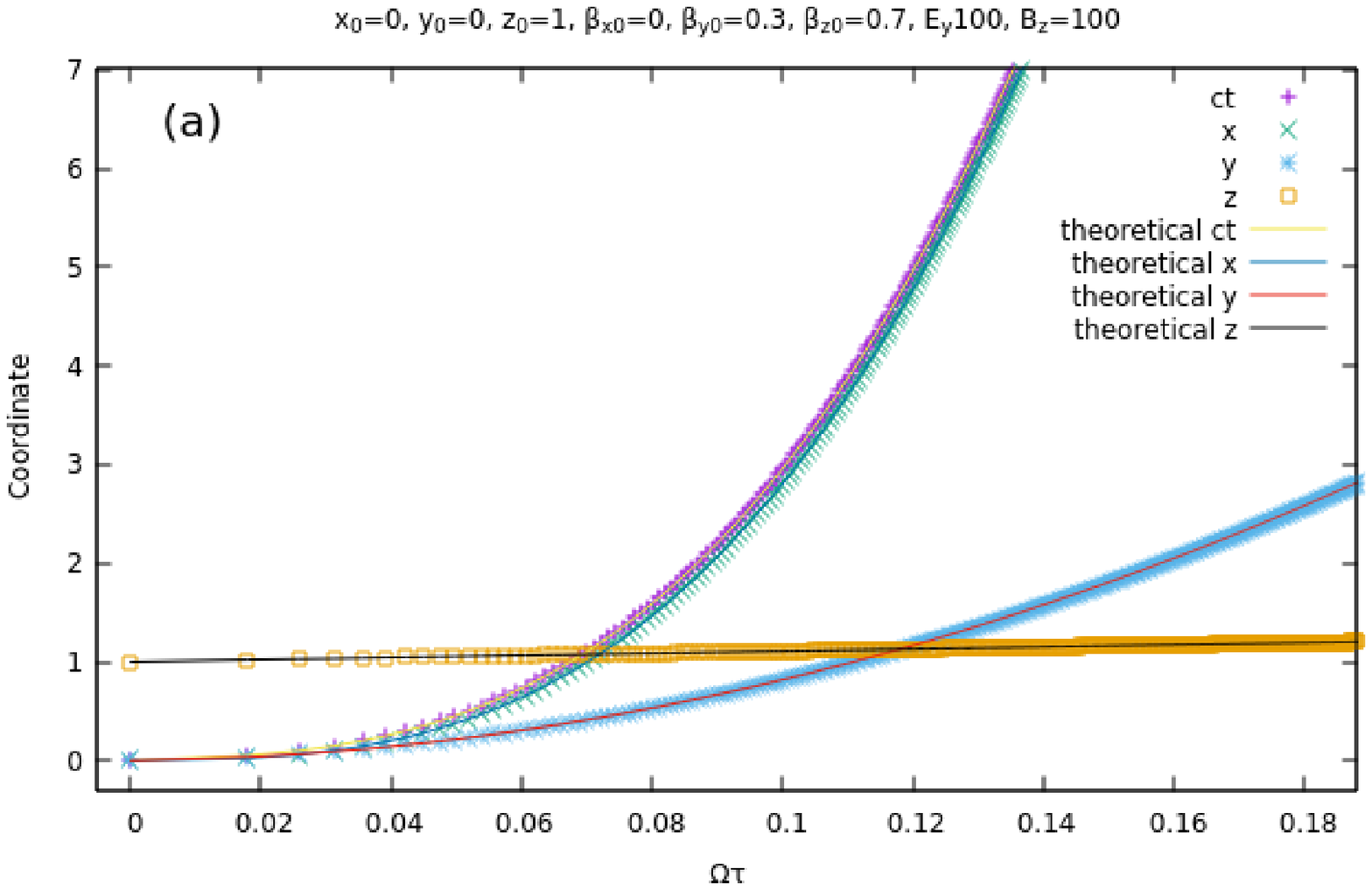}
%  \caption{Simulated and theoretical coordinates of a particle as function of its proper time $\tau$, normalised units}
\label{fig:Light_like_xyzt}
\end{subfigure}
\begin{subfigure}{.5\textwidth}
  \centering
  \includegraphics[scale=0.4]{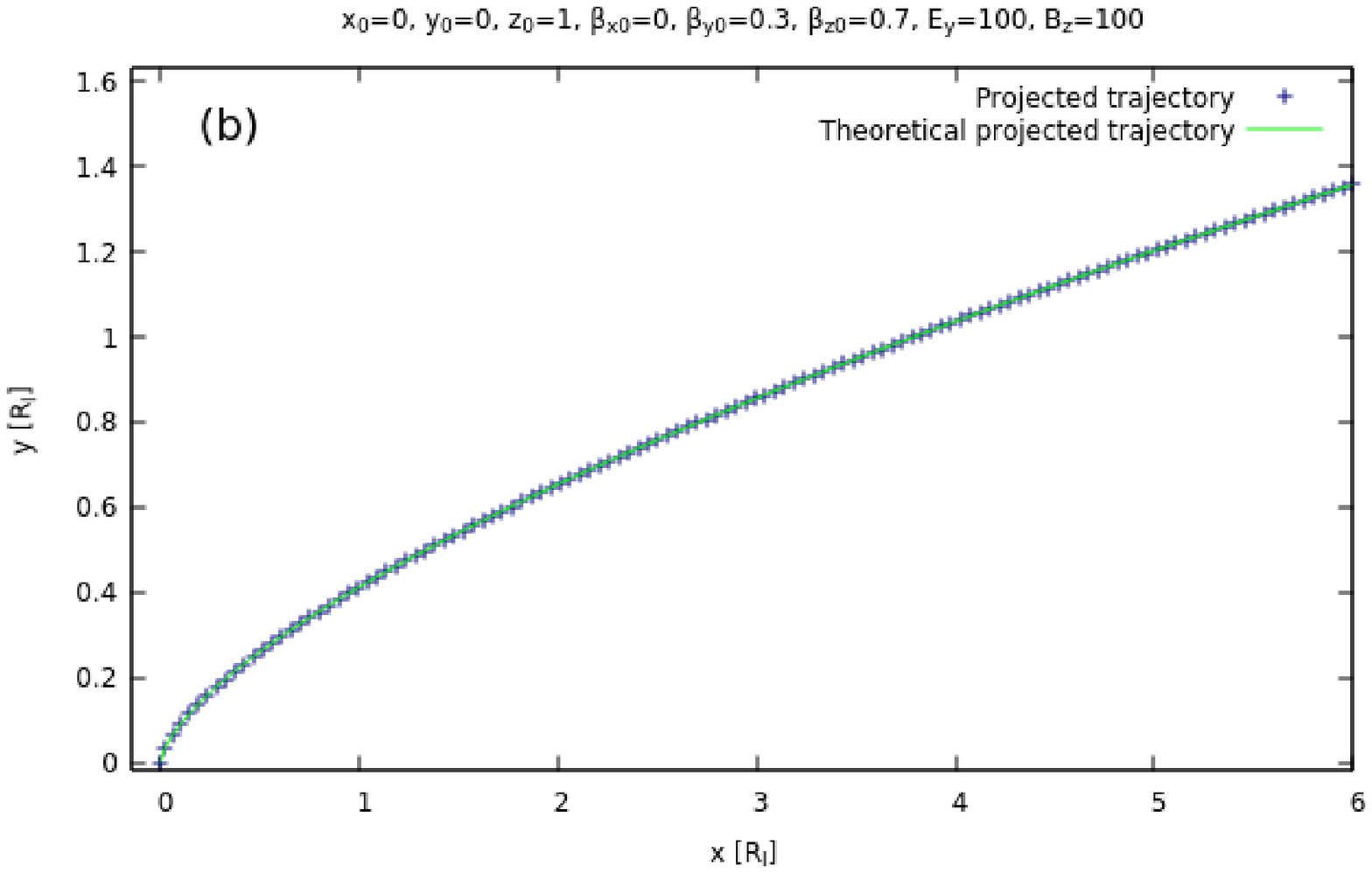}
%  \caption{Projected trajectory of a particle, normalised units}
\label{fig:Light_like}
\end{subfigure}
\caption{Evolution of a particle in a light-like field, $x_0=0$, $y_0=0$, $z_0=1$, $\beta^x_{0}=0$, $\beta^y_{0}=0.3$, $\beta^z_{0}=0.7$, $\frac{qE_y}{cm}=100$, $\frac{qB_z}{m}=100$. Theoretical and simulated 4-position as function of the proper time $\tau$ (a). Projection of the trajectory of the particle in the x-y plane (b).}
\label{fig:Light}
\end{figure}

If $\textit{\textbf{E}}$ and $\textit{\textbf{B}}$ are orthogonal (as in figure~\ref{fig:E_dom}, figure~\ref{fig:B_dom} and figure~\ref{fig:Light}, where $\textit{\textbf{E}}$ is along the y-axis and $\textit{\textbf{B}}$ along the z-axis), three cases are possible:
\begin{itemize}
\item $\textit{\textbf{E}}$ dominant shown in figure \ref{fig:E_dom} where the particle is accelerated along the $\textit{\textbf{E}}$ field but also drifts along $\textit{\textbf{E}} \times \textit{\textbf{B}}/E^2 $.
\item $\textit{\textbf{B}}$ dominant, as in figure \ref{fig:B_dom} where the particles moves in a cycloidal motion with the component of its speed along $\textit{\textbf{E}}$ switching between aligned and anti aligned, and the other component being again a drift along $\textit{\textbf{E}} \times \textit{\textbf{B}}/B^2$.
\item in the special case of the light-like field ($\mathbf{E} \cdot \mathbf{B} = 0$ and $E^2 = c^2 \, B^2$), in figure \ref{fig:Light}, where the trajectory of the particle is not periodic, in a similar fashion as the case $\textit{\textbf{E}}$ dominant with an increasing speed along $\textit{\textbf{B}}$ and again a drift along $\textit{\textbf{E}} \times \textit{\textbf{B}}$.
\end{itemize}

In addition to the tests in constant uniform fields, simulations were made with linearly and circularly polarised plane electromagnetic waves, for which exact solutions are known \citep{uzan_theories_2014}. 

Consider a linearly polarized plane wave propagating along the $\ex$ direction such that the 4-vector potential is $A^\alpha=(0,0,\frac{E}{\omega}\,\cos\xi,0)$. The phase is $\xi=\omega\,t-k\,x$. The electromagnetic field is given by 
\begin{subequations}
	\begin{align}
	\mathbf{E} & = E \, \sin \xi \, \ey \\
	\mathbf{B} & = \frac{E}{c} \, \sin \xi \, \ez .
	\end{align}
\end{subequations}
Initially the particle is at rest with a 4-velocity $u^\alpha_0=(c,\mathbf 0)$. Introducing the strength parameter of the wave by the ratio
\begin{equation}
a = \frac{q\,E}{m\,c\,\omega}
\end{equation}
the 4-velocity has components
\begin{subequations}
	\label{eq:vitesse_onde_lineaire}
	\begin{align}
	u^x & = \frac{a^2}{2} \, c \, (\cos \xi - 1 )^2 \\
	u^y & = -a \, c \, (\cos \xi - 1 ) \\
	\label{eq:vitesse_onde_lineaire_0}
	u^0 & = c + u^x .
	\end{align}
\end{subequations}
The Lorentz factor is deduced from $\gamma\,c = u^0$.

\begin{figure}
  \centering
  \includegraphics[scale=0.4]{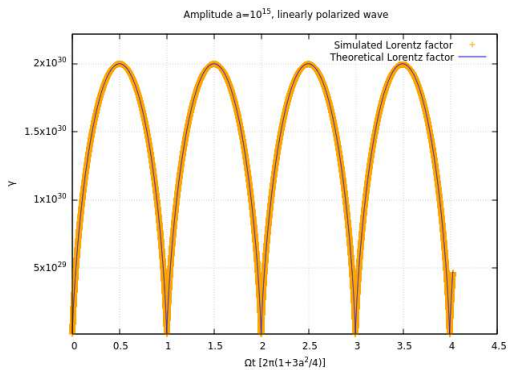}
\caption{Lorentz factor of a particle in a linearly polarised plane electromagnetic wave of strength parameter $a=10^{15}$.}
\label{fig:Linear_wave}
\end{figure}

\begin{figure}
  \centering
  \includegraphics[scale=0.4]{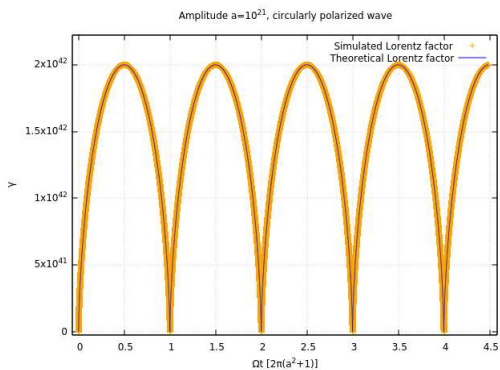}
\caption{Lorentz factor of a particle in a circularly polarised plane electromagnetic wave of strength parameter $a=10^{21}$.}
\label{fig:Circular_wave}
\end{figure}

Consider now a circularly  polarized plane wave propagating in the $\ex$ direction such that the vector potential has components $A^\alpha=(0,0,\frac{E}{\omega}\,\cos\xi,\frac{E}{\omega}\,\sin\xi)$ and the phase $\xi=\omega\,t-k\,x$. The electromagnetic field is then given by 
	\begin{subequations}
		\begin{align}
		\mathbf{E} & = E \, ( \sin \xi \, \ey - \cos \xi \, \ez ) \\
		\mathbf{B} & = \frac{E}{c} \, ( \sin \xi \, \ez + \cos \xi \, \ey ) .
		\end{align}
	\end{subequations}
	Initially the particle is at rest with 4-velocity $u^\alpha_0=(c,\mathbf 0)$. The time evolution of the components of this 4-velocity will be
	\begin{subequations}
		\label{eq:vitesse_onde_circulaire}
		\begin{align}
		u^x & = a^2 \, c \, (1 - \cos \xi ) = a \, u^y \\
		u^y & = a \, c \, (1 - \cos \xi ) \\
		u^z & = - a \, c \, \sin \xi \\
		u^0 & = c + u^x .
		\end{align}
	\end{subequations}
Simulations of linearly and circularly polarized plane waves were carried out as shown in figure \ref{fig:Linear_wave} and figure~\ref{fig:Circular_wave}. In these waves, the particle undergoes a series of acceleration and braking.

%\com{JE NE VOIS PAS DE REF À LA FIG 4.}

%\revone{2. Analytical solutions for the plane wave with linearly or circularly polarization should be more
%referenced, and may be given in their analytical forms in an appendix.}
%
%\revone{
%where $\gamma$ follows
%\begin{equation}\label{eq:Gamma_circular}
%\gamma=1+a^2(1-\cos(\phi))=1+\left(\frac{qB}{m\Omega} \right)^2(1-\cos(\phi))
%\end{equation}
%}
%\com{Met plutôt les expressions de la 4-vitesse.}
%for a circularly polarized wave, with
%\textcolor{red}{
%$\phi=\Omega (t - x/c)$
%}
The particle's Lorentz factor reaches a maximum value of 
\begin{equation}\label{eq:gammaMax}
\gamma=1+2\,a^2 %= 1 + 2 \left(\frac{qB}{m\Omega} \right)^2
\end{equation}
%at the peaks, where $q$ is the charge of the particle, $m$ its mass, $\Omega$ the pulsation of the wave and $B$ the amplitude of the magnetic field in the wave.
This shows that the code works well in ultra-strong fields with little to no error. 

In addition, the cases of elliptically polarized waves can be treated in the code. The equation of the fields of such waves is given by
	\begin{subequations}
	\begin{align}
	\mathbf{E} & = E \, ( \sin \xi \, \ey - \alpha \, \cos \xi \, \ez ) \\
	\mathbf{B} & = \frac{E}{c} \, ( \sin \xi \, \ez + \alpha \, \cos \xi \, \ey ) .
	\end{align}
\end{subequations}
%\begin{subequations}
%	\label{eq:Plane_wave_elliptical}
%	\begin{align}
%	B_x=0\\
%	B_y= \alpha B\cos(\Omega (t-x/c))\\
%	B_z=B\sin(\Omega (t-x/c))\\
%	E_x=0\\
%	E_y=cB\sin(\Omega (t-x/c))\\
%	E_z=- \alpha cB\cos(\Omega (t-x/c))     .
%	\end{align}
%\end{subequations}
With $0 \leq \alpha \leq 1$. The case $\alpha=1$ gives the circular wave whereas the case $\alpha=0$ returns the linearly polarized wave. For any kind of polarization, the maximum Lorentz factor of the particle is still $\gamma=1+2a^2$ however the evolution of the particle's Lorentz factor is the following %\com{$\phi$ est $\xi$ ici je pense??}
\begin{equation}\label{eq:Gamma_elliptical}
\gamma=1+a^2\,(1-\cos\xi)-\dfrac{a^2}{2} \, (1-\alpha^2)\sin^2\xi
\end{equation}
which is equation~(157) of \cite{MichelLiElec} except for a typo in the sign of the last term.
%\textbf{ATTENTION, IN MICHEL AND LI ARTICLE, IT SEEMS THAT A MISTAKE WAS MADE ON ONE SIGN OF THIS EQUATION} (\eqref{eq:Gamma_elliptical}) 
More details about plane waves and how to find these equations can be found in \cite{MichelLiElec}.

%\revtwo{
%Can the authors mention more specifically the advantage and disadvantage of the present approach, e.g. in the test cases with uniform fields as shown here: how good/bad would standard schemes fair?  I missed a clear description as to why the tests in section 3 pose challenges that are simply not possible with already known schemes.
%}
In the case of plane electromagnetic waves other algorithms have trouble at strength parameters above $a=25$ \cite{arefiev_temporal_2015} whereas our scheme easily exceeds this value of $a$ with short computational times.

%\revone{3. For analytical test cases of Section 3, it sould be interesting to see numerical error estimates $(in supremum - L^\infty-or L^2 norm)$ versus the discretisation parameters (number of time steps and/or number of particles) to
%see the order of this scheme. In particular such analysis is important for the case of plane wave, because
%in this case the electromagnetic field is no more constant. Since the present algorithm is based on analytical
%solutions for constant electromagnetic fields, when the electromagnetic fields is no more constant
%(like plane waves) a  space-time discretization of the (E,B) field is carried out.}

\section{Convergence and error of the algorithm}
\label{sec:convergence}

\begin{figure}
  \centering
  \begin{subfigure}[t]{0.03\textwidth}
    \textbf{(a)}
  \end{subfigure}
\begin{subfigure}{.45\textwidth}
  \centering
  \includegraphics[scale=0.4, valign=t]{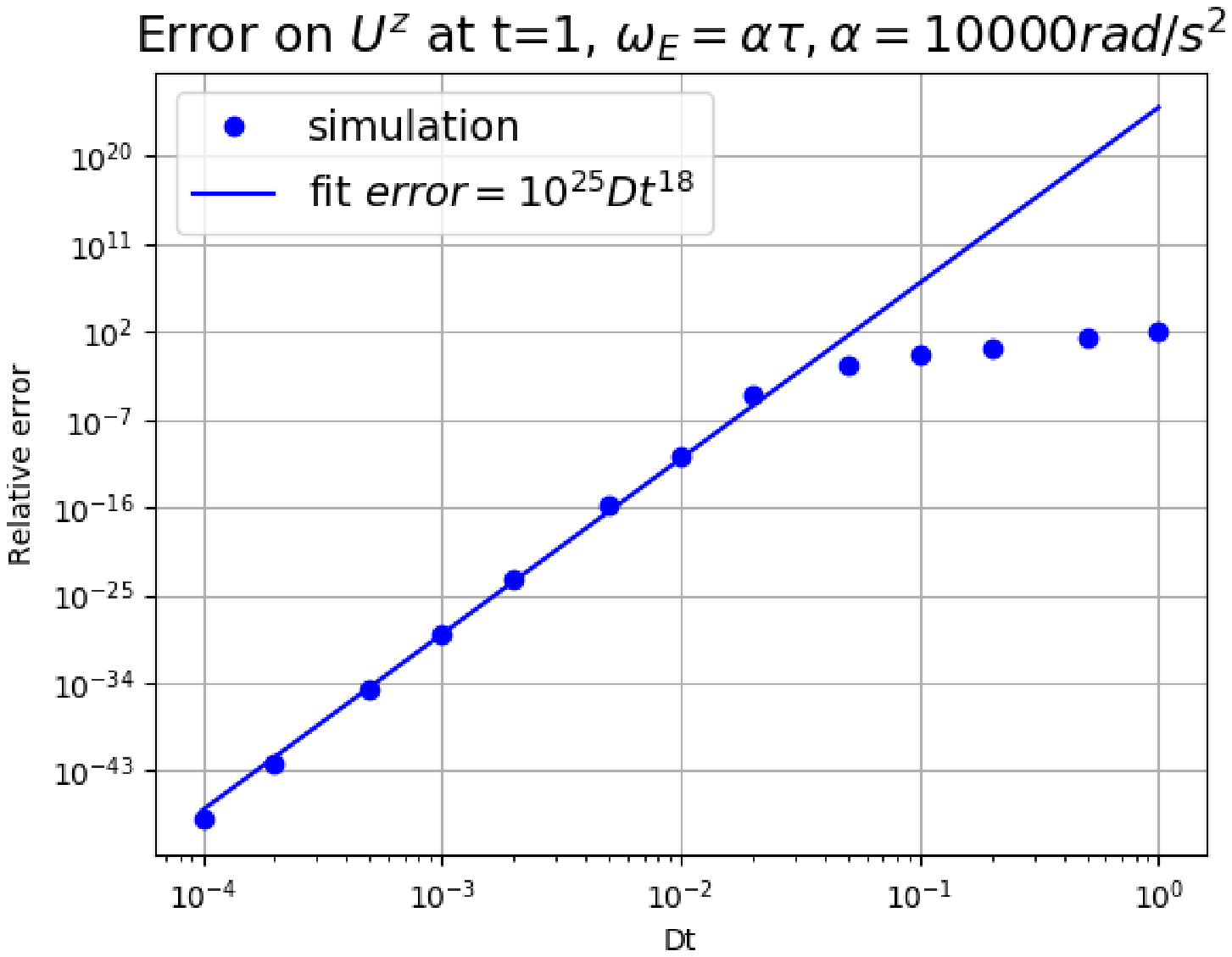}
\label{fig:Error_speed}
   % \textbf{A}
\end{subfigure}
  \begin{subfigure}[t]{0.03\textwidth}
    \textbf{(b)}
  \end{subfigure}
\begin{subfigure}{.45\textwidth}
  \centering
  \includegraphics[scale=0.4, valign=t]{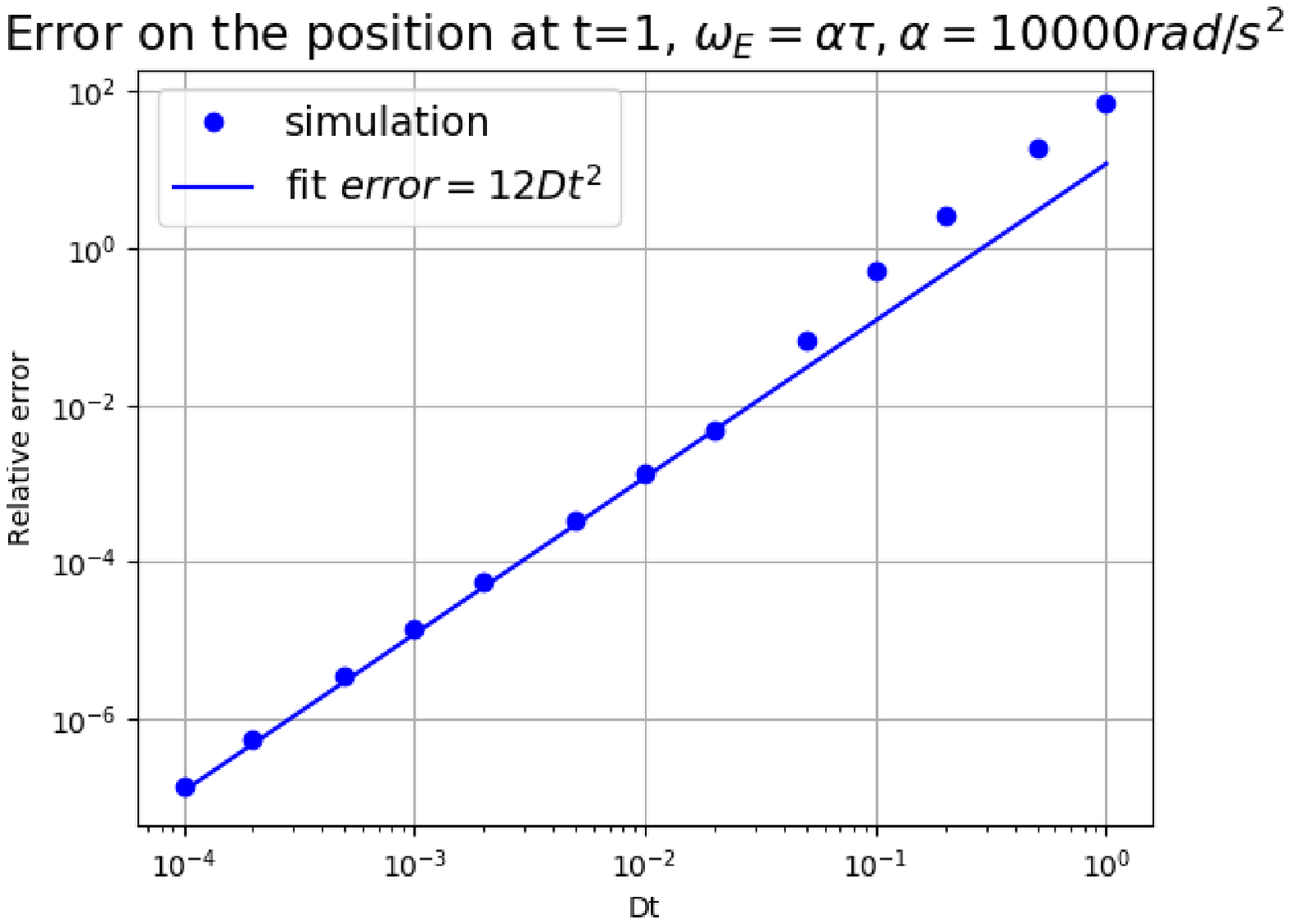}
\label{fig:Error_position}
    %\textbf{B}
\end{subfigure}
\caption{Relative error on the speed and on the position of the particle in an electric field aligned with $\textbf{e}_z$ where $\omega_E = \alpha \tau$, $\alpha=10^4$. %\com{Il faut encore déterminer l'ordre de la méthode et tracer la ligne sur les courbes.}
}
\label{fig:Error}
\end{figure}

%\subsection{Configuration}
We study the convergence of our algorithm in simple test cases where analytical solutions are known. For instance, let us track a particle of charge~$q$ and mass~$m$ subject to an acceleration in an electric field $\textbf{E}$ assumed to grow linearly in proper time~$\tau$ such that
\begin{equation}
\textbf{E}(\tau) = E(\tau) \textbf{e}_z
\end{equation}
with
\begin{equation}
\dfrac{q}{mc}E(\tau) = \alpha \tau .
\end{equation}
We choose this form of field since a constant field will give no error, for the code is based on analytical solutions in constant fields. At $\tau=0$, the particle is at $x^{\mu}=\begin{pmatrix} 0 & 0 & 0 & 0 \end{pmatrix}$ and its 4-velocity is $u^{\mu}=\begin{pmatrix} c & 0 & 0 & 0 \end{pmatrix}$.
%\subsection{Motion of the particle}
In this configuration, the Faraday tensor is
\begin{equation}
F^{\mu \nu}=
\begin{pmatrix}
0 & 0 & 0 & -E(\tau)/c \\
0 & 0 & 0 & 0 \\
0 & 0 & 0 & 0 \\
E(\tau)/c & 0 & 0 & 0 \\
\end{pmatrix}.
\end{equation}
Starting from the equation of motion:
\begin{equation}
\dfrac{d u^{\mu}}{d \tau}=\dfrac{q}{m}F^{\mu \nu}u_{\nu}
\end{equation}
When looking only at $u^t$ and $u^z$ we find
\begin{subequations}
\begin{align}
\dfrac{d u^t}{d \tau} &= - \frac{q E(\tau)}{mc} u_z = -\alpha \tau u_z = \alpha \tau u^z \label{eq:Dut} \\
\dfrac{d u^z}{d \tau} &= \frac{q E(\tau)}{mc} u_t =\alpha \tau u_t = \alpha \tau u^t
\label{eq:Dux}
\end{align}
\end{subequations}
We then define
\begin{subequations}
\begin{align}
\Delta = u^t-u^z \\
\Sigma = u^t+u^z
\end{align}
\end{subequations}
So we can obtain with \eqref{eq:Dut}-\eqref{eq:Dux} and \eqref{eq:Dut}+\eqref{eq:Dux}
\begin{subequations}
\begin{align}
\dfrac{d \Delta}{d \tau} &= -\alpha \tau \Delta \\
\dfrac{d \Sigma}{d \tau} &= \alpha \tau \Sigma
\end{align}
\end{subequations}
Which solves into
\begin{subequations}
\begin{align}
\Delta (\tau) &= \delta e^{-\alpha \tau^2/2}\\
\Sigma (\tau) &= \sigma e^{\alpha \tau^2/2}
\end{align}
\end{subequations}
At $\tau=0$, we find:
\begin{equation}
\delta = \sigma = \gamma_0
\end{equation}
Noticing that 
\begin{subequations}
\begin{align}
u^t &= \dfrac{\Sigma + \Delta }{2} = \gamma_0 \dfrac{ e^{\alpha \tau^2/2} + e^{-\alpha \tau^2/2} }{2} \\
u^z &= \dfrac{\Sigma - \Delta }{2} = \gamma_0 \dfrac{ e^{\alpha \tau^2/2} - e^{-\alpha \tau^2/2} }{2}
\end{align}
\end{subequations}
So the expression of the Lorentz factor and the four velocity along z are:
\begin{subequations}
\begin{align}
u^t &= \gamma_0 c \cosh(\alpha \tau^2/2) \\
u^z &= \gamma_0 c \sinh(\alpha \tau^2/2)
\label{eq:Ux_lin}
\end{align}
\end{subequations}
After another integration, we get the position of the particle:
\begin{subequations}
\begin{align}
ct &= \gamma_0 c \sqrt{\dfrac{\upi}{8 \alpha}} [ \text{erf}(\tau \sqrt{\alpha / 2}) + \text{erfi}(\tau \sqrt{\alpha / 2}) ] \\
z &= \gamma_0 c \sqrt{\dfrac{\upi}{8 \alpha}} [ \text{erfi}(\tau \sqrt{\alpha / 2}) - \text{erf}(\tau \sqrt{\alpha / 2}) ]
\label{eq:x_lin}
\end{align}
\end{subequations}
Where $\text{erf}$ and $\text{erfi}$ are defined (with $i$ the imaginary unit):
\begin{subequations}
\begin{align}
\text{erf}(\tau)=\dfrac{2}{\sqrt{\upi}}\int_{0}^{\tau} \exp(-t^2) dt \\
\text{erfi}(\tau)=\dfrac{\text{erf}(i\tau)}{i}
\label{eq:erf}
\end{align}
\end{subequations}
%\com{Explique la signification/définition des fonctions \text{erf/i}.} 
When comparing \eqref{eq:Ux_lin} and \eqref{eq:x_lin} to the outcome of the simulations, we can get the error $\delta \alpha$ thanks to the folowing formula:
\begin{equation}
\delta \alpha=\dfrac{\vert \alpha_{measured} - \alpha_{theoretical} \vert}{ \vert \alpha_{theoretical} \vert},
\end{equation}
when $\alpha$ is taken to be $u^z$ or $z$ we obtain the the plots of figure~\ref{fig:Error}, highlighting a fast decrease on the error on the 4-velocity however the error on the 4-position shows that the scheme of the simulation is of order~2. %\com{Il faut aussi revoir cette affirmation.}

\begin{figure}
  \centering
  \begin{subfigure}[t]{0.03\textwidth}
    \textbf{(a)}
  \end{subfigure}
\begin{subfigure}{.45\textwidth}
  \centering
  \includegraphics[scale=0.4, valign=t]{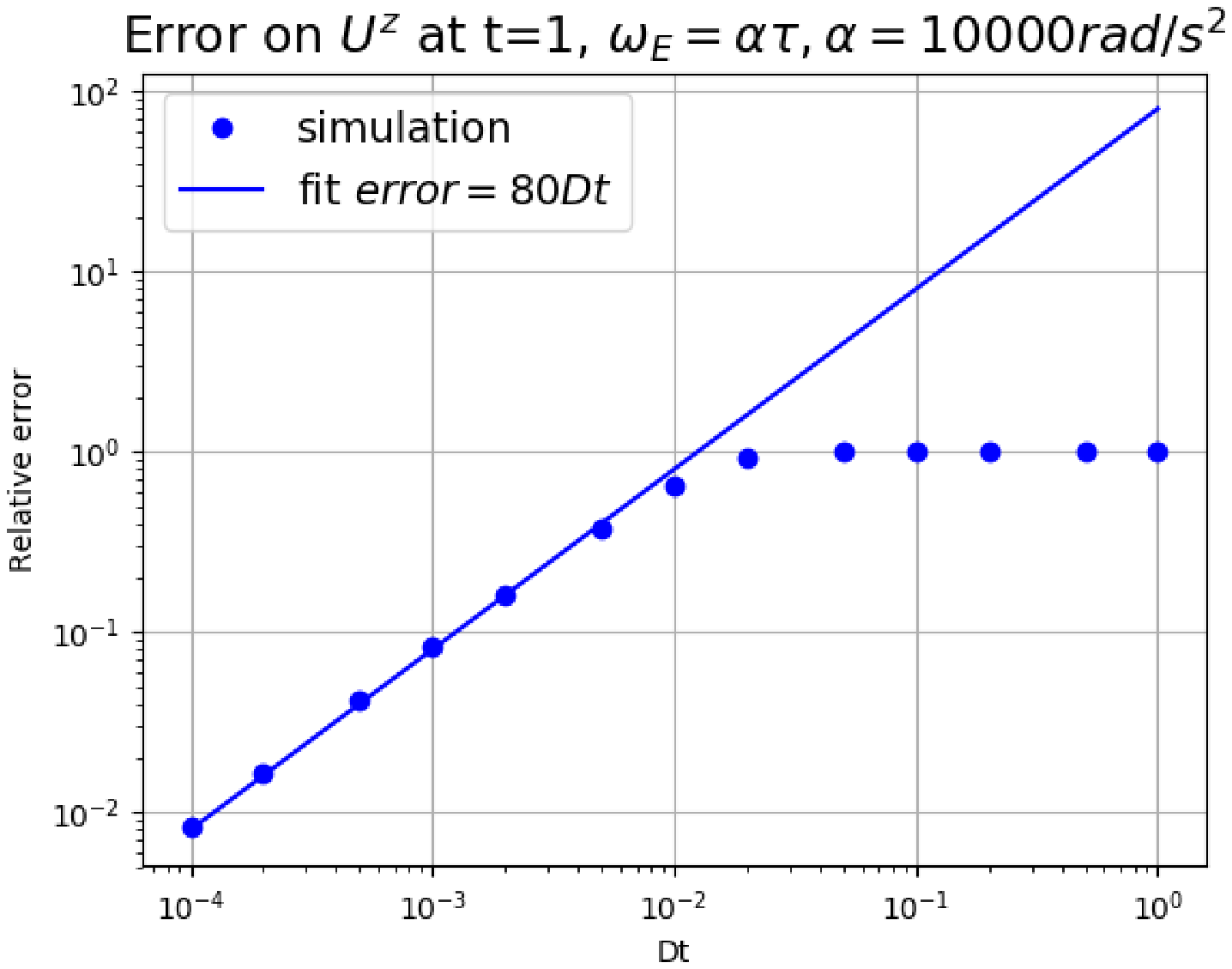}
\label{fig:Error_speed_no_loop}
   % \textbf{A}
\end{subfigure}
  \begin{subfigure}[t]{0.03\textwidth}
    \textbf{(b)}
  \end{subfigure}
\begin{subfigure}{.45\textwidth}
  \centering
  \includegraphics[scale=0.4, valign=t]{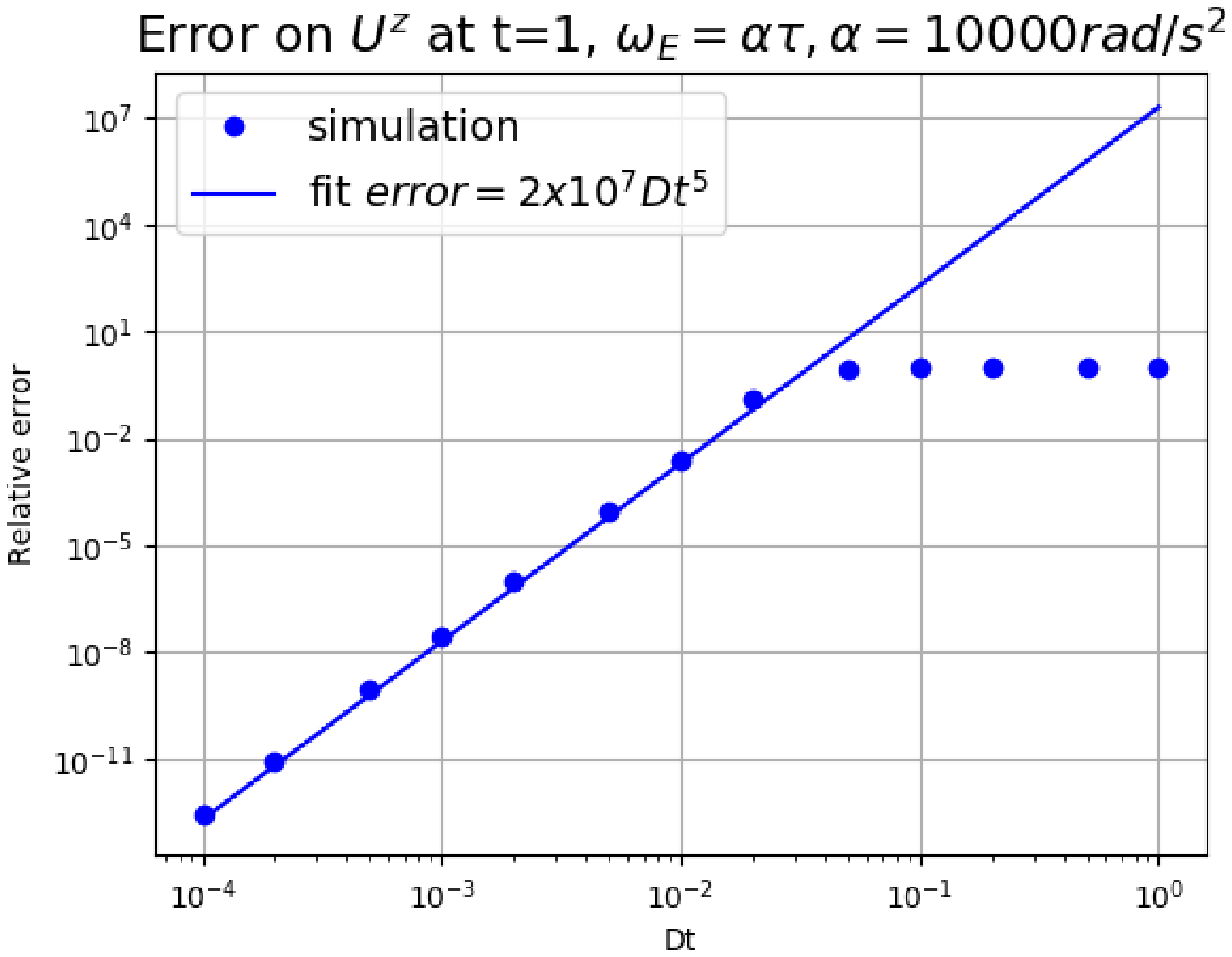}
\label{fig:Error_speed_3_loops}
    %\textbf{B}
\end{subfigure}
\caption{Relative error on the speed of the particle in an electric field aligned with $\textbf{e}_z$ where $\omega_E = \alpha \tau$, $\alpha=10^4$. (a) No convergence loop on the field. (b) At most three run in the convergence loop. }
\label{fig:Error_change_loop}
\end{figure}

A more in-depth analysis showed that the number of run through the field convergence loop discussed in \ref{sec:Algo} has an influence on the order of the algorithm, explaining why such a high order is found on figure~\ref{fig:Error}. Simulations where no convergence is applied or where at most three runs through the convergence loop are made are respectively of order 1 or 5, as shown by figure~\ref{fig:Error_change_loop}.

Other studies of the rate of convergence to the analytical solutions can be found in \cite{petri_relativistic_2020} for a linearly and a circularly polarized wave. There it is shown that the error is second order in time.

\section{Rotating neutron star fields}
\label{sec:Deutsch}

\subsection{Vacuum rotating electromagnetic field solution}

\begin{figure}
  \centering
  \includegraphics[scale=0.4]{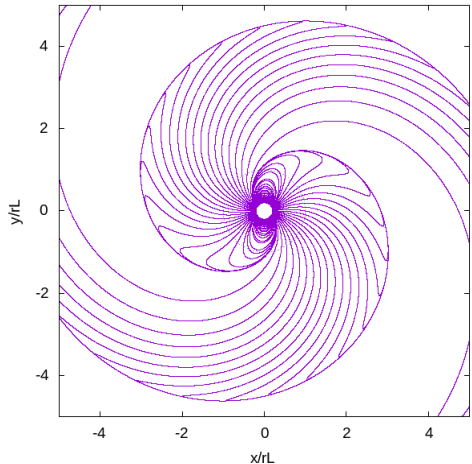}
\caption{Magnetic field line of an orthogonal rotator with  $\rlight=5\,R$.}
\label{fig:Deutsch_field}
\end{figure}

Eventually we implemented the Deutsch field in our code in order to mimic a rotating magnet in vacuum.
The exact solution for the rotating dipole in vacuum is given by \cite{deutsch_electromagnetic_1955}. The corresponding electromagnetic field, written in a complex form, reads
\begin{subequations}
	\begin{align}
	\label{eq:DeutschEM}
	B_r(\mathbf{r},t) & = 2 \, B \, \left[ \frac{R^3}{r^3} \, \cos\chi \, \cos\vartheta + 
	\frac{R}{r} \, \frac{h^{(1)}_1(k\,r)}{h^{(1)}_1(k\,R)} \, 
	\sin\chi \, \sin \vartheta \, e^{i\,\psi} \right] \\
	B_\vartheta(\mathbf{r},t) & = B \, \left[ \frac{R^3}{r^3} \, \cos\chi \, \sin\vartheta + \right. \\ 
	& \left. \left( \frac{R}{r} \, 
	\frac{\frac{d}{dr} \left( r \, h^{(1)}_1(k\,r) \right)}{h^{(1)}_1(k\,R)} + \frac{R^2}{\rlight^2} \, 
	\frac{h^{(1)}_2(k\,r)}{\frac{d}{dr} \left( r \, h^{(1)}_2(k\,r) \right) |_{R}} \right) \, 
	\sin\chi \, \cos \vartheta \, e^{i\,\psi} \right] \\
	B_\varphi(\mathbf{r},t) & = B \, \left[ \frac{R}{r} \, 
	\frac{\frac{d}{dr} ( r \, h^{(1)}_1(k\,r) )}{h^{(1)}_1(k\,R)} \, 
	+ \frac{R^2}{\rlight^2}
	\frac{h^{(1)}_2(k\,r)}{\frac{d}{dr} \left( r \, h^{(1)}_2(k\,r) \right) |_{R}}
	\, \cos 2\,\vartheta \right] \, i \, \sin\chi \, \, e^{i\,\psi} \\
	E_r(\mathbf{r},t) & = \Omega \, B \, R \, 
	\left[ \left( \frac{2}{3} - \frac{R^2}{r^2} ( 3 \, \cos^2\vartheta - 1 ) \right) \
	\, \frac{R^2}{r^2} \, \cos\chi \right. \nonumber \\
	& + \left. 3 \, \sin\chi\, \sin 2\,\vartheta \, e^{i\,\psi}  \,
	\frac{R}{r} \, \frac{ h^{(1)}_2(k\,r)}
	{\frac{d}{dr} \left( r \, h^{(1)}_2(k\,r) \right) |_{R}} \right] \\
	E_\vartheta(\mathbf{r},t) & = \Omega \, B \, R \, 
	\left[ - \frac{R^4}{r^4} \sin 2\,\vartheta \, \cos\chi + 
	\sin\chi\, e^{i\,\psi}  \, \left(
	\frac{R}{r} \, \frac{\frac{d}{dr} \left( r \, h^{(1)}_2(k\,r) \right)}
	{\frac{d}{dr} \left( r \, h^{(1)}_2(k\,r) \right)|_{R}} \, \cos 2\,\vartheta -
	\frac{h^{(1)}_1(k\,r)}{h^{(1)}_1(k\,R)} \right) \right] \\
	E_\varphi(\mathbf{r},t) & = \Omega \, B \, R \, \left[ \frac{R}{r} \, 
	\frac{\frac{d}{dr} \left( r \, h^{(1)}_2(k\,r) \right)}
	{\frac{d}{dr} \left( r \, h^{(1)}_2(k\,r) \right)|_{R}} -
	\frac{h^{(1)}_1(k\,r)}{h^{(1)}_1(k\,R)} \right] \, i \sin\chi \, \cos\vartheta \, e^{i\,\psi}
	\end{align}
\end{subequations}
where $h^{(1)}_\ell$ are the spherical Hankel functions for outgoing waves \citep{arfken_mathematical_2005}, $k\,\rlight=1$ and the phase is $\psi=-\Omega\,t$. An example of magnetic field lines for a perpendicular rotator is shown in figure~\ref{fig:Deutsch_field}.

Our aim for these simulations was to investigate the behaviour of particles close to the neutron star, especially to find the final Lorentz factor the particles were able to reach as well as the influence of their initial position on their final position and energy.

As expected, the particles could either be ejected far from the neutron star reaching an almost ballistic trajectory or crash onto the neutron star surface or be trapped around the pulsar. 

In all of our simulations, we work with a millisecond pulsar for which $\rlight=10\,R=100\,\text{km}$ so $\Omega=3000$~$\text{rad.s}^{-1}$ and the period is about $P\approx2.1~\text{ms}$. We choose the magnetic field strength at the equator to be $B=10^{5}~\text{T}$.

%\subsection{Van Allen belts}
%
%\revtwo{
%4) Figure 5 shows an interesting case of a trapped particle in a slowly rotating, low inclination case, but the case is "not deeply investigated". This must be remedied: either a more serious discussion is provided, or this case should simply be omitted from the paper. Why is such trapping not possible at higher rotation?
%}
%
%Interestingly, in some of the first test simulations in the Deutsch field, the code was able to highlight structures similar to Van-Allen belts, as in figure~\ref{fig:Van_Allen}, but this result only occurred with a slowly rotating ($\rlight=10^4\,R$), low inclination ($\chi=18^{\circ} $) neutron star, which is a case we did not deeply investigate. \com{A creuser comme le demande le referee.}
% \begin{figure}
%  \centering
%  \includegraphics[scale=0.5]{VanAllen52.png}% Images in 100% size
%  \caption{Trajectory of a particle forming a Van Allen like belt around the neutron star.}
%\label{fig:Van_Allen}
%\end{figure}

%For each inclination, we injected 2.048 test particles at rest. \com{Essaie avec plus de particules.} 
% and there were no interaction between particles (as every particle was injected separately). 

\subsection{Settings}
\label{ssec:Settings}

The main purpose of our simulations was to investigate the evolution of five types of particles: electrons, protons, antiprotons, fully ionized iron \ce{^56Fe} and fictive particles with the charge of a proton, but $10^6$ times its mass (mentioned in \ref{sssec:Symmetry}). The inclination angle of the neutron stars is taken in the set $\chi=\{0^{\circ}, 30^{\circ}, 60^{\circ}, 90^{\circ}, 120^{\circ}, 150^{\circ}, 180^{\circ}\}$.

We tried three different initial set ups by imposing the starting positions of the particles
\begin{enumerate}
\item The first set up used 2.048~particles for each obliquity and type of charges, following the same regular pattern. These particles were distributed uniformly in spherical coordinates, radius ($r=(2+k)R$, $k \in [\![ 0;7 ]\!]$), colatitude ($\theta=\frac{l}{9}\upi$ , $l \in [\![ 1;8 ]\!]$) and azimuth ($\phi = \frac{m}{16} \upi$, $m \in [\![ 0;31 ]\!]$). (Note that this does not correspond to a uniform distribution in space.)
\item The second set up also used 2.048~particles per obliquity. However, they were randomly placed in space. The radius follows an uniform distribution~$r\in[2R;10R]$, the azimuth $\phi$ also follows an uniform distribution $\phi\in[0;2 \upi]$, however, for the colatitude, we use another variable~$u$, following an uniform probability $u\in[0;1]$ and then we set the initial colatitude of the particle to be $\theta = \arccos(1-2\,u)$ so as to avoid artificial particles overdensities close to the rotation axis because of the singularity along this axis.
\item The third set up only used protons, but with 16.384~particles for each inclination, with the same random distribution as in point~(ii).

\end{enumerate}
Unless they crashed onto the neutron star, these test particles were free to evolve for a time $T=15\,\Omega^{-1}$ in normalised units which is $T=5~\text{ms} \simeq 2.39$ periods.

\subsection{Regularly placed particles}

\subsubsection{Particle distribution functions and symmetries}
\label{sssec:Symmetry}

\begin{figure}
  \begin{subfigure}[t]{0.02\textwidth}
    \textbf{(a)}
  \end{subfigure}
\begin{subfigure}{.46\textwidth}
  \centering
  \includegraphics[scale=0.36]{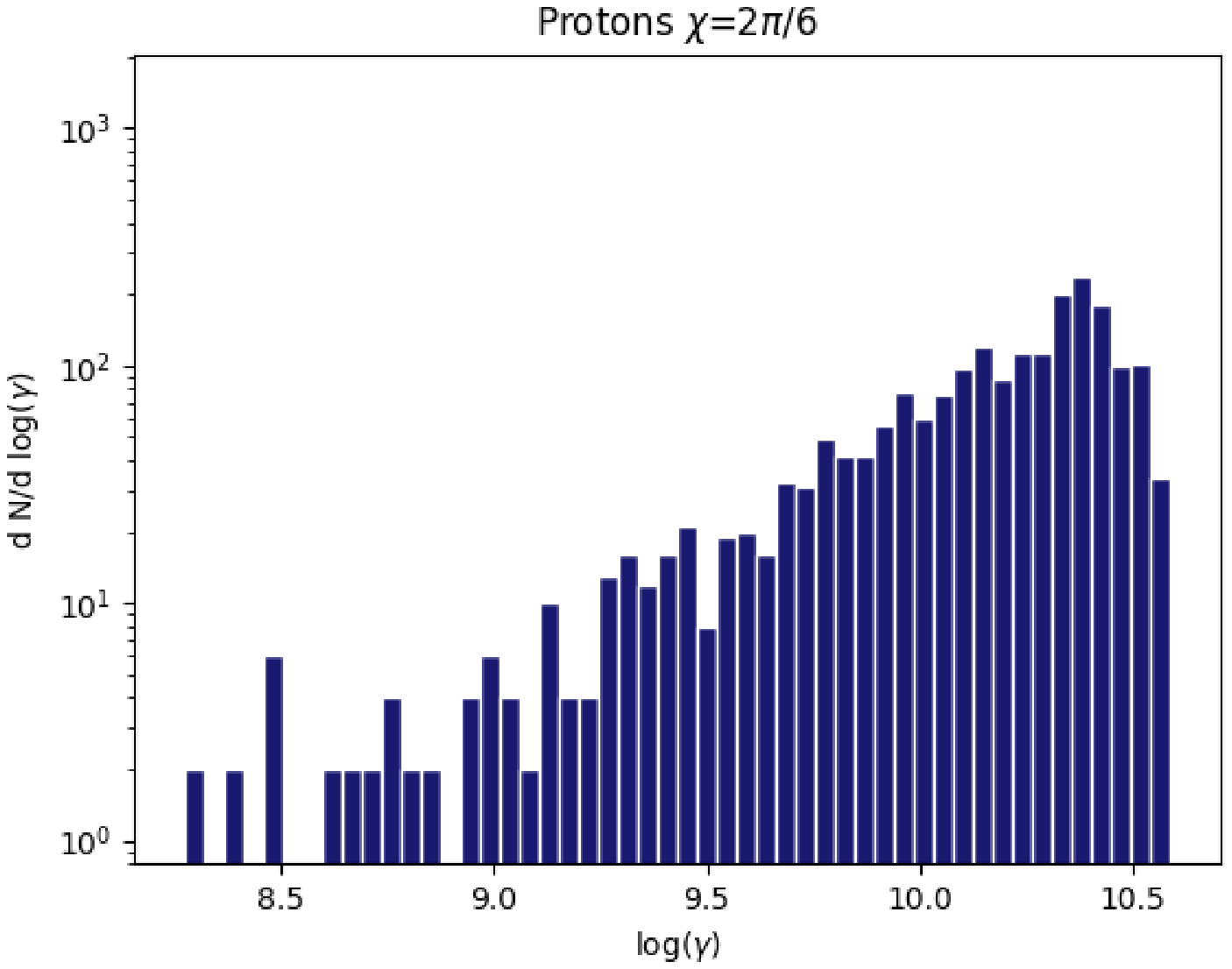}% Images in 100% size
\label{fig:Spectre_prot_chi2sur6}
\end{subfigure}
  \begin{subfigure}[t]{0.02\textwidth}
    \textbf{(b)}
  \end{subfigure}
\begin{subfigure}{.46\textwidth}
  \centering
  \includegraphics[scale=0.36]{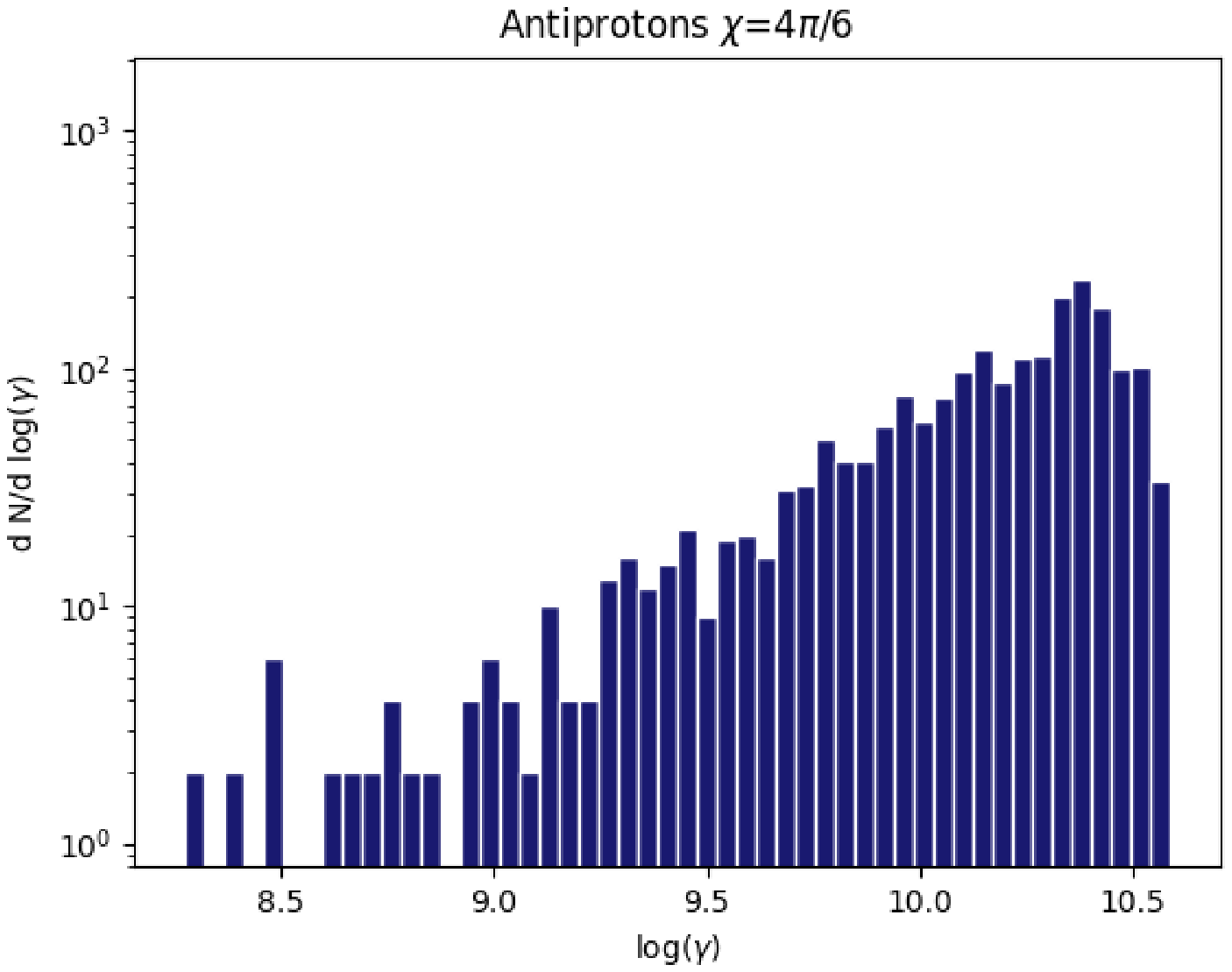}% Images in 100% size
\label{fig:Spectre_aprot_chi4sur6}
\end{subfigure}
  \begin{subfigure}[t]{0.02\textwidth}
    \textbf{(c)}
  \end{subfigure}
\begin{subfigure}{.47\textwidth}
  \centering
  \includegraphics[scale=0.36]{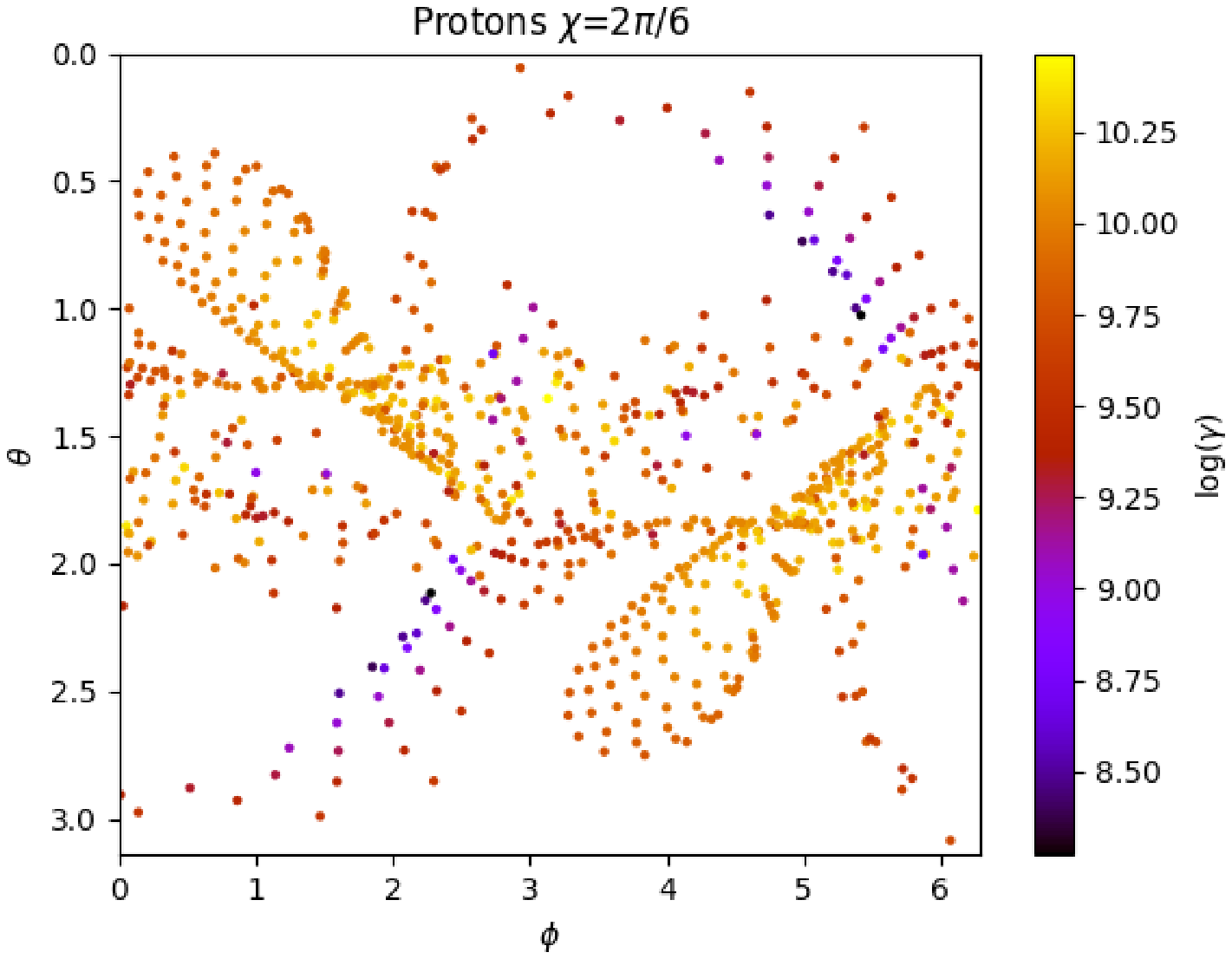}% Images in 100% size
\label{fig:Eject_prot_chi2sur6}
\end{subfigure}
  \begin{subfigure}[t]{0.04\textwidth}
    \textbf{(d)}
  \end{subfigure}
\begin{subfigure}{.47\textwidth}
  \centering
  \includegraphics[scale=0.36]{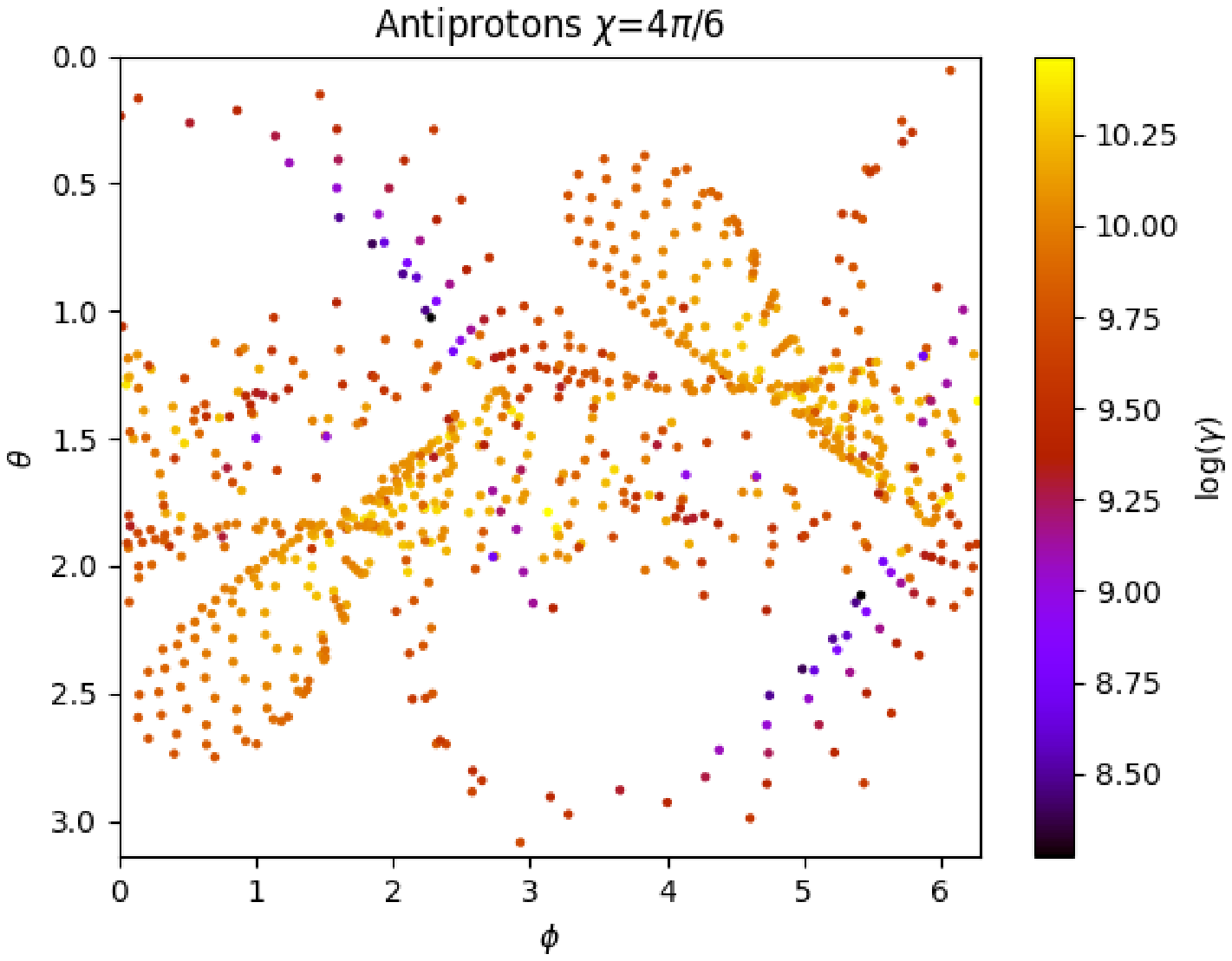}% Images in 100% size
\label{fig:Eject_aprot_chi4sur6}
\end{subfigure}
%  \begin{subfigure}[t]{0.01\textwidth}
%  \end{subfigure}
%\begin{subfigure}{.5\textwidth}
%  \centering
%  \includegraphics[scale=0.4]{ejection_proton_B5_CHI2sur6.png}% Images in 100% size
%\label{fig:Eject_prot_chi2sur6}
%\end{subfigure}
%  \begin{subfigure}[t]{0.01\textwidth}
%  \end{subfigure}
%\begin{subfigure}{.5\textwidth}
%  \centering
%  \includegraphics[scale=0.4]{ejection_antiproton_B5_CHI4sur6.png}% Images in 100% size
%\label{fig:Eject_aprot_chi4sur6}
%\end{subfigure}
  \caption{Comparison between protons around a pulsar of inclination $\chi=60^{\circ}$, and antiprotons around a pulsar of inclination $\chi=120^{\circ}$ to highlight their symmetrical behaviour. (a) Proton Lorentz factor spectra. (b) Antiproton Lorentz factor spectra. (c) Proton ejection map. (d) Antiproton ejection map. The ejection maps are the final latitudes $\theta$ and azimuths $\phi$ of particles that went beyond the light cylinder. These particles have an almost radial trajectory.}
\label{fig:Comparaison_prot_aprot}
\end{figure}

The particle distribution functions~$f(\gamma) = dN(\gamma)/d\gamma$, where $dN(\gamma)$ is the number of particles with Lorentz factor in the range~$[\gamma,\gamma+d\gamma]$, and final positions are shown in figure \ref{fig:Comparaison_prot_aprot}. For particles of the same mass~$m$ but opposite charge~$\pm q$, the behaviour is symmetric, meaning that
if for an inclination~$\chi$ a proton starts at position~$(r_i,\theta_i,\phi_i)$ and ends at position~$(r_f,\theta_f,\phi_f)$, for an inclination such that $\chi'=\upi-\chi$ an antiproton with an initial position~$(r_i' = r_i,\theta_i' = \upi-\theta_i,\phi_i'=\phi_i)$ will end at $(r_f'=r_f,\theta_f'=\upi-\theta_f,\phi_f'=\phi_f)$. 
Thus we only need to investigate particles of one charge, let it be leptons or their antiparticles, which is expected according to symmetry arguments discussed in \cite{laue_acceleration_1986}. 
Another symmetry relative to the centre of the neutron star is also noticeable in figure~\ref{fig:Comparaison_prot_aprot}, figure~\ref{fig:Ejections} and figure~\ref{fig:Crash} and tells us that for an inclination~$\chi$, if a particle is injected in $(r_i,\theta_i,\phi_i)$ and ends at $(r_f,\theta_f,\phi_f)$, a particle injected in $(r_i,\upi-\theta_i,\upi+\phi_i)$ will end at $(r_f,\upi-\theta_f,\upi+\phi_f)$. Again this was discussed in \cite{laue_acceleration_1986}.

\subsubsection{Final Lorentz factors}
\label{sssec:Lorentz}

In the simulations, we explored the dependence of the final Lorentz factor with respect to the particle charge to mass ratio~$q/m$. Results of our computations showed that the iron~\ce{^56Fe} reaches Lorentz factors of $\gamma_\textrm{Fe} \sim 10^{10.5} $, protons and antiprotons up to $\gamma_\textrm{p} \sim 10^{10.7}$ while electrons up to $\gamma_\textrm{e} \sim 10^{14}$. As a check for the scaling we also tried fictive particles of artificially very high mass $m_\textrm{Mp}=10^6\, m_p$ and charge $q_\textrm{Mp} = q_\textrm{p} = +e$ reaching only Lorentz factors up to $\gamma_\textrm{Mp} \sim 10^{4.7}$.

\begin{figure}
  \centering
  \begin{subfigure}[t]{0.03\textwidth}
    \textbf{(a)}
  \end{subfigure}
\begin{subfigure}{.45\textwidth}
  \centering
  \includegraphics[scale=0.37]{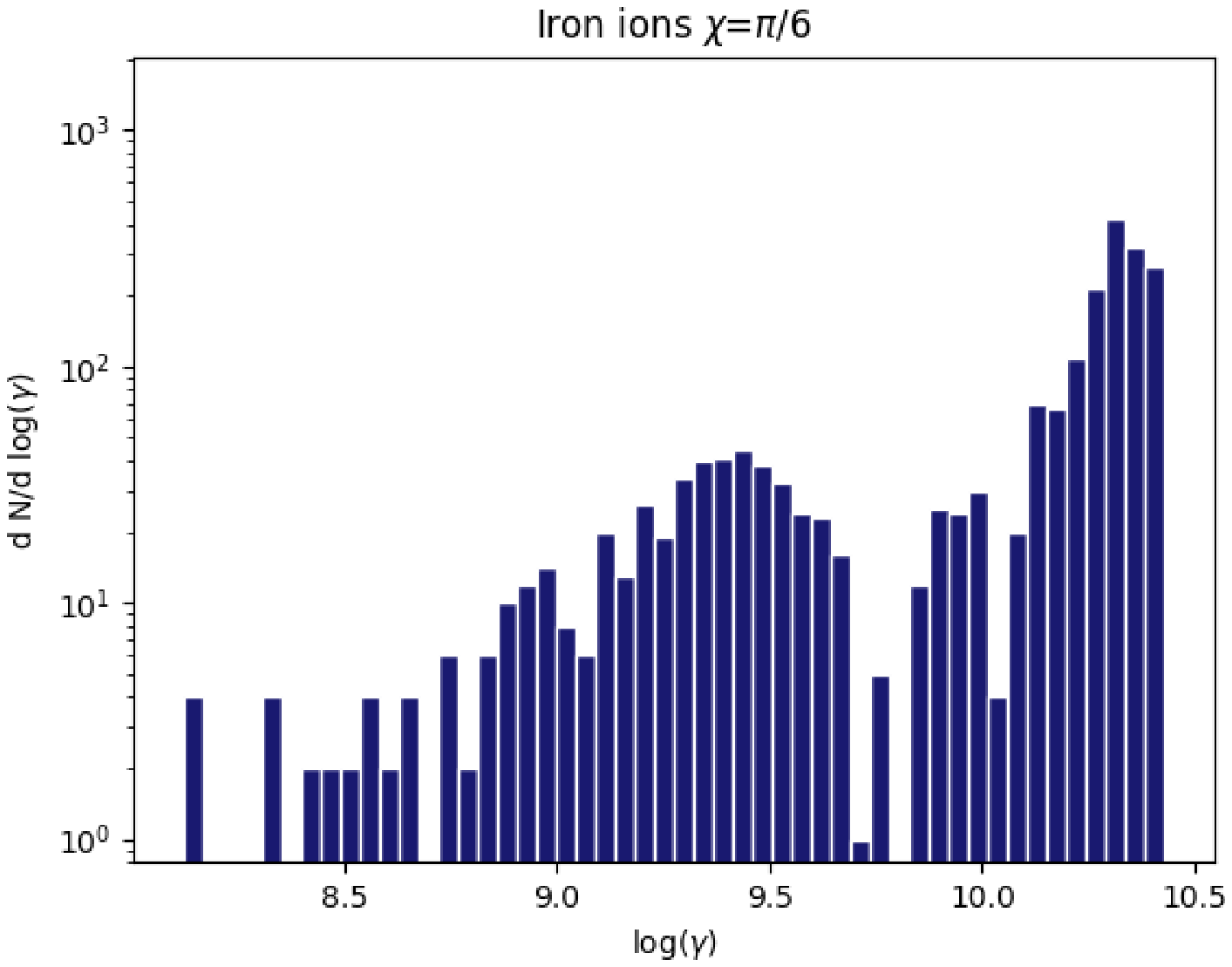}% Images in 100% size
  \caption{}
%  \caption{Lorentz factor spectrum for iron ions around a pulsar of inclination $\chi=\upi/6$}
\label{fig:Spectre_fer_chi1sur6}
\end{subfigure}
  \begin{subfigure}[t]{0.03\textwidth}
    \textbf{(b)}
  \end{subfigure}
\begin{subfigure}{.45\textwidth}
  \centering
  \includegraphics[scale=0.37]{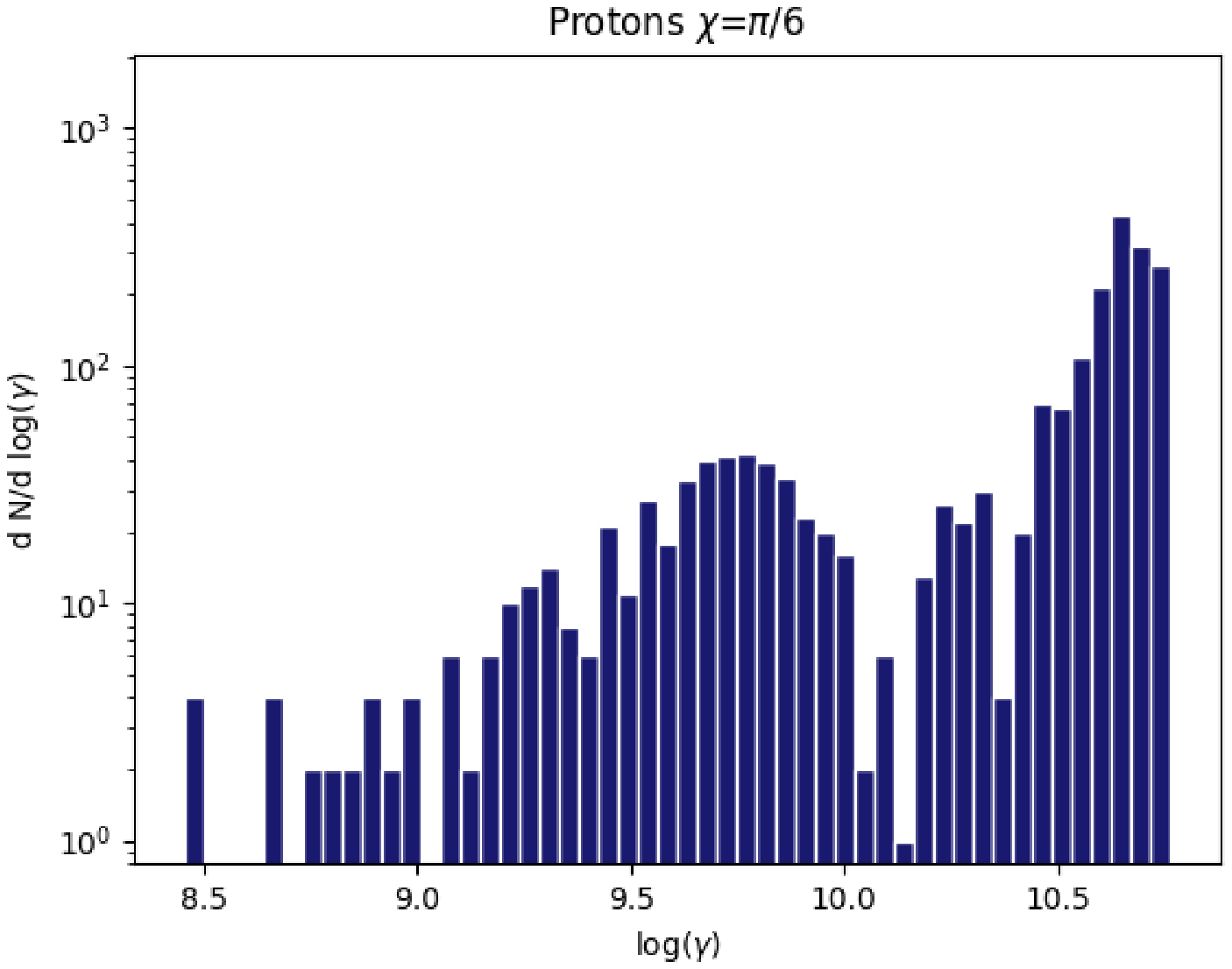}% Images in 100% size
  \caption{}
%  \caption{Lorentz factor spectrum for }
\label{fig:Spectre_prot_chi1sur6}
\end{subfigure}
  \begin{subfigure}[t]{0.03\textwidth}
    \textbf{(c)}
  \end{subfigure}
\begin{subfigure}{.45\textwidth}
  \centering
  \includegraphics[scale=0.37]{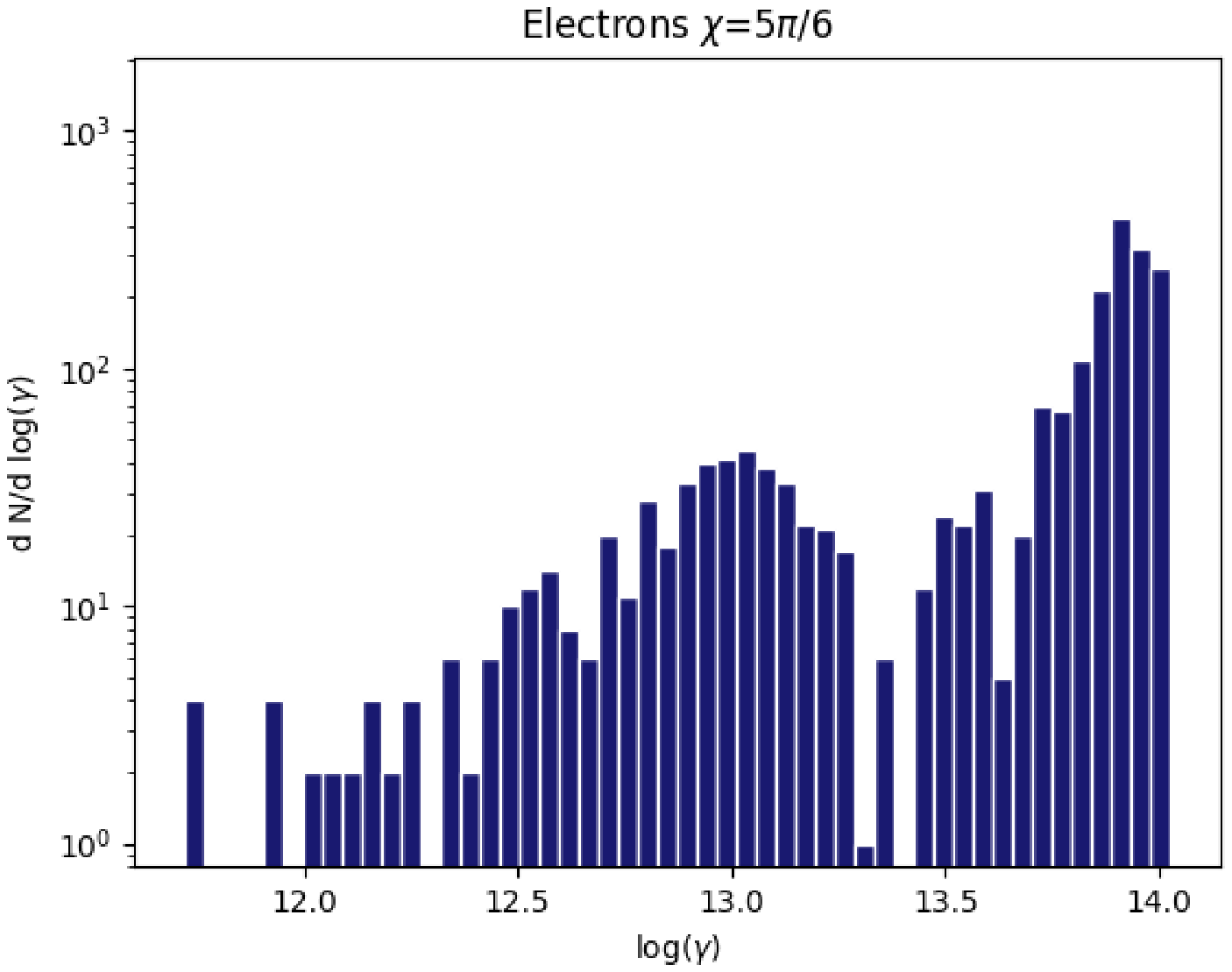}% Images in 100% size
  \caption{}
%  \caption{Lorentz factor spectrum for electrons around a pulsar of inclination $\chi=5\upi/6$}
\label{fig:Spectre_elec_chi5sur6}
\end{subfigure}
  \begin{subfigure}[t]{0.03\textwidth}
    \textbf{(d)}
  \end{subfigure}
\begin{subfigure}{.45\textwidth}
  \centering
  \includegraphics[scale=0.37]{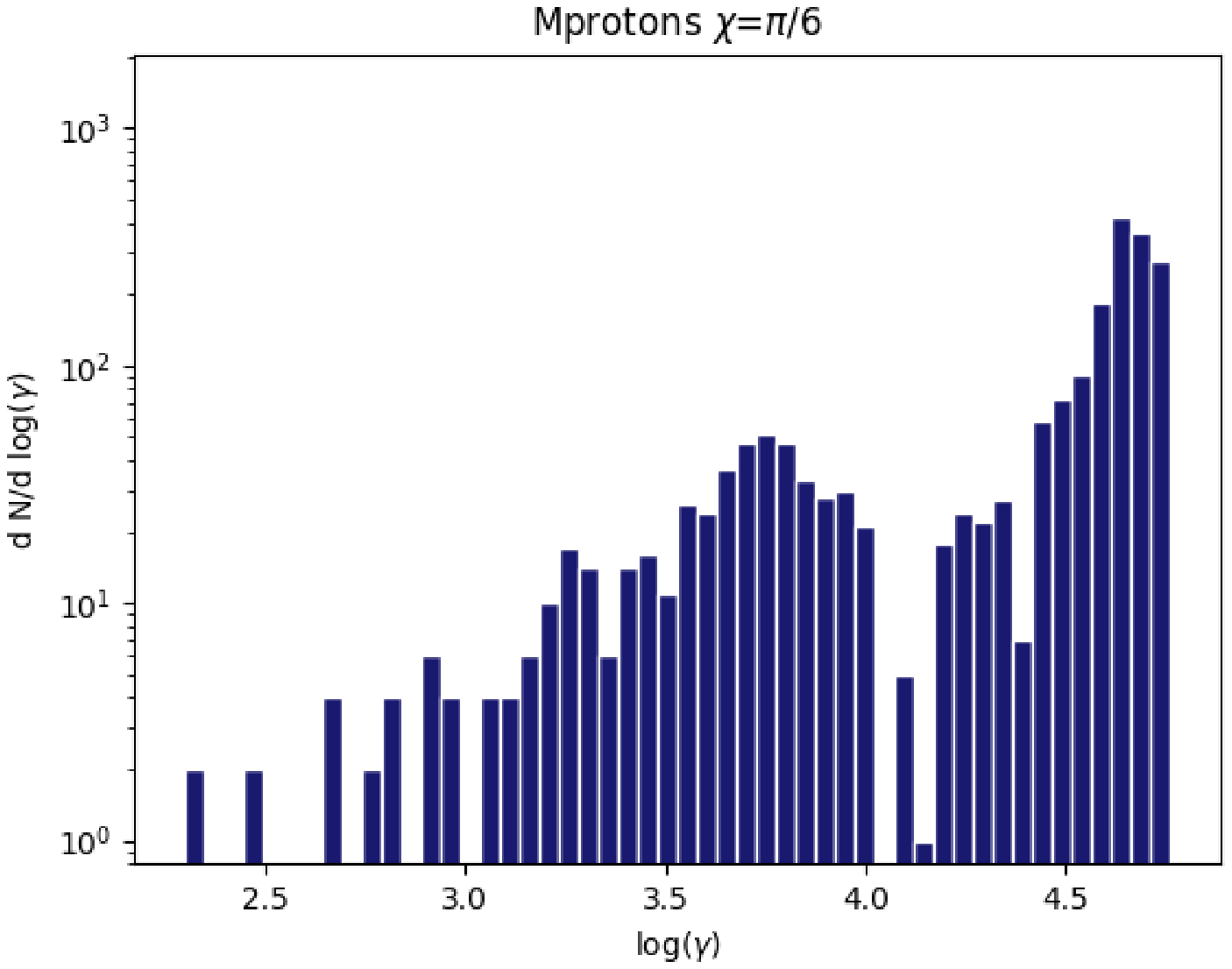}% Images in 100% size
  \caption{}
%  \caption{Lorentz factor spectrum for a fictive particle ($q=q_p$, $m=10^{6}m_p$) around a pulsar of inclination $\chi=\upi/6$}
\label{fig:Spectre_lourd_chi1sur6}
\end{subfigure}
\caption{Comparison of different particle distribution functions following the regular placement: (a) \ce{^56Fe} ions around a pulsar of inclination $\chi=30^{\circ}$, (b) Protons around a pulsar of inclination $\chi=30^{\circ}$, (c) Electrons around a pulsar of inclination $\chi=150^{\circ}$, (d) Fictive particle ($q_{Mp}=q_p$, $m_{Mp}=10^{6}m_p$) around a pulsar of inclination $\chi=30^{\circ}$.}
\label{fig:Spectre_comparison}
\end{figure}

As anticipated particles with high charge to mass ratio are accelerated more efficiently. In fact the different spectra have similar shapes simply shifted along the $\log(\gamma)$ axis. 
This is visible by comparing for instance figure~\ref{fig:Spectre_fer_chi1sur6} and figure~\ref{fig:Spectre_prot_chi1sur6} showing the distribution functions for iron and protons respectively. For these spectra, the difference is $\Delta \log(\gamma) \sim 0.33 \sim \log(2.15)$ meaning that $\gamma_{p}\sim 2.15 \gamma_{Fe}$. %($\gamma_p$ being the Lorentz factor of protons and $\gamma_{Fe}$ that of iron nuclei. 
When comparing figure~\ref{fig:Spectre_prot_chi1sur6} to figure~\ref{fig:Spectre_lourd_chi1sur6} for protons and massive particles, the shift is $\Delta \log(\gamma) \sim 6$ so $\gamma_{p} \sim 10^{6} \gamma_{Mp}$.%, with $\gamma_{Mp}$ the Lorentz factor of the fictive particles.

Just like for protons and iron nuclei, and taking into account the symmetrical behaviour of particles of opposite charge, Lorentz factor distribution functions of protons and electrons show alike features for inclination such as $\chi_e = \upi - \chi_p$ ($\chi_e$ is the inclination of the pulsar fo the electron and $\chi_p$ for the proton) but shifted by an amount $\Delta \log(\gamma) \sim 3.3 \sim \log(1836)$, compare figure~\ref{fig:Spectre_elec_chi5sur6} with figure~\ref{fig:Spectre_prot_chi1sur6}.

As an example, in figure \ref{fig:Spectre_prot_chi1sur6} with protons, when $\chi=\frac{\upi}{6}$, two lobes are visible in the spectrum: one from $\log(\gamma)=8.3$ to $\log(\gamma)=10.1$ and the other from $\log(\gamma)=10.2$ to $\log(\gamma)=10.7$. For electrons in figure~\ref{fig:Spectre_elec_chi5sur6}, with $\chi=\frac{5\upi}{6}$ the lobes also appear but from $\log(\gamma)=11.6$ to $\log(\gamma)=13.5$ and from $\log(\gamma)=13.5$ to $\log(\gamma)=14$. This difference show that $\gamma_e \sim 1836 \gamma_p$ ($\gamma_e$ is the Lorentz factor of electrons) for $\chi_e= \upi - \chi_p$.

Noticing the following ordering
\begin{equation}\label{eq:Ordering}
\dfrac{q_{p}/m_{p}}{q_{Fe}/m_{Fe}}=\dfrac{56}{26} \sim 2.154 \quad , \quad \dfrac{|q_{e}/m_{e}|}{|q_{p}/m_{p}|} \sim 1836 \quad , \quad \dfrac{q_{p}/m_{p}}{q_{Mp}/m_{Mp}} = 10^6
\end{equation}
we conclude that the final Lorentz factor of the test particles is proportional to the strength parameter of these particles, as long as they are relativistic.

\begin{figure}
  \begin{subfigure}[t]{0.03\textwidth}
    \textbf{(a)}
  \end{subfigure}
\begin{subfigure}{.5\textwidth}
  \centering
  \includegraphics[scale=0.4]{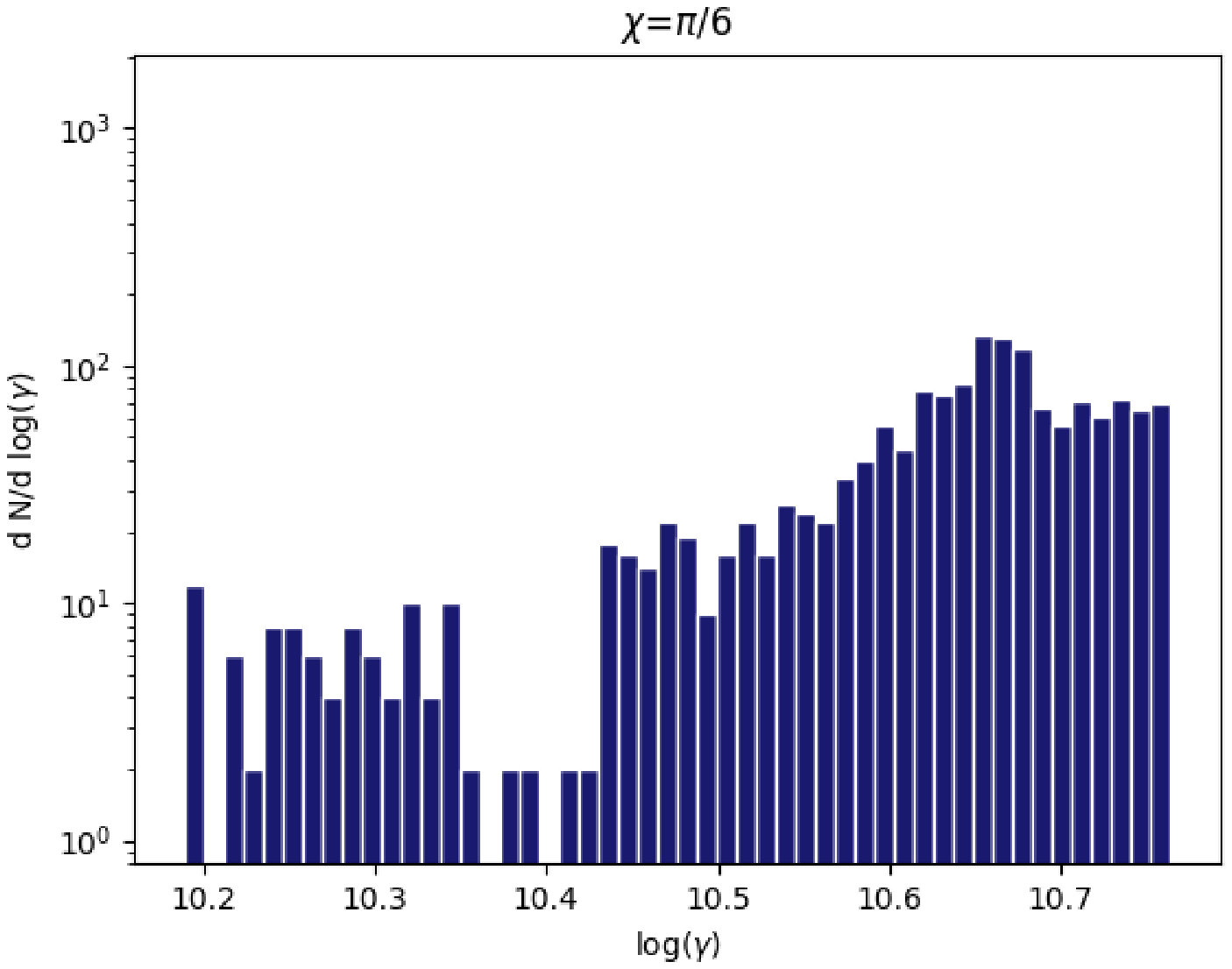}% Images in 100% size
%  \caption{Lorentz factor spectrum for crashed protons around a pulsar of inclination $\chi=\upi/6$}
\label{fig:Spectre_prot_chi1sur6_crash}
\end{subfigure}
  \begin{subfigure}[t]{0.03\textwidth}
    \textbf{(b)}
  \end{subfigure}
\begin{subfigure}{.5\textwidth}
  \centering
  \includegraphics[scale=0.4]{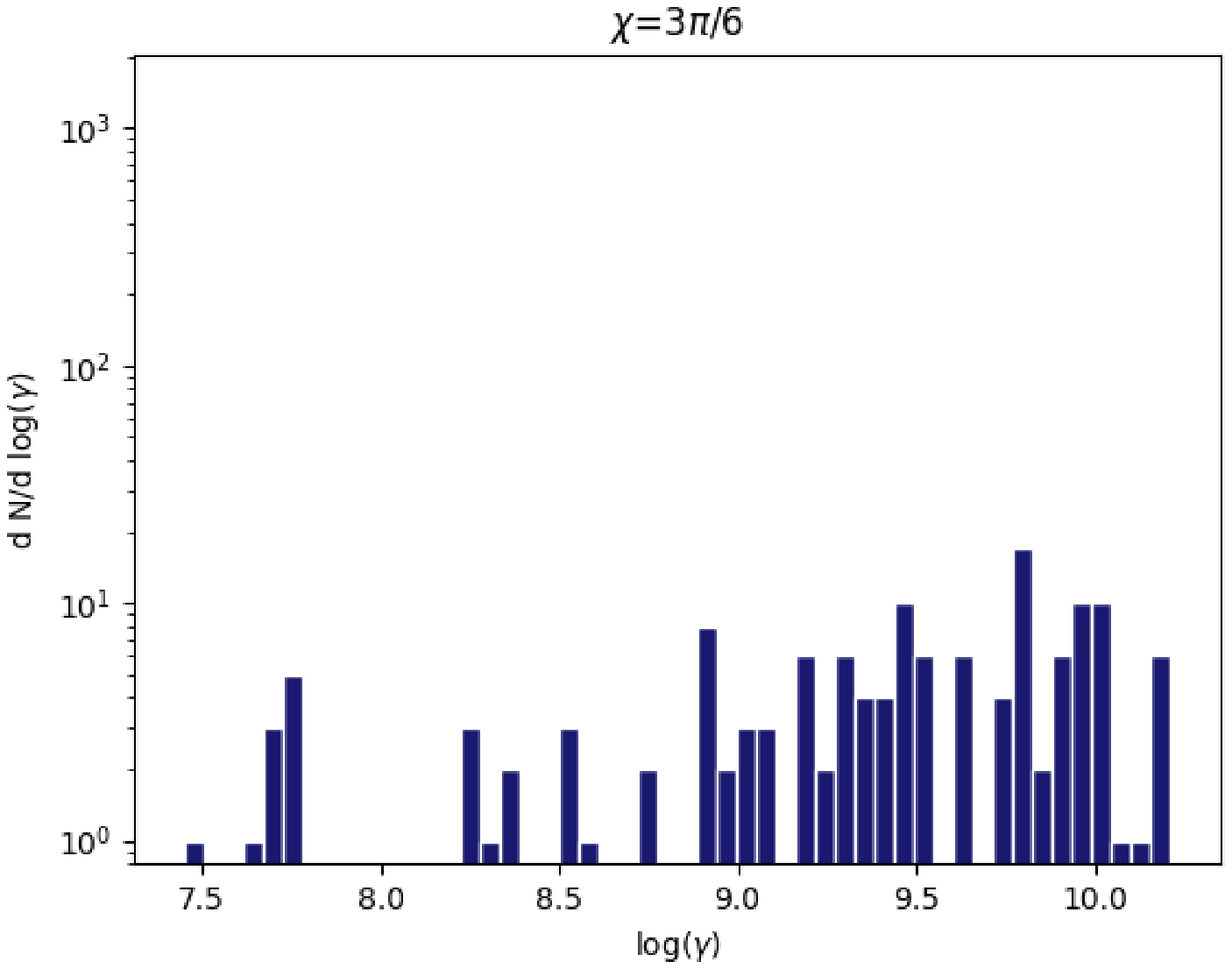}% Images in 100% size
%  \caption{Lorentz factor spectrum for crashed protons around a pulsar of inclination $\chi=\upi/2$}
\label{fig:Spectre_prot_chi3sur6_crash}
\end{subfigure}

  \begin{subfigure}[t]{0.03\textwidth}
    \textbf{(c)}
  \end{subfigure}
\begin{subfigure}{.5\textwidth}
  \centering
  \includegraphics[scale=0.4]{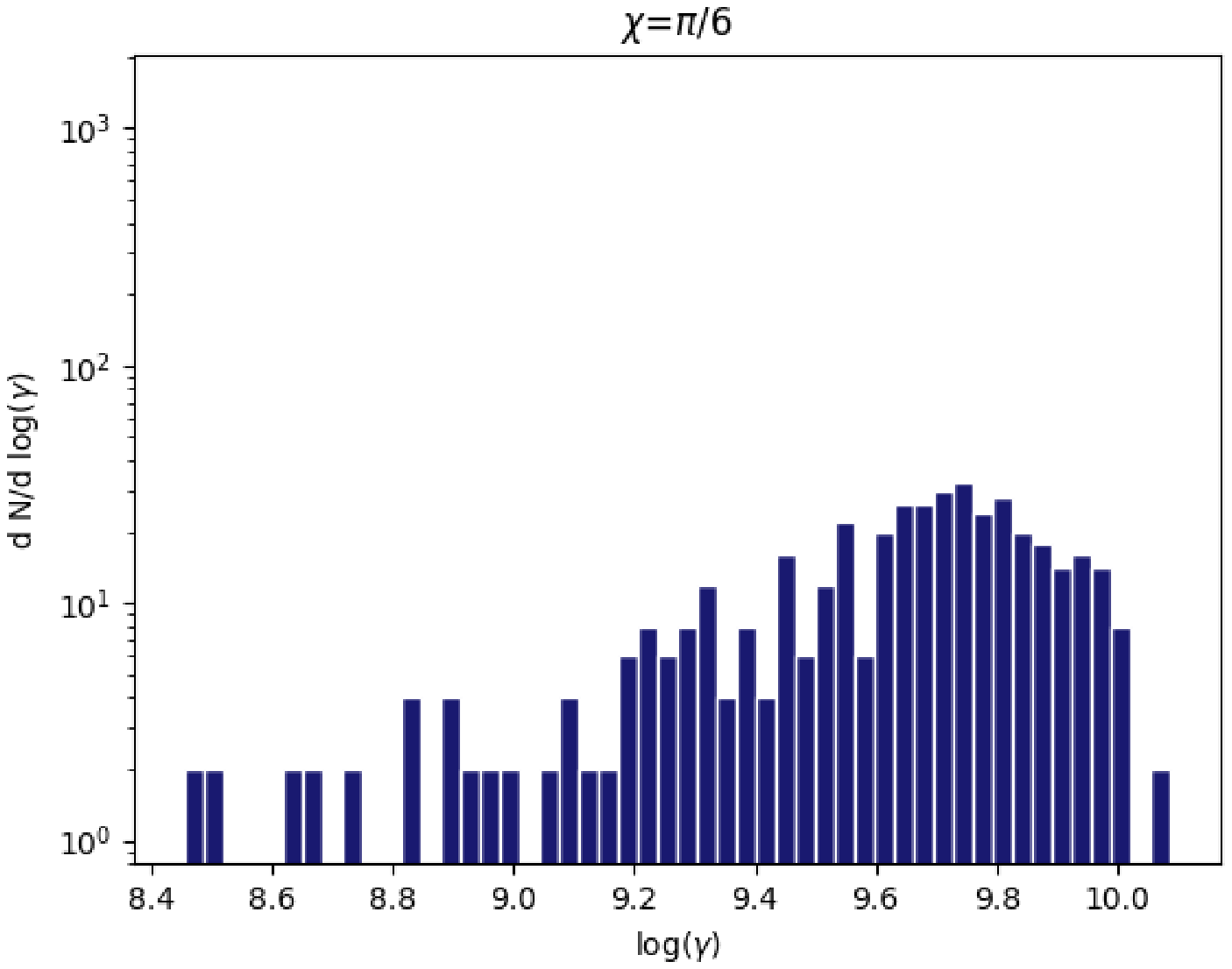}% Images in 100% size
%  \caption{Lorentz factor spectrum for ejected protons around a pulsar of inclination $\chi=\upi/6$}
\label{fig:Spectre_prot_chi1sur6_eject}
\end{subfigure}
  \begin{subfigure}[t]{0.03\textwidth}
    \textbf{(d)}
  \end{subfigure}
\begin{subfigure}{.5\textwidth}
  \centering
  \includegraphics[scale=0.4]{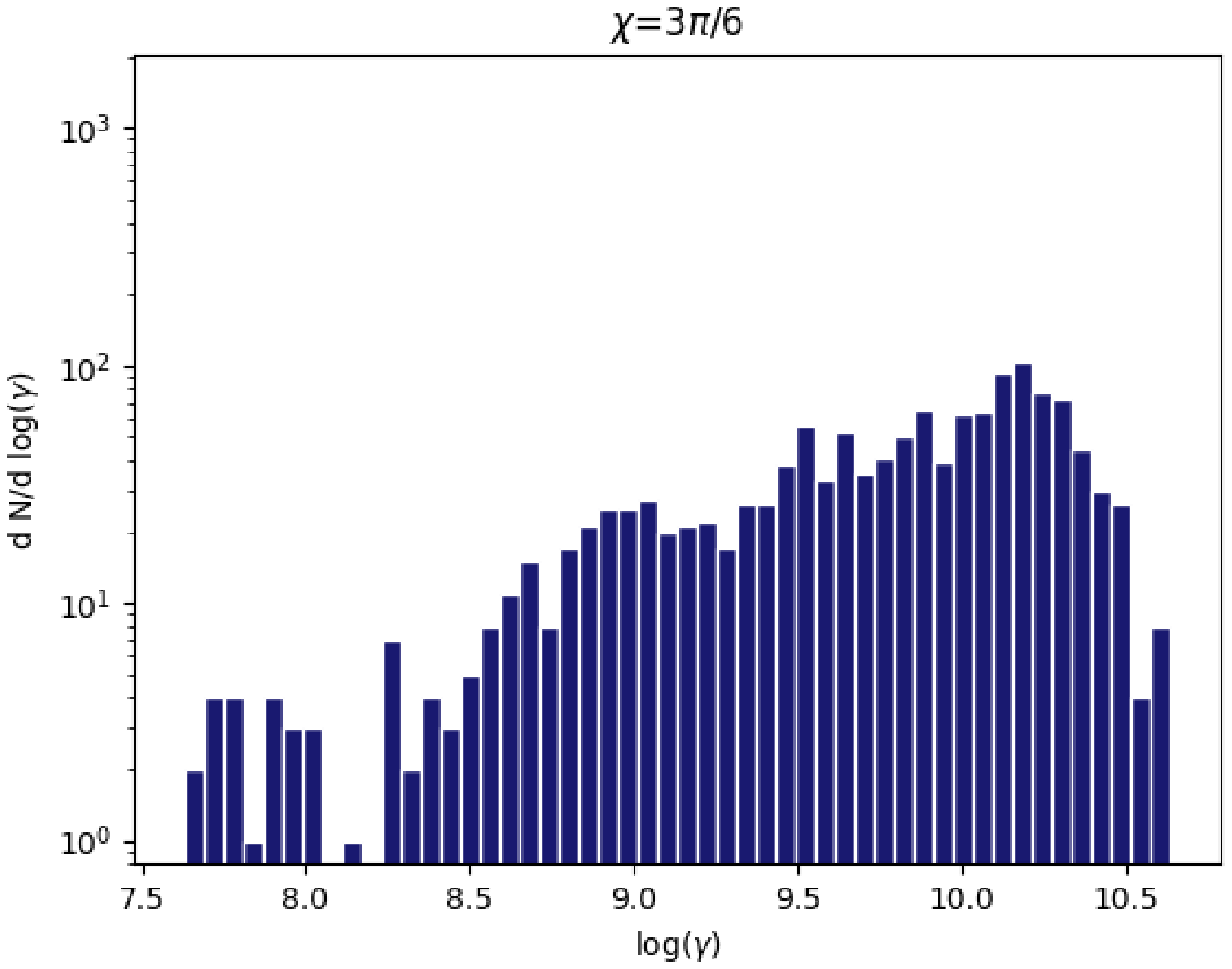}% Images in 100% size
%  \caption{Lorentz factor spectrum for ejected protons around a pulsar of inclination $\chi=\upi/2$}
\label{fig:Spectre_prot_chi3sur6_eject}
\end{subfigure}

  \begin{subfigure}[t]{0.03\textwidth}
    \textbf{(e)}
  \end{subfigure}
\begin{subfigure}{.5\textwidth}
  \centering
  \includegraphics[scale=0.4]{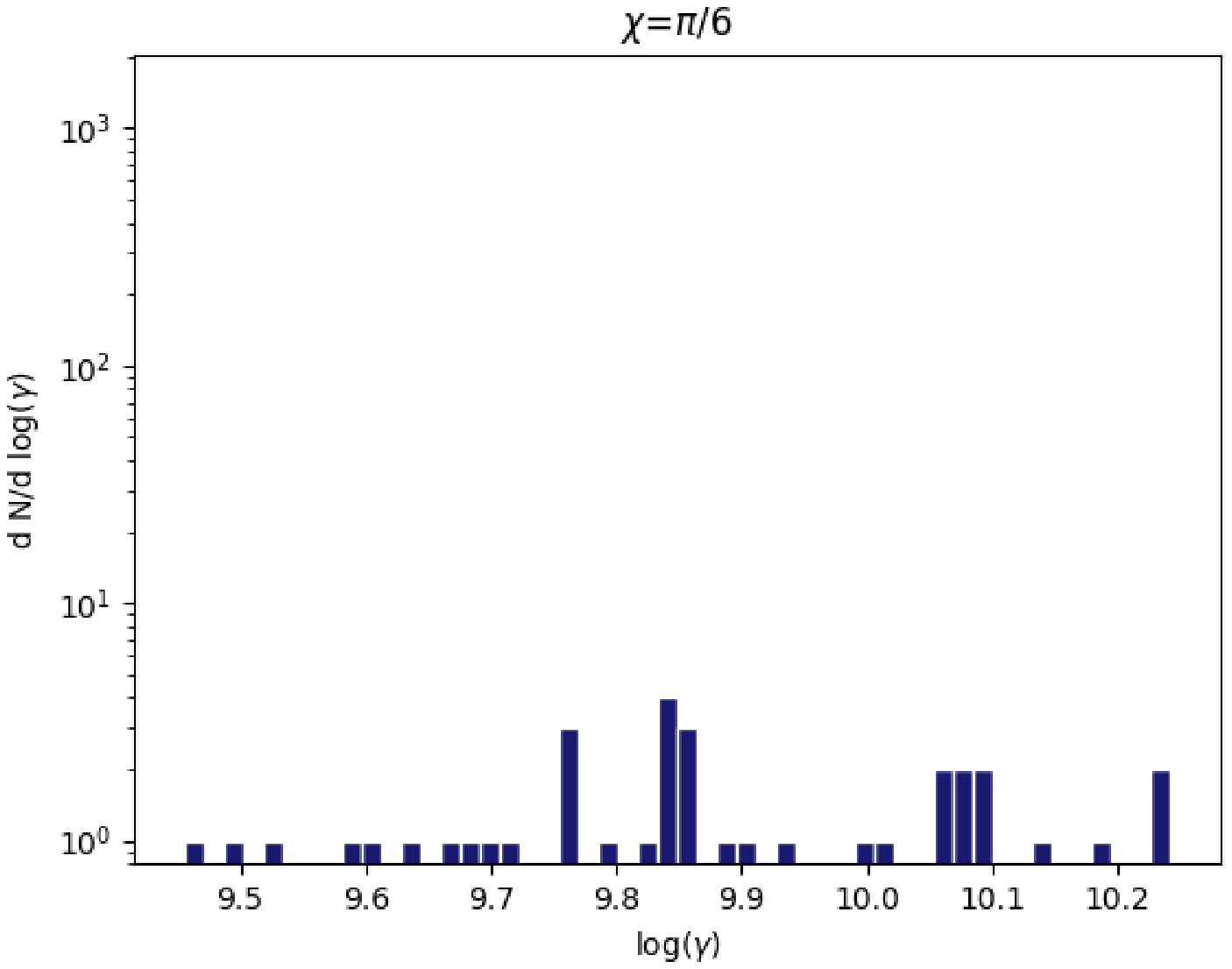}% Images in 100% size
%  \caption{Lorentz factor spectrum for trapped protons around a pulsar of inclination $\chi=\upi/6$}
\label{fig:Spectre_prot_chi1sur6_trap}
\end{subfigure}
  \begin{subfigure}[t]{0.03\textwidth}
    \textbf{(f)}
  \end{subfigure}
\begin{subfigure}{.5\textwidth}
  \centering
  \includegraphics[scale=0.4]{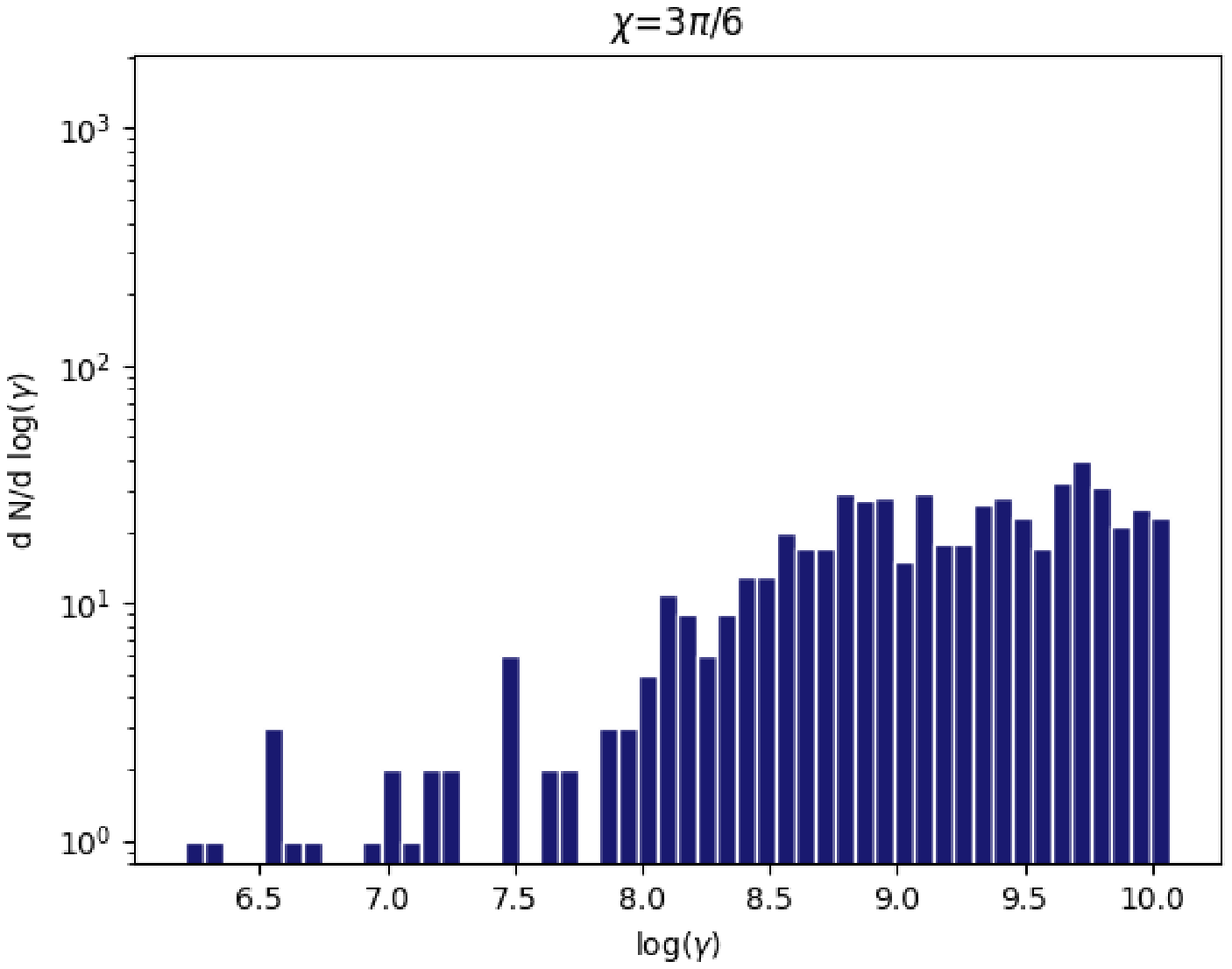}% Images in 100% size
%  \caption{Lorentz factor spectrum for trapped protons around a pulsar of inclination $\chi=\upi/2$}
\label{fig:Spectre_prot_chi3sur6_trap}
\end{subfigure}
\caption{Proton distribution functions depending on their final state. (a) Crashed, $\chi=30^{\circ}$, (b) Crashed, $\chi=90^{\circ}$, (c) Ejected, $\chi=30^{\circ}$, (d) Ejected, $\chi=90^{\circ}$, (e) Trapped, $\chi=30^{\circ}$, (f) Trapped, $\chi=90^{\circ}$, for a regular pattern of particles. %\com{Tu n'as pas mis de lettre a,b,c,... pour les figures.}
}
\label{fig:Spectre_prot_state}
\end{figure}

In complement, the spectra were analysed depending on whether the particles were ejected, trapped or crashed like in figure \ref{fig:Spectre_prot_state}. For protons, at inclinations $\chi=\{0, \frac{\upi}{6}, \frac{\upi}{3} \} $ the most energetic particles were those falling on the neutron star (at $\chi=0^{\circ}$ all particles crashed with $10.3 \leq \log(\gamma) \leq 10.8$). For $\chi=30^{\circ}$ falling particles are responsible for the high Lorentz factor lobe of the spectra. At $\chi=90^{\circ}$ however, crashing particles had similar energies as the ejected or trapped ones. 

\subsubsection{Positions of particles}
\label{sssec:Positions}

\begin{figure}
  \begin{subfigure}[t]{0.03\textwidth}
    \textbf{(a)}
  \end{subfigure}
\begin{subfigure}{.5\textwidth}
  \centering
  \includegraphics[scale=0.4]{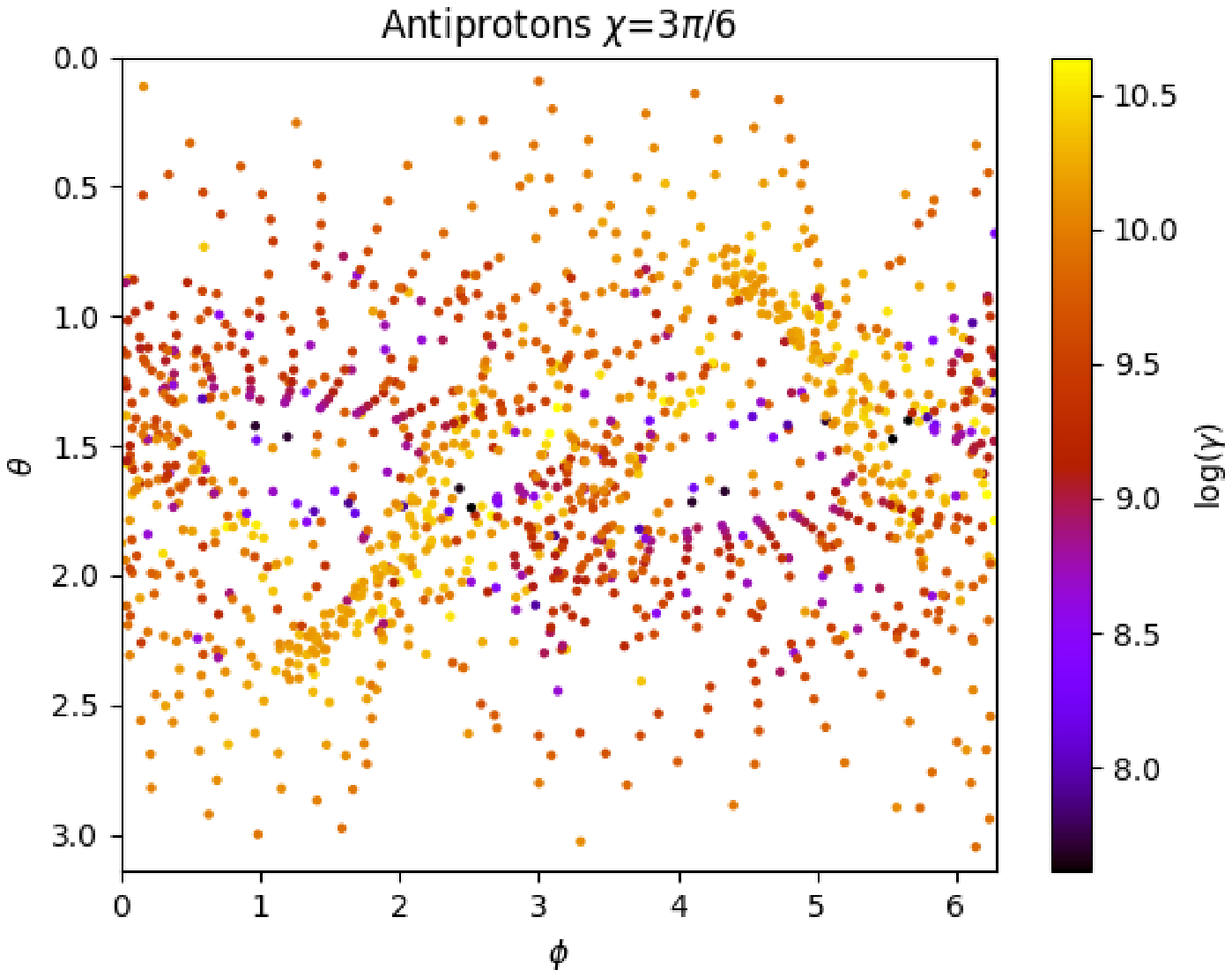}% Images in 100% size
%  \caption{Antiprotons}
\label{fig:Eject_aprot_chi3sur6}
\end{subfigure}
  \begin{subfigure}[t]{0.03\textwidth}
    \textbf{(b)}
  \end{subfigure}
\begin{subfigure}{.5\textwidth}
  \centering
  \includegraphics[scale=0.4]{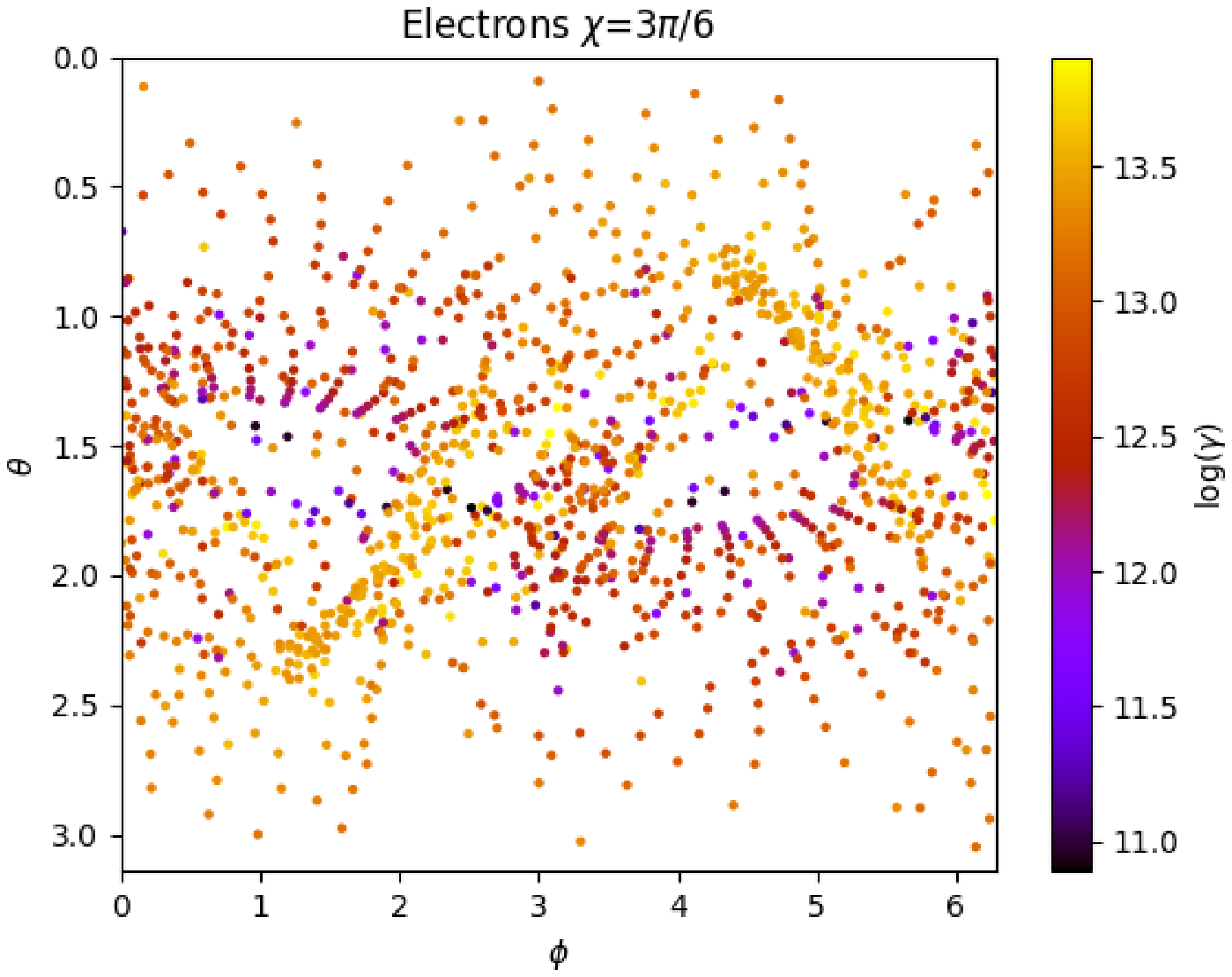}% Images in 100% size
%  \caption{Electrons}
\label{fig:Eject_elec_chi3sur6}
\end{subfigure}
  \caption{Ejection maps of antiprotons (a) and electrons (b) for $\chi=90^{\circ}$. Regular pattern placement.}
\label{fig:Ejections}
\end{figure}

The same trajectory of antiprotons and electrons despite their mass difference is visible in figure~\ref{fig:Ejections}. Our guess is that the magnetic field is so strong that particles are forced to follow the same magnetic field lines and only a low charge to mass ratio particle could have a different trajectory since it would be accelerated less efficiently and take more time to reach a speed close to~$c$.

\begin{figure}
  \begin{subfigure}[t]{0.03\textwidth}
    \textbf{(a)}
  \end{subfigure}
\begin{subfigure}{.5\textwidth}
  \centering
  \includegraphics[scale=0.4]{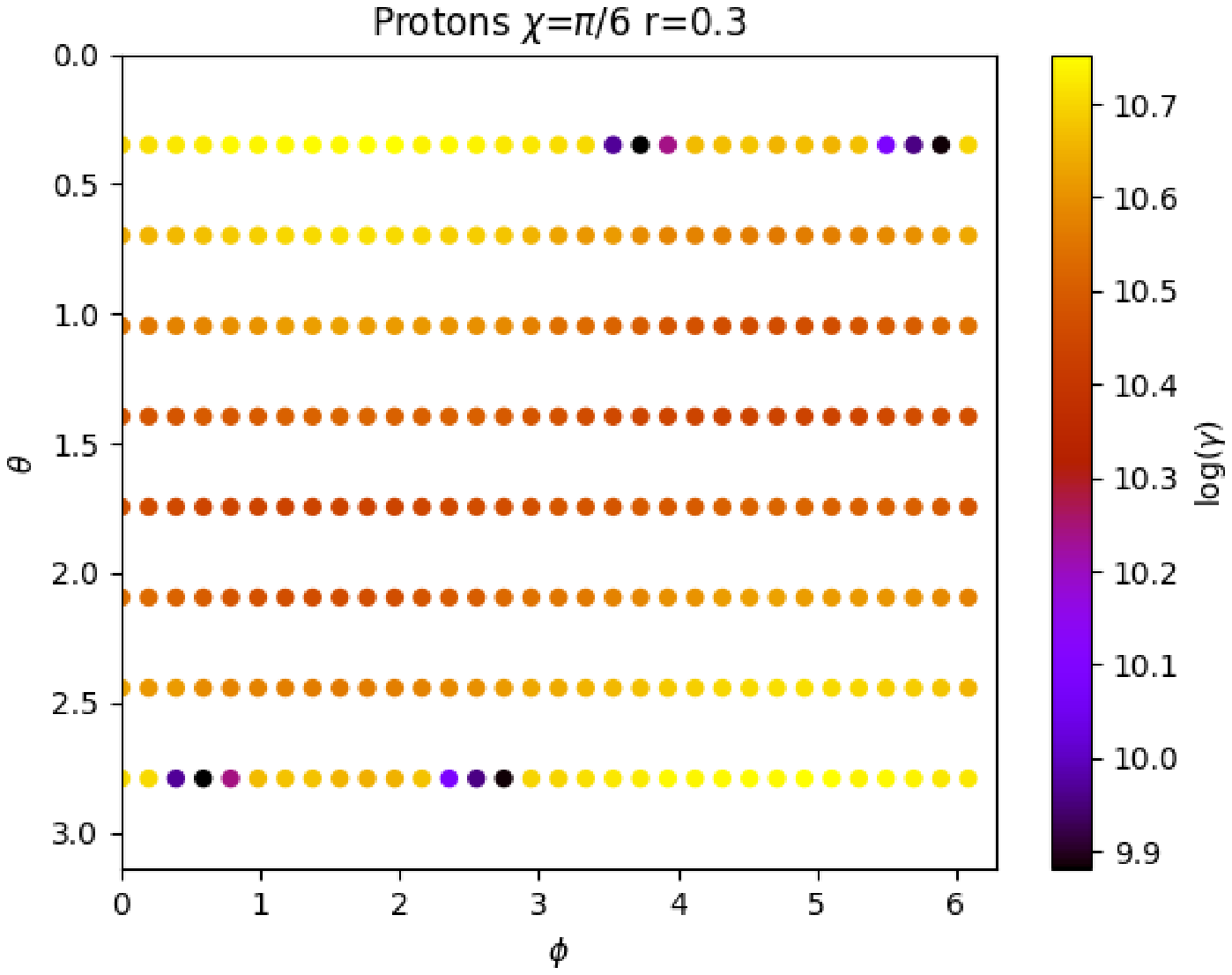}% Images in 100% size
%  \caption{}
\label{fig:Depart_prot_chi1sur6_r3}
\end{subfigure}
  \begin{subfigure}[t]{0.03\textwidth}
    \textbf{(b)}
  \end{subfigure}
\begin{subfigure}{.5\textwidth}
  \centering
  \includegraphics[scale=0.4]{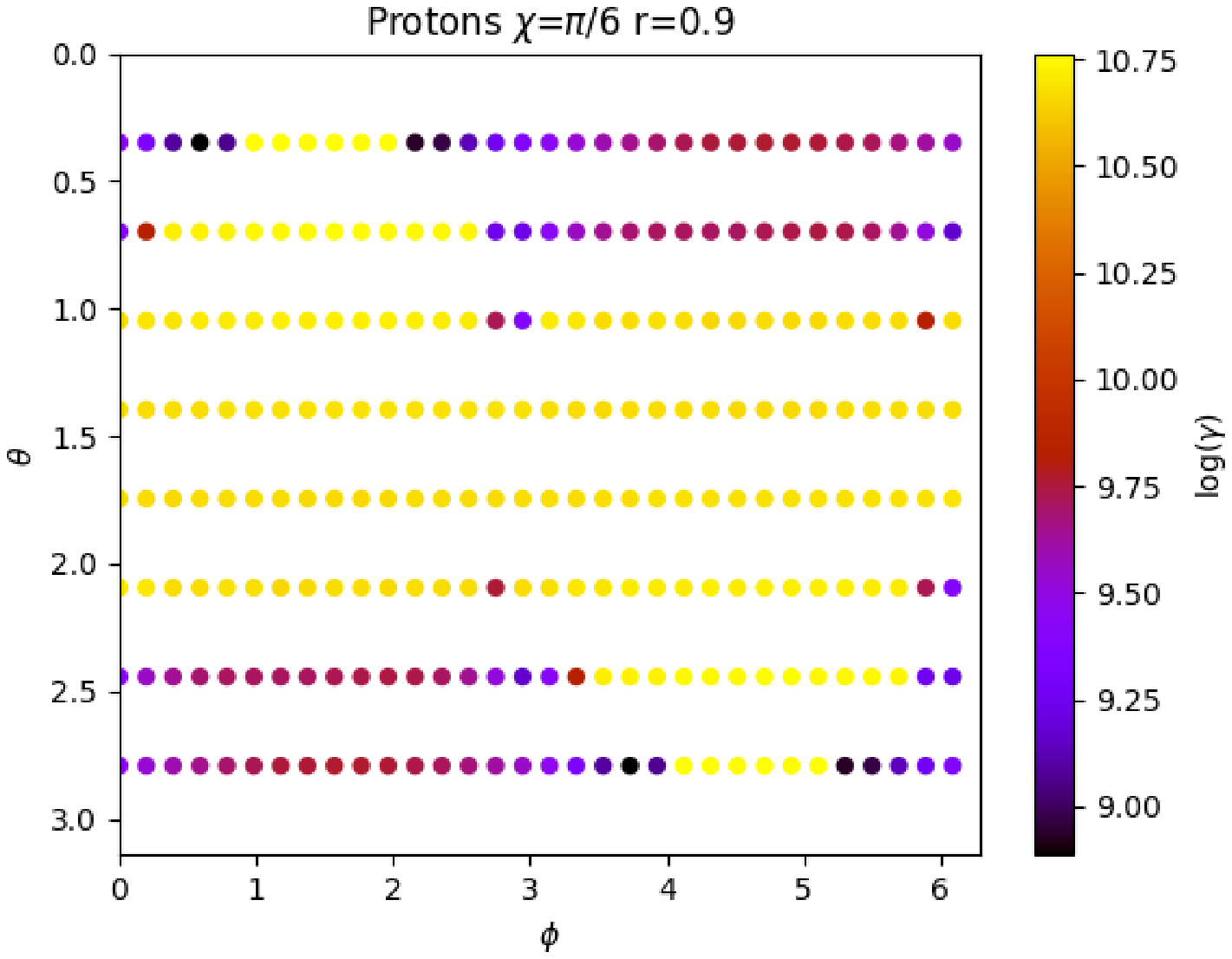}% Images in 100% size
%  \caption{}
\label{fig:Depart_prot_chi1sur6_r9}
\end{subfigure}

  \begin{subfigure}[t]{0.03\textwidth}
    \textbf{(c)}
  \end{subfigure}
\begin{subfigure}{.5\textwidth}
  \centering
  \includegraphics[scale=0.4]{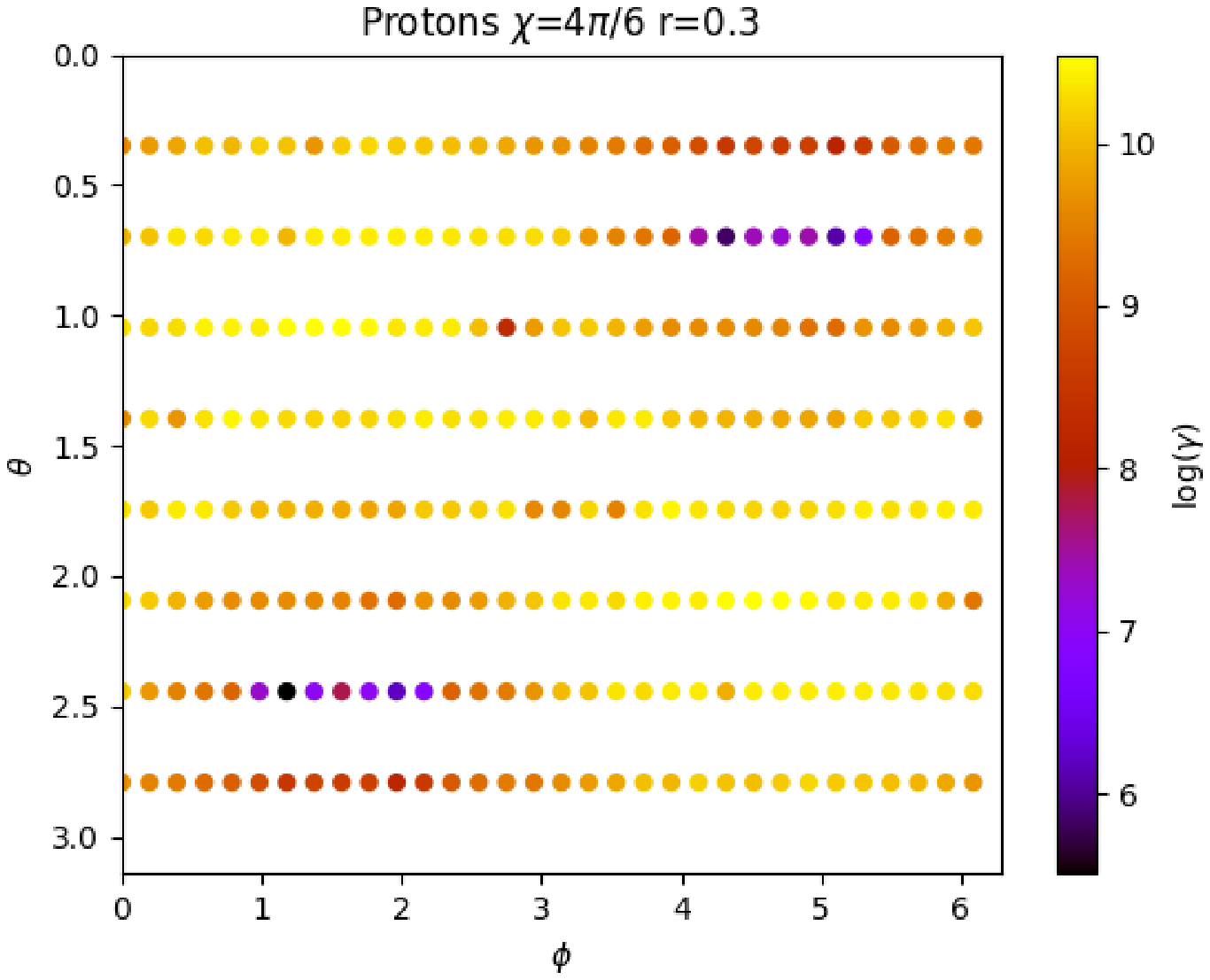}% Images in 100% size
%  \caption{}
\label{fig:Depart_prot_chi4sur6_r3}
\end{subfigure}
  \begin{subfigure}[t]{0.03\textwidth}
    \textbf{(d)}
  \end{subfigure}
\begin{subfigure}{.5\textwidth}
  \centering
  \includegraphics[scale=0.4]{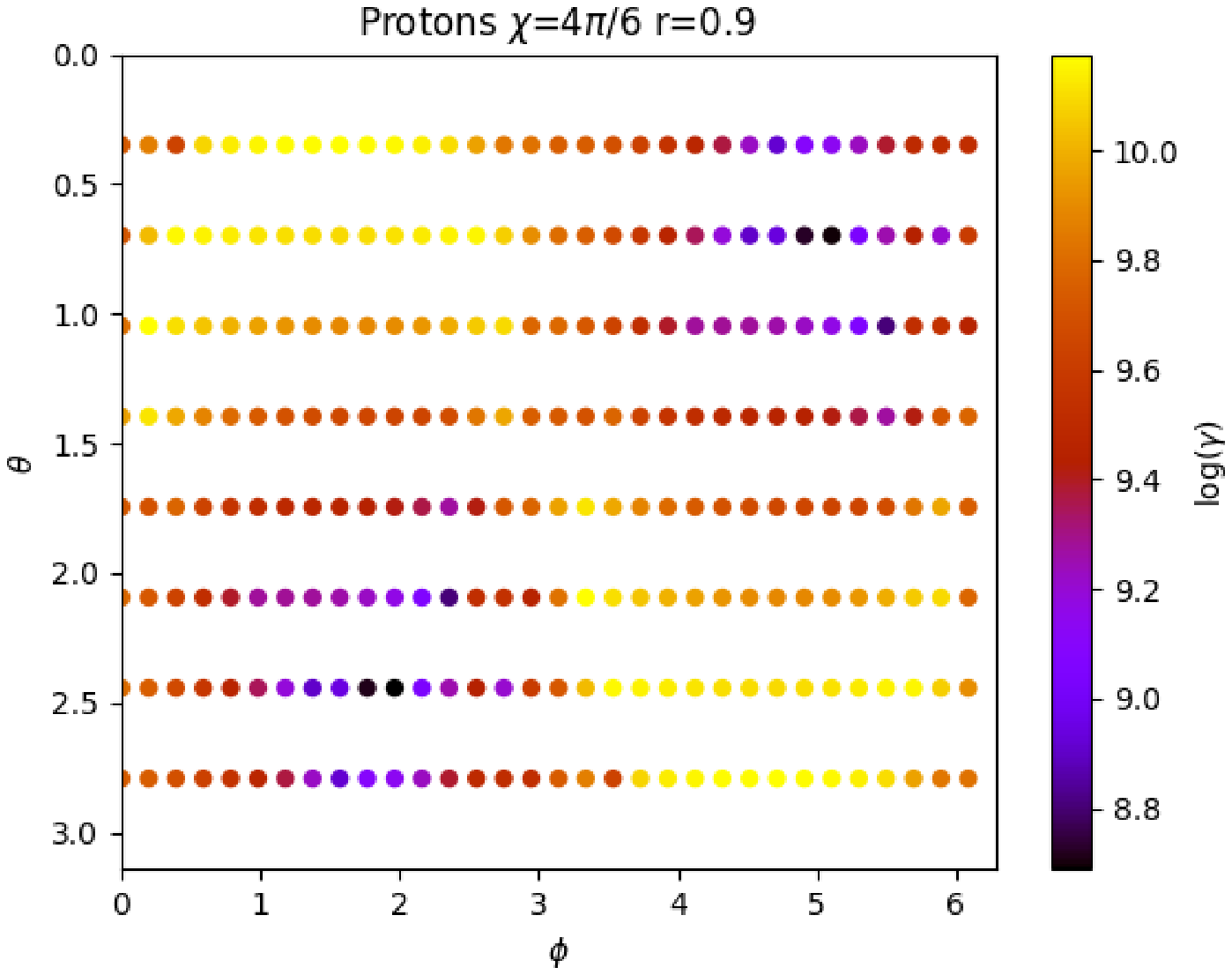}% Images in 100% size
%  \caption{}
\label{fig:Depart_prot_chi4sur6_r9}
\end{subfigure}
  \caption{Final Lorentz factor of the protons depending on their initial positions, for several pulsar inclinations and a given radius of injection of the particles: (a) $r_0=0.3\rlight$, $\chi=30^{\circ}$, (b) $r_0=0.9\rlight$, $\chi=30^{\circ}$, (c) $r_0=0.3\rlight$, $\chi=120^{\circ}$, (d) $r_0=0.9\rlight$, $\chi=120^{\circ}$.}
\label{fig:Departs_gamma}
\end{figure}

In addition, plotting the Lorentz factor map of the particles depending on their initial position shows that some areas are more prone to particle acceleration and that these regions are not randomly distributed and follow well a central symmetry, see figure~\ref{fig:Departs_gamma}. These maps highlighted however some issues since a few particles did not follow exactly the central symmetry, meaning that computationally speaking, some errors tend to appear during the simulations.  

\begin{figure}
  \begin{subfigure}[t]{0.03\textwidth}
    \textbf{(a)}
  \end{subfigure}
\begin{subfigure}{.5\textwidth}
  \centering
  \includegraphics[scale=0.4]{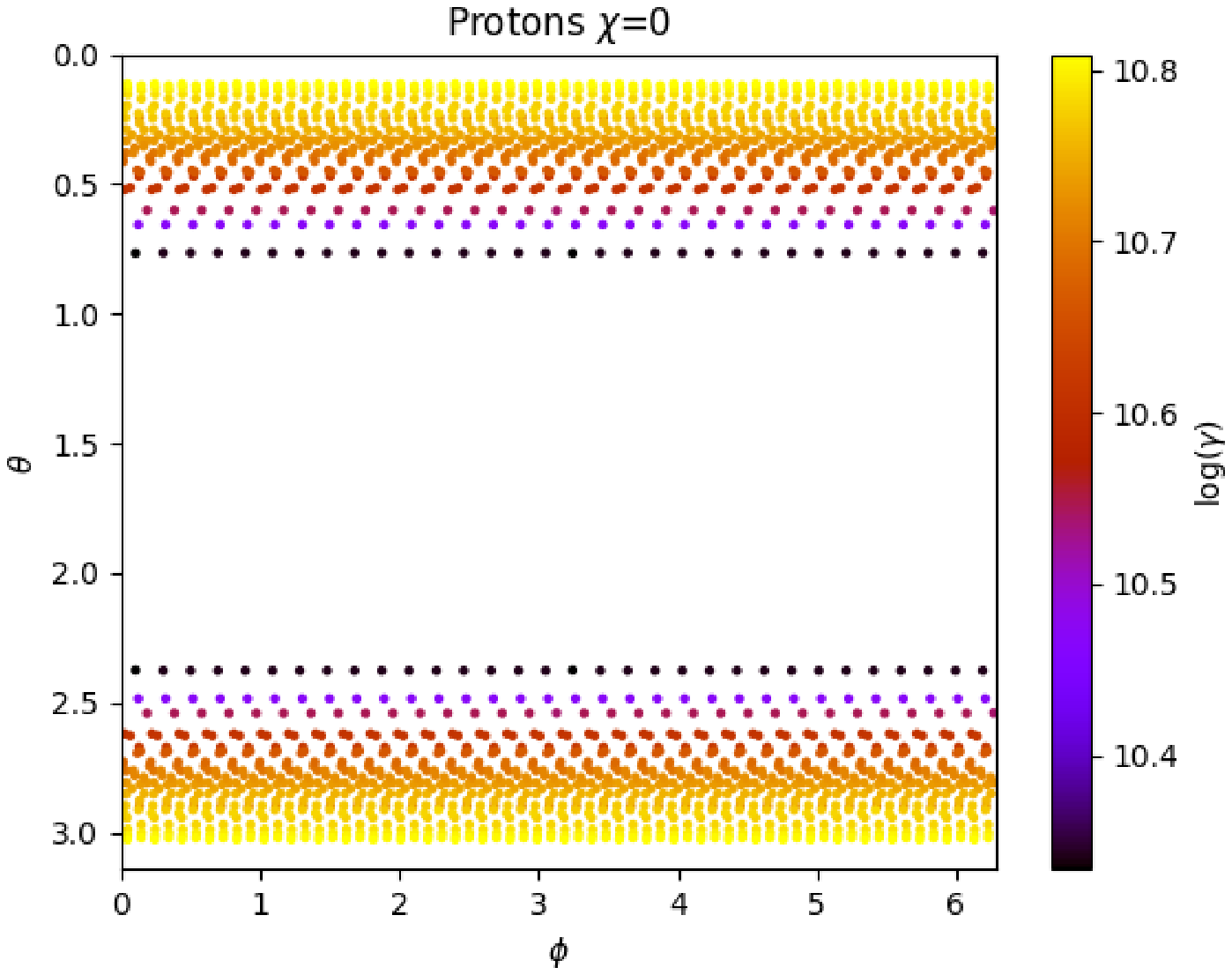}% Images in 100% size
  %\caption{$\chi=0\upi/6$}
\label{fig:Crash_prot_chi0sur6}
\end{subfigure}
  \begin{subfigure}[t]{0.03\textwidth}
    \textbf{(b)}
  \end{subfigure}
\begin{subfigure}{.5\textwidth}
  \centering
  \includegraphics[scale=0.4]{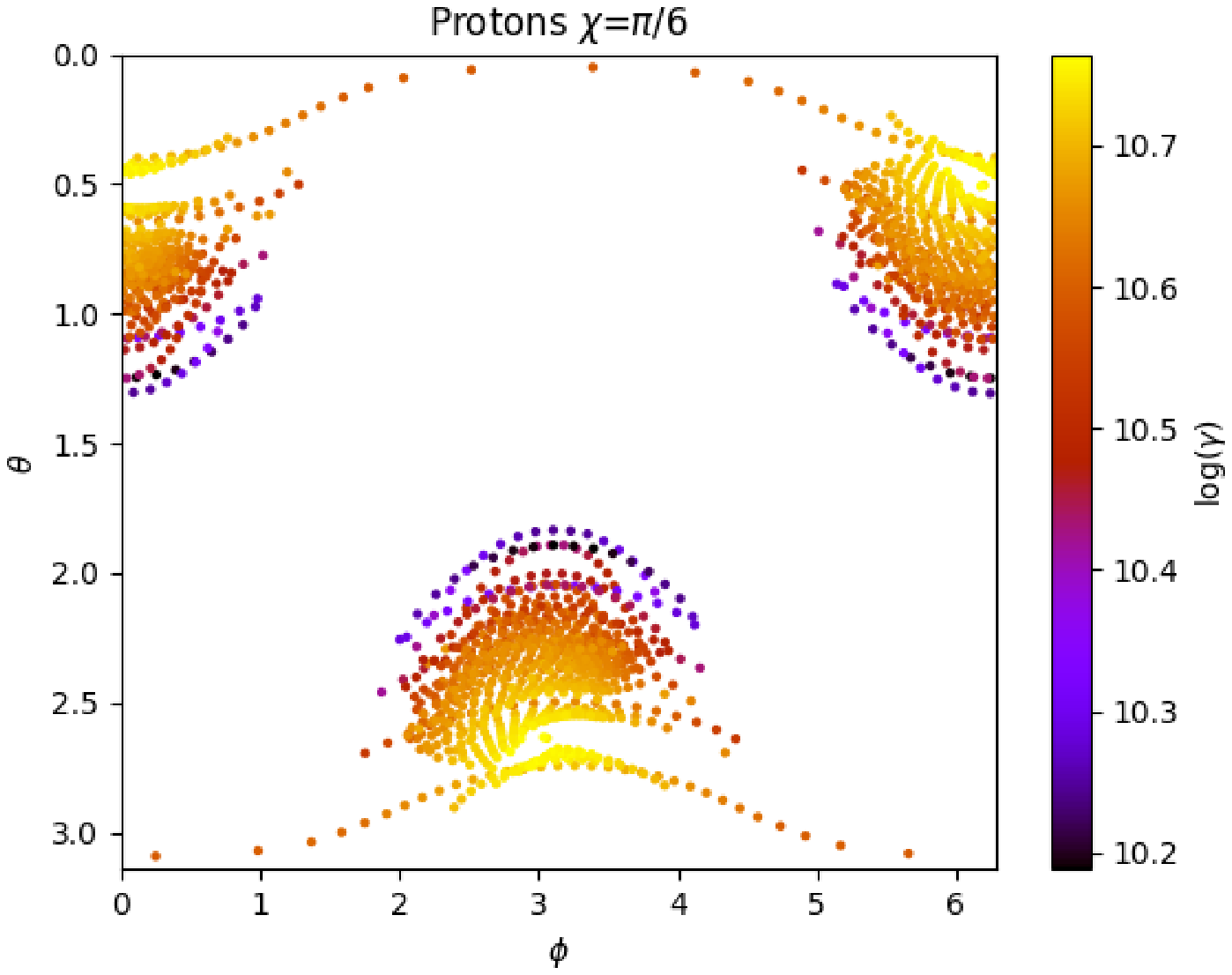}% Images in 100% size
  %\caption{$\chi=\upi/6$}
\label{fig:Crash_prot_chi1sur6}
\end{subfigure}

  \begin{subfigure}[t]{0.03\textwidth}
    \textbf{(c)}
  \end{subfigure}
\begin{subfigure}{.5\textwidth}
  \centering
  \includegraphics[scale=0.4]{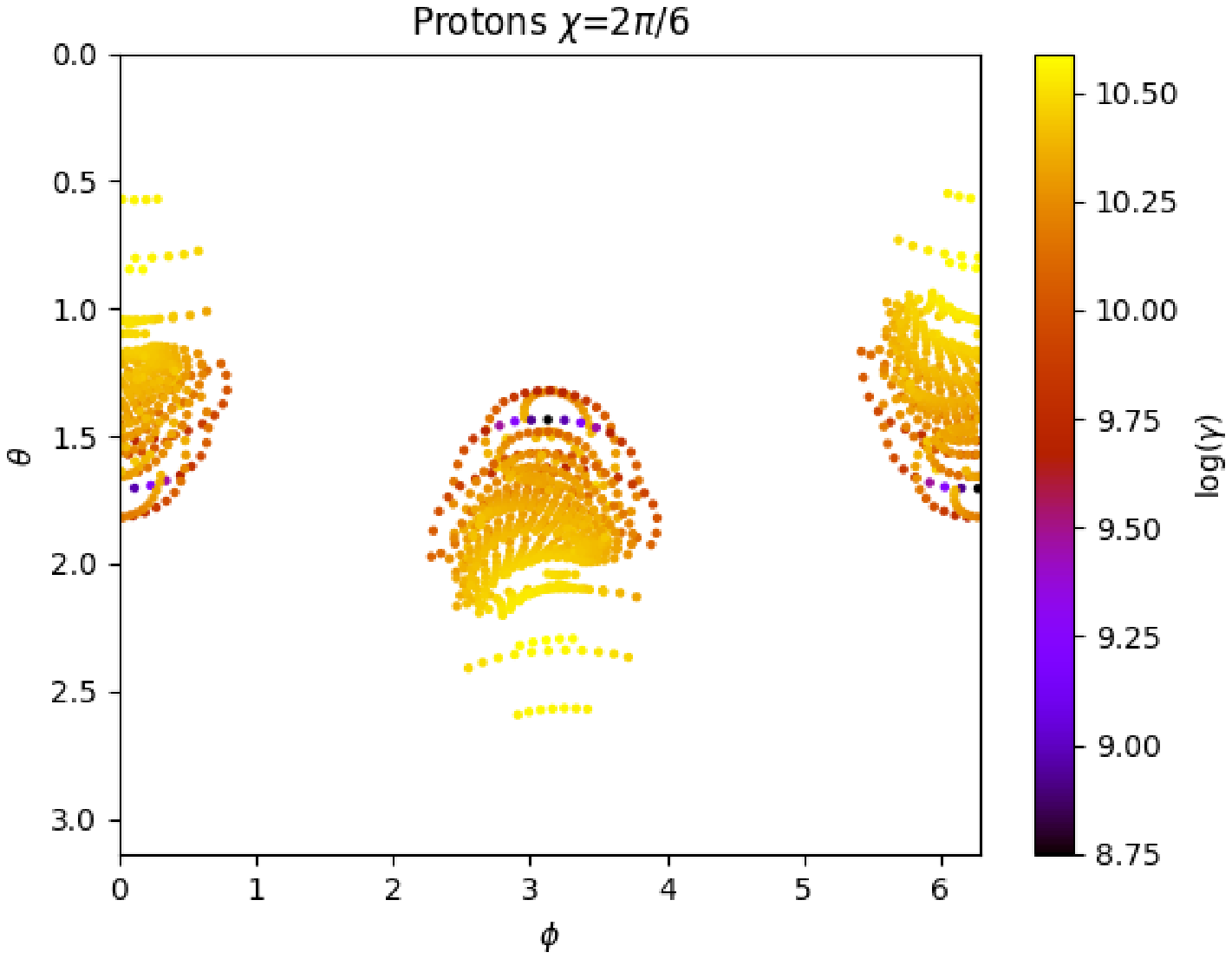}% Images in 100% size
  %\caption{$\chi=2\upi/6$}
\label{fig:Crash_prot_chi2sur6}
\end{subfigure}
  \begin{subfigure}[t]{0.03\textwidth}
    \textbf{(d)}
  \end{subfigure}
\begin{subfigure}{.5\textwidth}
  \centering
  \includegraphics[scale=0.4]{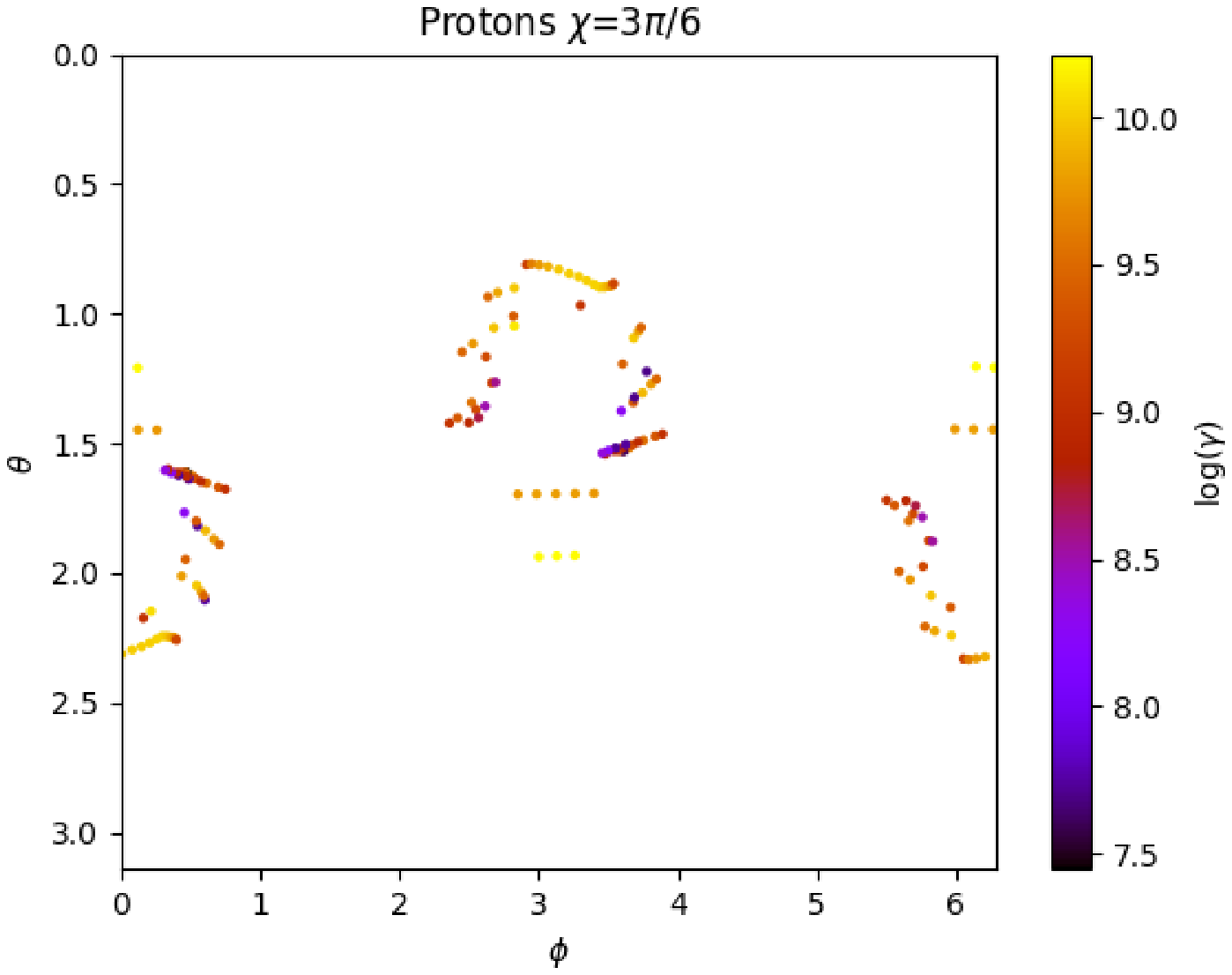}% Images in 100% size
  %\caption{$\chi=3\upi/6$}
\label{fig:Crash_prot_chi3sur6}
\end{subfigure}
  \caption{Impact maps of protons on the neutron star surface for $\chi=0^{\circ}$ (a), $\chi=30^{\circ}$ (b), $\chi=60^{\circ}$ (c) and $\chi=90^{\circ}$ (d). Regular placement of particles.}
\label{fig:Crash}
\end{figure}

Finally, aside when $\chi=90^{\circ}$ only particles with a charge of a given sign could fall onto the neutron star. Indeed particles with positive charge crashed for $0^{\circ} \leq \chi \leq 90^{\circ}$ and those with negative charge for $90^{\circ}, \leq \chi \leq 180^{\circ}$.
When investigating their final coordinates, and taking into account the delay between the start of the simulation and the time the particles hit the surface, the impact maps show that the particles form hotspots around the magnetic axis instead of randomly hitting the surface, as figure~\ref{fig:Crash} shows.
 
\subsection{Randomly placed particles}

%\revtwo{
%5) The results then discuss particle distributions, always based on 2.048 particles that start at rest in specified locations. It would be relevant to see how in one of the cases shown, increasing the coverage (now 8x8x32 in 3D space) gives a clearly similar result as discussed here: perhaps some of the 'gaps' close up when increasing particle numbers, and so the actual distributions may still modify?
%}

\subsubsection{Lorentz factor of the particles}

\begin{figure}
  \begin{subfigure}[t]{0.03\textwidth}
    \textbf{(a)}
  \end{subfigure}
\begin{subfigure}{.46\textwidth}
  \centering
  \includegraphics[scale=0.4]{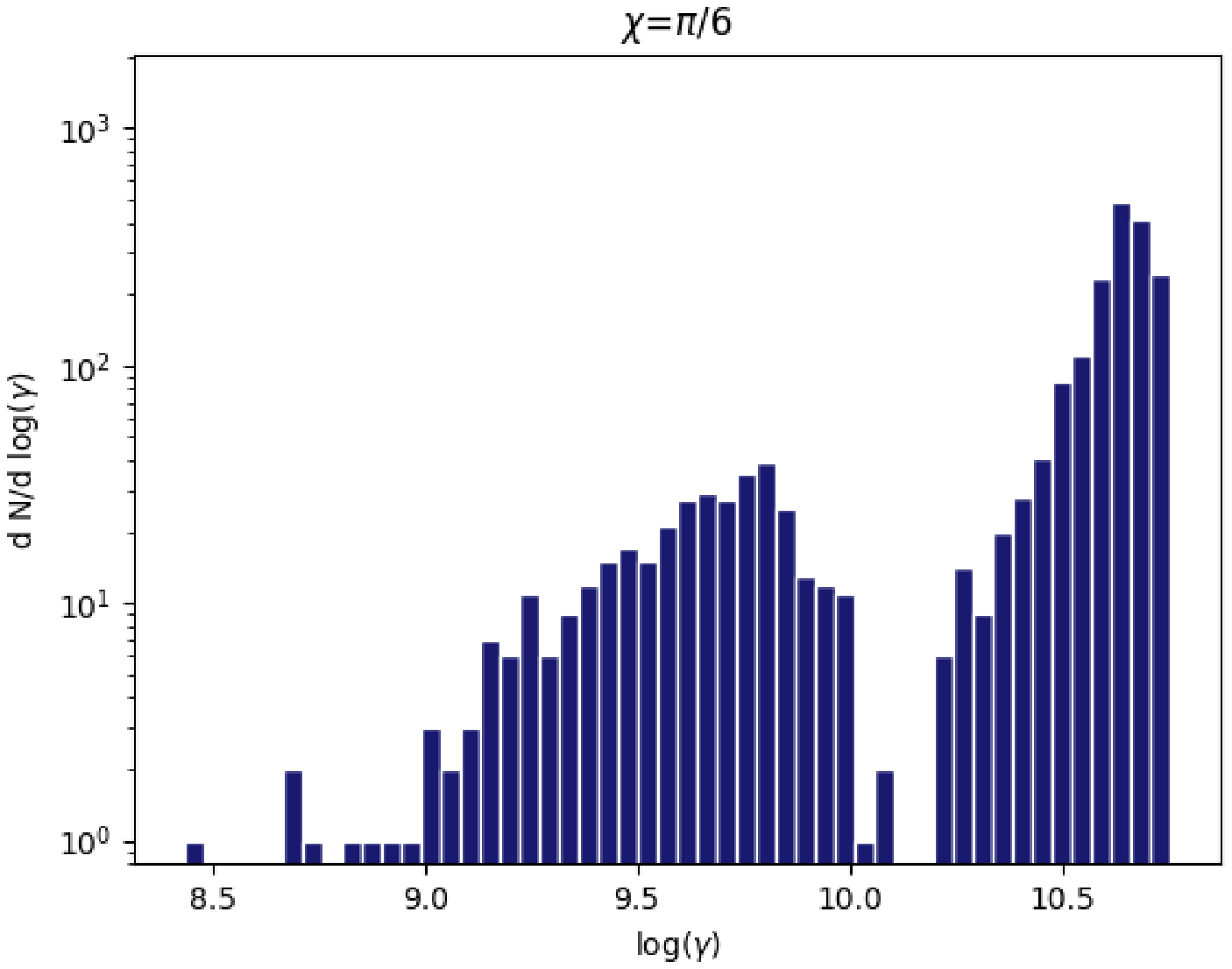}% Images in 100% size
  \caption{}
%  \caption{Lorentz factor spectrum for iron ions around a pulsar of inclination $\chi=\upi/6$}
\label{fig:Spectre_prot_chi1sur6_rand}
\end{subfigure}
  \begin{subfigure}[t]{0.03\textwidth}
    \textbf{(b)}
  \end{subfigure}
\begin{subfigure}{.46\textwidth}
  \centering
  \includegraphics[scale=0.4]{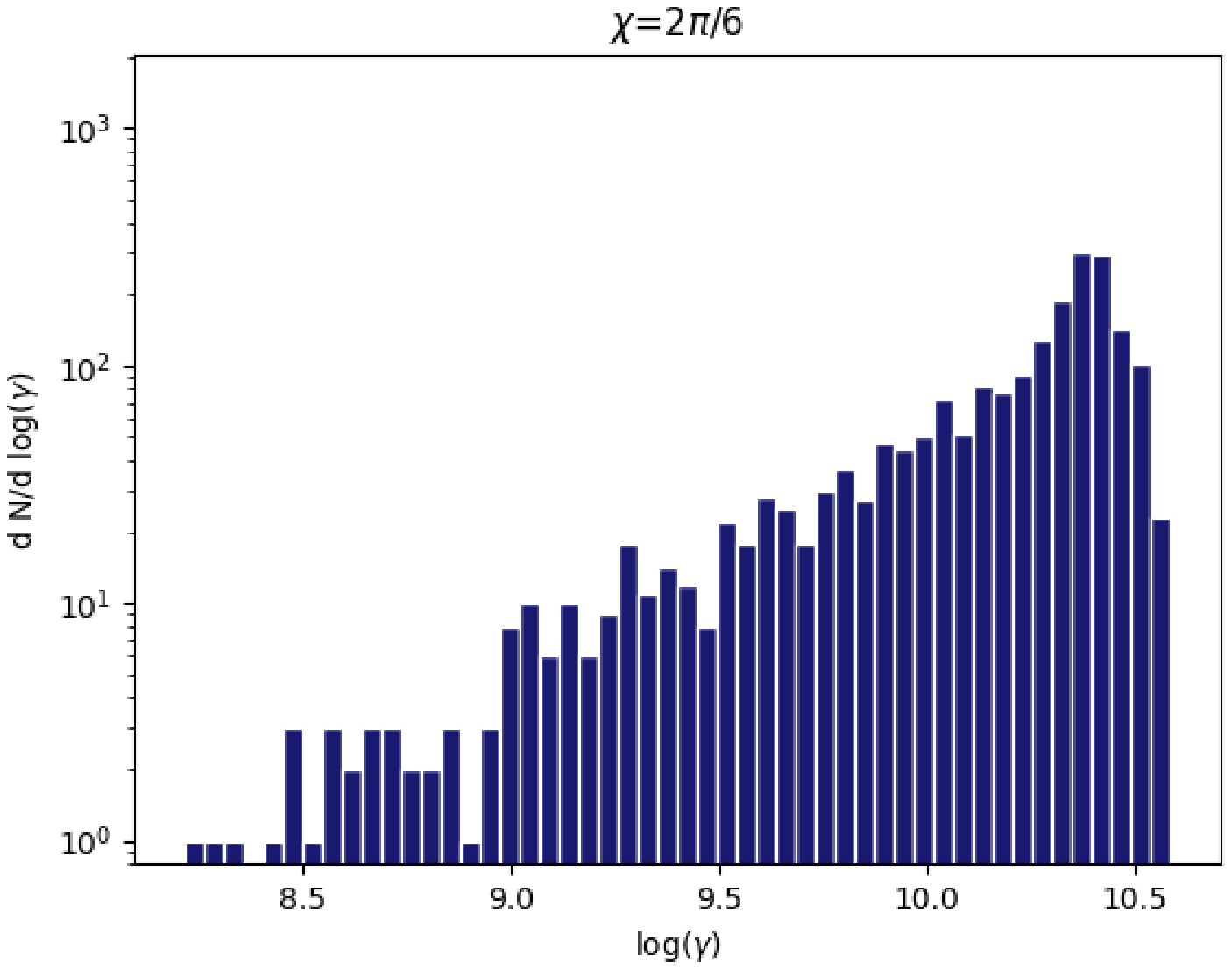}% Images in 100% size
  \caption{}
%  \caption{Lorentz factor spectrum for }
\label{fig:Spectre_prot_chi2sur6_rand}
\end{subfigure}
  \begin{subfigure}[t]{0.03\textwidth}
    \textbf{(c)}
  \end{subfigure}
\begin{subfigure}{.46\textwidth}
  \centering
  \includegraphics[scale=0.4]{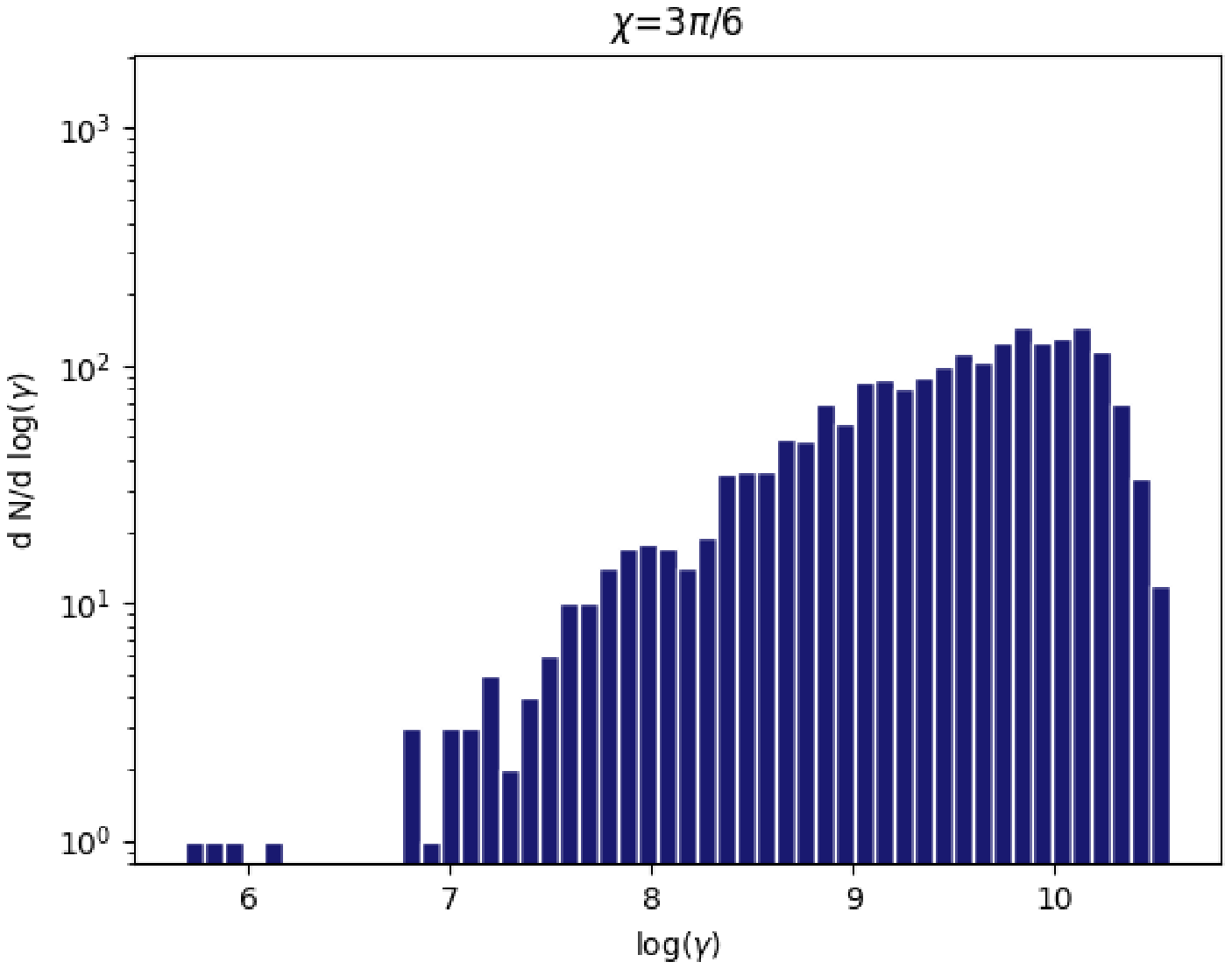}% Images in 100% size
  \caption{}
%  \caption{Lorentz factor spectrum for electrons around a pulsar of inclination $\chi=5\upi/6$}
\label{fig:Spectre_prot_chi3sur6_rand}
\end{subfigure}
  \begin{subfigure}[t]{0.03\textwidth}
    \textbf{(d)}
  \end{subfigure}
\begin{subfigure}{.46\textwidth}
  \centering
  \includegraphics[scale=0.4]{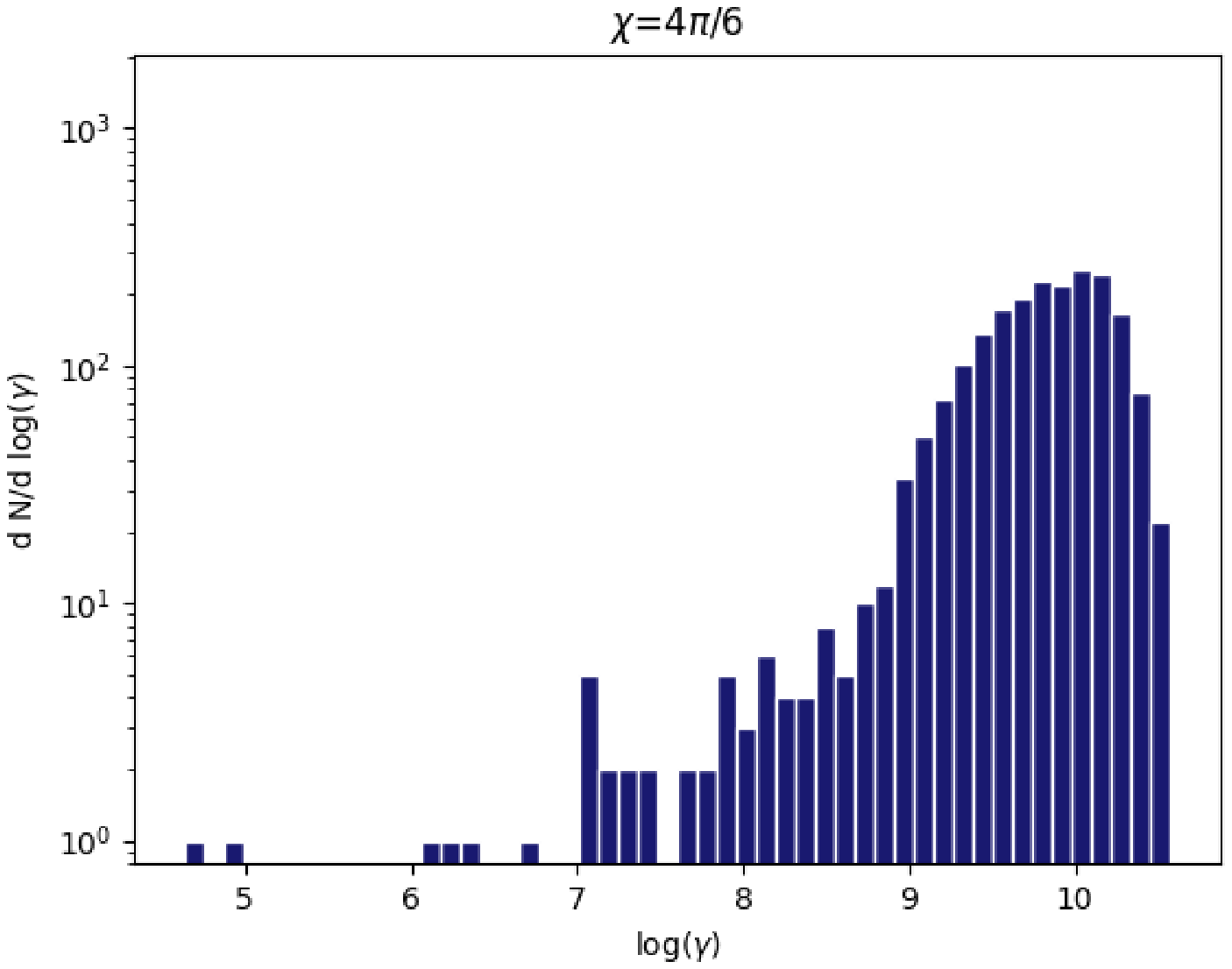}% Images in 100% size
  \caption{}
%  \caption{Lorentz factor spectrum for a fictive particle ($q=q_p$, $m=10^{6}m_p$) around a pulsar of inclination $\chi=\upi/6$}
\label{fig:Spectre_prot_chi4sur6_rand}
\end{subfigure}
\caption{Comparison of the distribution functions for 2.048 protons following the random placement: (a) inclination $\chi=30^{\circ}$, (b) inclination $\chi=60^{\circ}$, (c) inclination $\chi=90^{\circ}$, (d) inclination $\chi=120^{\circ}$.}
\label{fig:Spectre_comparison_rand}
\end{figure}

\begin{figure}
  \begin{subfigure}[t]{0.03\textwidth}
    \textbf{(a)}
  \end{subfigure}
\begin{subfigure}{.46\textwidth}
  \centering
  \includegraphics[scale=0.4]{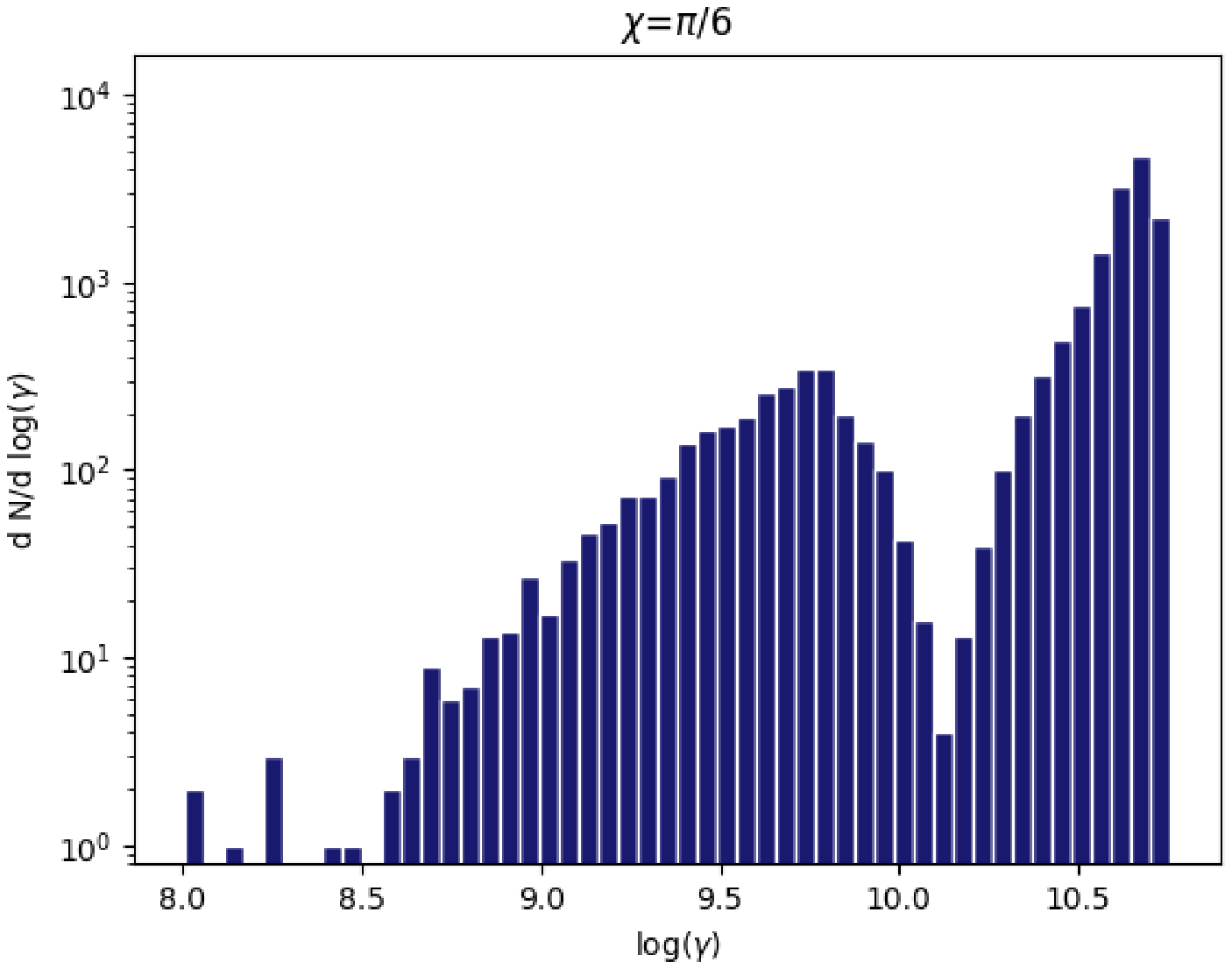}% Images in 100% size
  \caption{}
%  \caption{Lorentz factor spectrum for iron ions around a pulsar of inclination $\chi=\upi/6$}
\label{fig:Spectre_fer_chi1sur6_rand_16384}
\end{subfigure}
  \begin{subfigure}[t]{0.03\textwidth}
    \textbf{(b)}
  \end{subfigure}
\begin{subfigure}{.46\textwidth}
  \centering
  \includegraphics[scale=0.4]{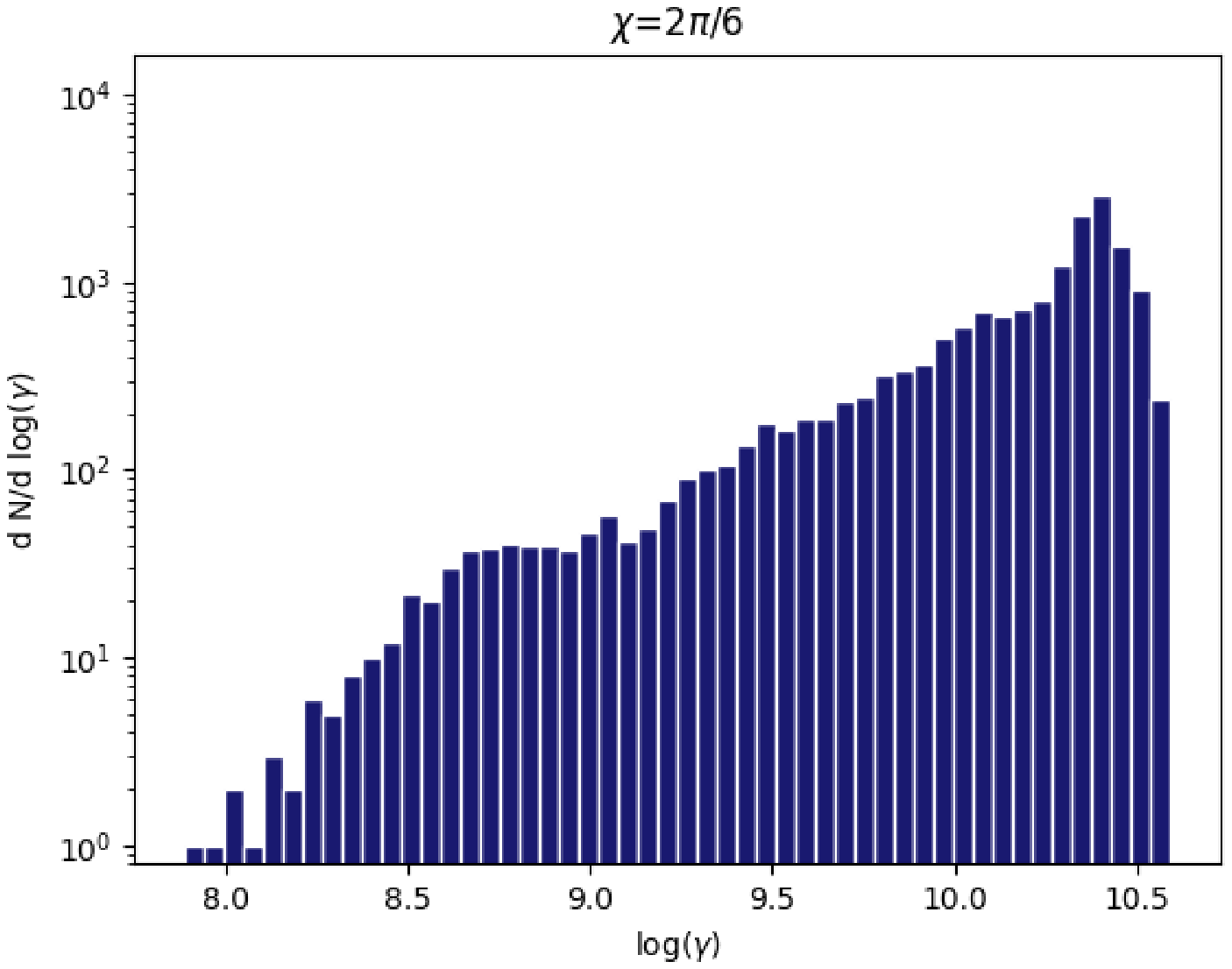}% Images in 100% size
  \caption{}
%  \caption{Lorentz factor spectrum for }
\label{fig:Spectre_prot_chi2sur6_rand_16384}
\end{subfigure}
  \begin{subfigure}[t]{0.03\textwidth}
    \textbf{(c)}
  \end{subfigure}
\begin{subfigure}{.46\textwidth}
  \centering
  \includegraphics[scale=0.4]{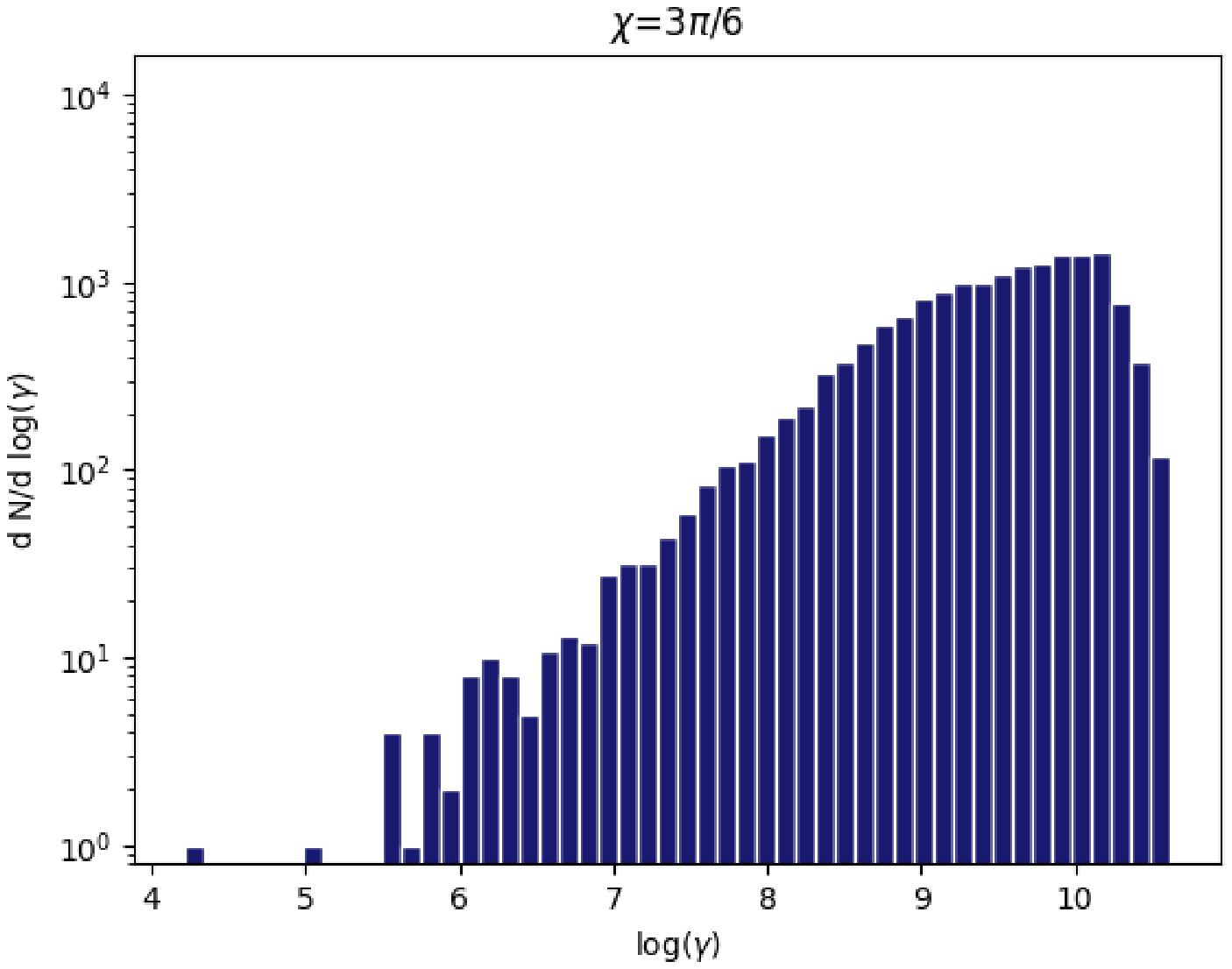}% Images in 100% size
  \caption{}
%  \caption{Lorentz factor spectrum for electrons around a pulsar of inclination $\chi=5\upi/6$}
\label{fig:Spectre_prot_chi3sur6_rand_16384}
\end{subfigure}
  \begin{subfigure}[t]{0.03\textwidth}
    \textbf{(d)}
  \end{subfigure}
\begin{subfigure}{.46\textwidth}
  \centering
  \includegraphics[scale=0.4]{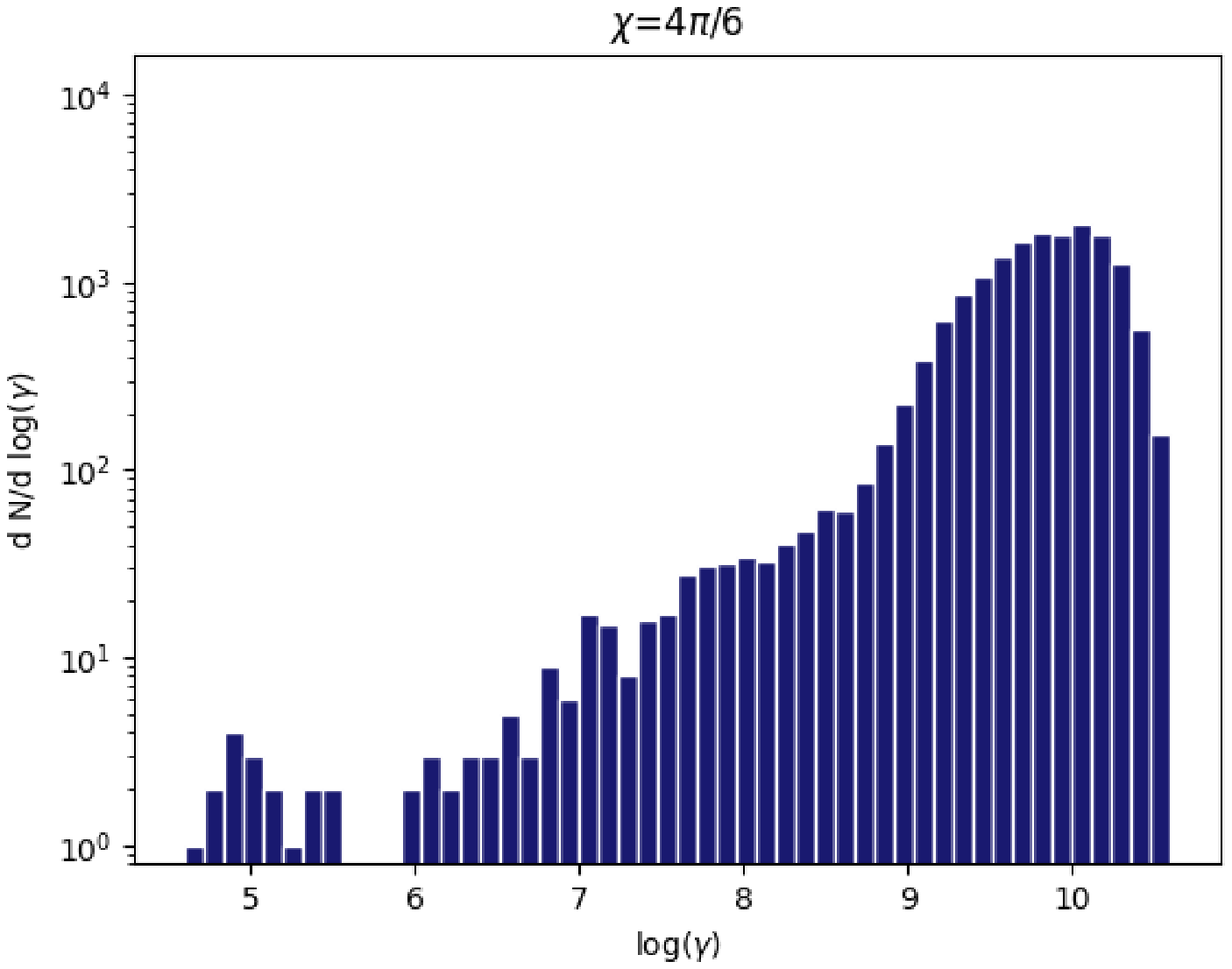}% Images in 100% size
  \caption{}
%  \caption{Lorentz factor spectrum for a fictive particle ($q=q_p$, $m=10^{6}m_p$) around a pulsar of inclination $\chi=\upi/6$}
\label{fig:Spectre_prot_chi4sur6_rand_16384}
\end{subfigure}
\caption{Comparison of the distribution functions for 16.384 protons following the random placement: (a) inclination $\chi=30^{\circ}$, (b) inclination $\chi=60^{\circ}$, (c) inclination $\chi=90^{\circ}$, (d) inclination $\chi=120^{\circ}$.}
\label{fig:Spectre_comparison_rand_16384}
\end{figure}

The simulations with randomly placed particles were useful for the spectral analysis in complement to those of \ref{sssec:Lorentz}. Indeed, figure \ref{fig:Spectre_prot_chi1sur6_rand} allows us to get rid of the "gap" in the high energy lobe that is visible otherwise in figure \ref{fig:Spectre_prot_chi1sur6} while keeping the two distinct lobes %\com{Revoir les références aux figures, je ne suis pas sûr qu'elles sont toutes bonnes.}
. Otherwise, the overall tendencies of spectra were those discussed in \ref{sssec:Lorentz}
A deeper analysis with more particle yields the results shown in figure \ref{fig:Spectre_comparison_rand_16384} with even smoother spectra and filling more the low energy end of the spectra.

\begin{figure}
  \begin{subfigure}[t]{0.03\textwidth}
    \textbf{(a)}
  \end{subfigure}
\begin{subfigure}{.46\textwidth}
  \centering
  \includegraphics[scale=0.4]{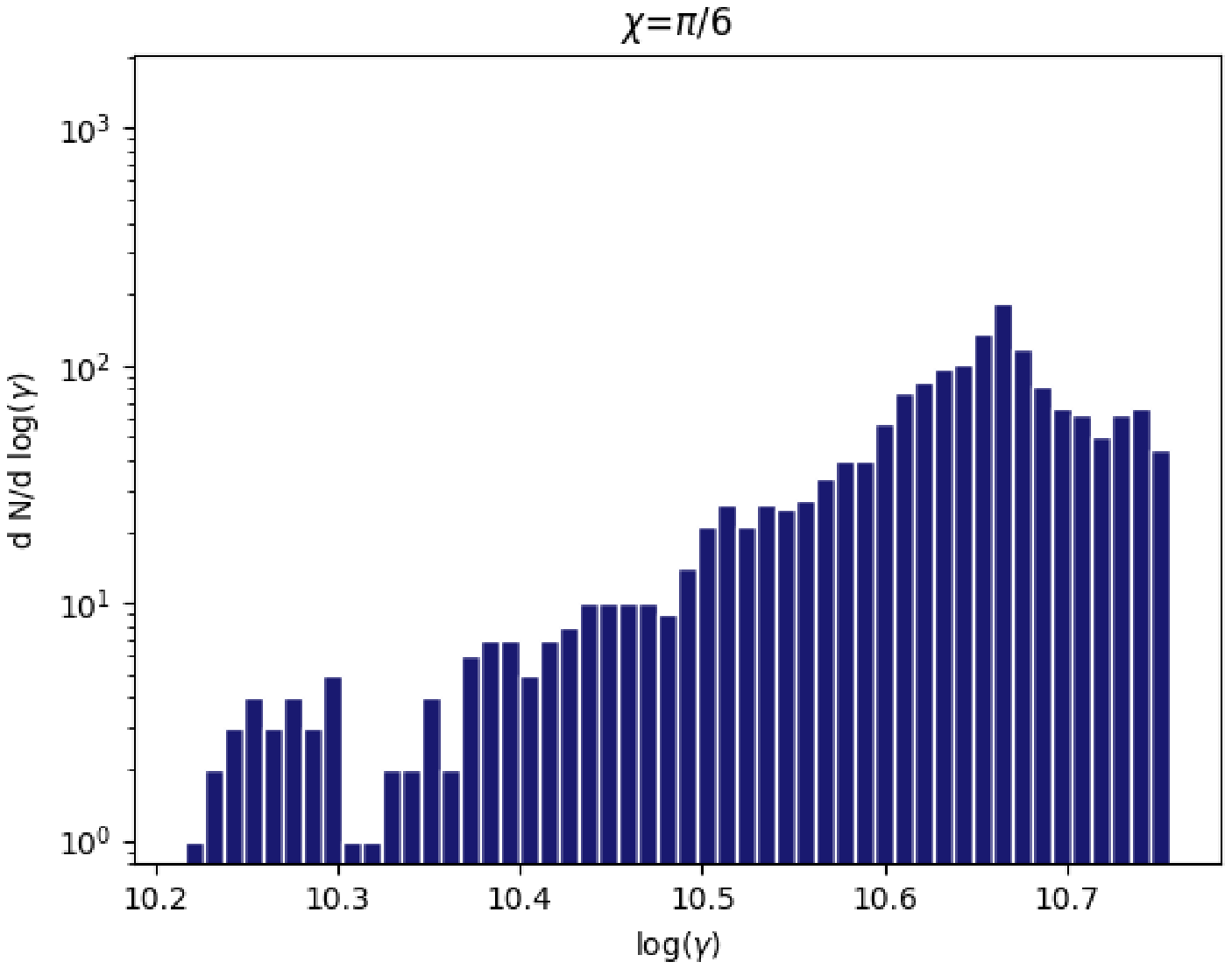}% Images in 100% size
%  \caption{Lorentz factor spectrum for crashed protons around a pulsar of inclination $\chi=\upi/6$}
\label{fig:Spectre_prot_chi1sur6_crash_rand}
\end{subfigure}
  \begin{subfigure}[t]{0.03\textwidth}
    \textbf{(b)}
  \end{subfigure}
\begin{subfigure}{.46\textwidth}
  \centering
  \includegraphics[scale=0.4]{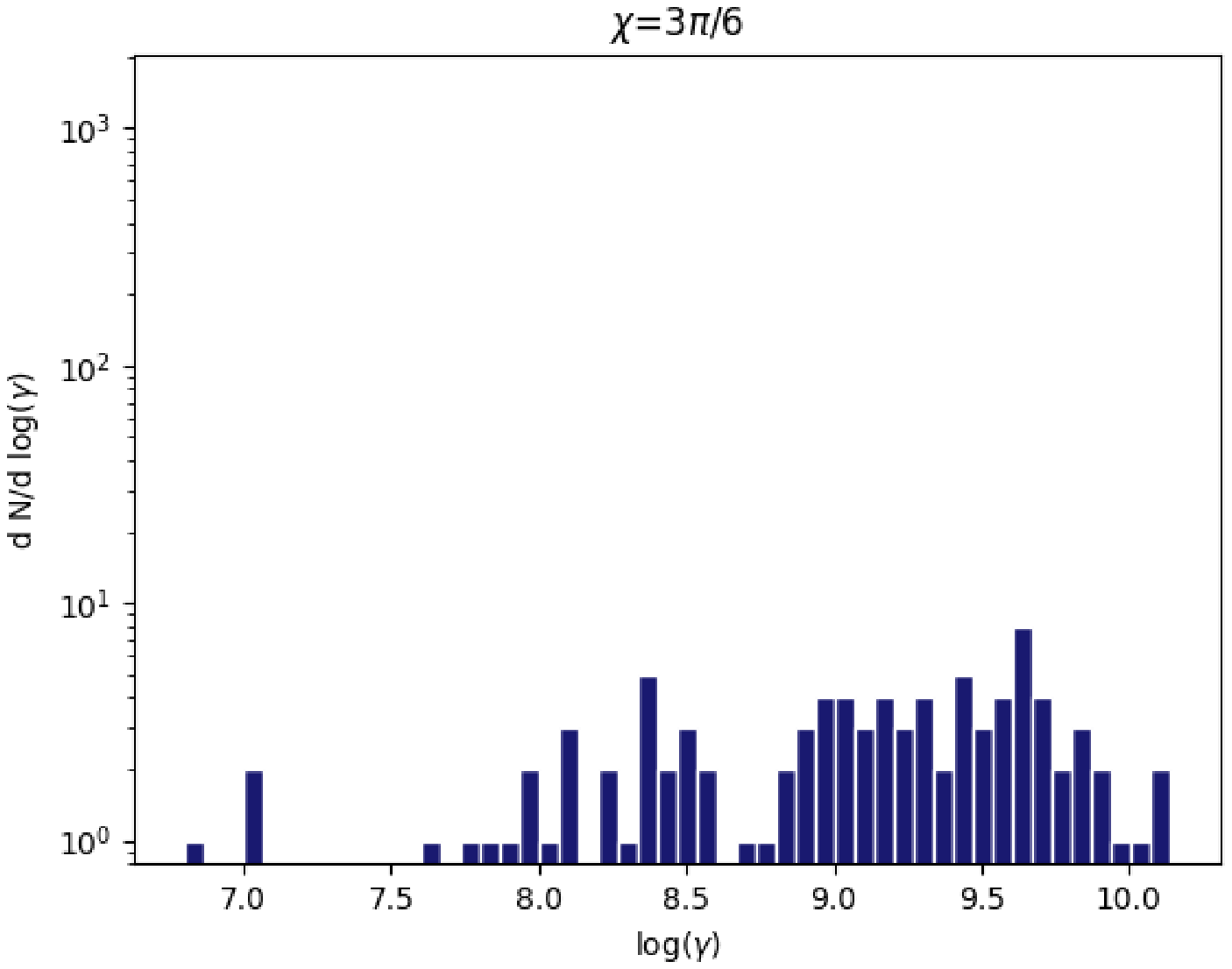}% Images in 100% size
%  \caption{Lorentz factor spectrum for crashed protons around a pulsar of inclination $\chi=\upi/2$}
\label{fig:Spectre_prot_chi3sur6_crash_rand}
\end{subfigure}
  \begin{subfigure}[t]{0.03\textwidth}
    \textbf{(c)}
  \end{subfigure}
\begin{subfigure}{.46\textwidth}
  \centering
  \includegraphics[scale=0.4]{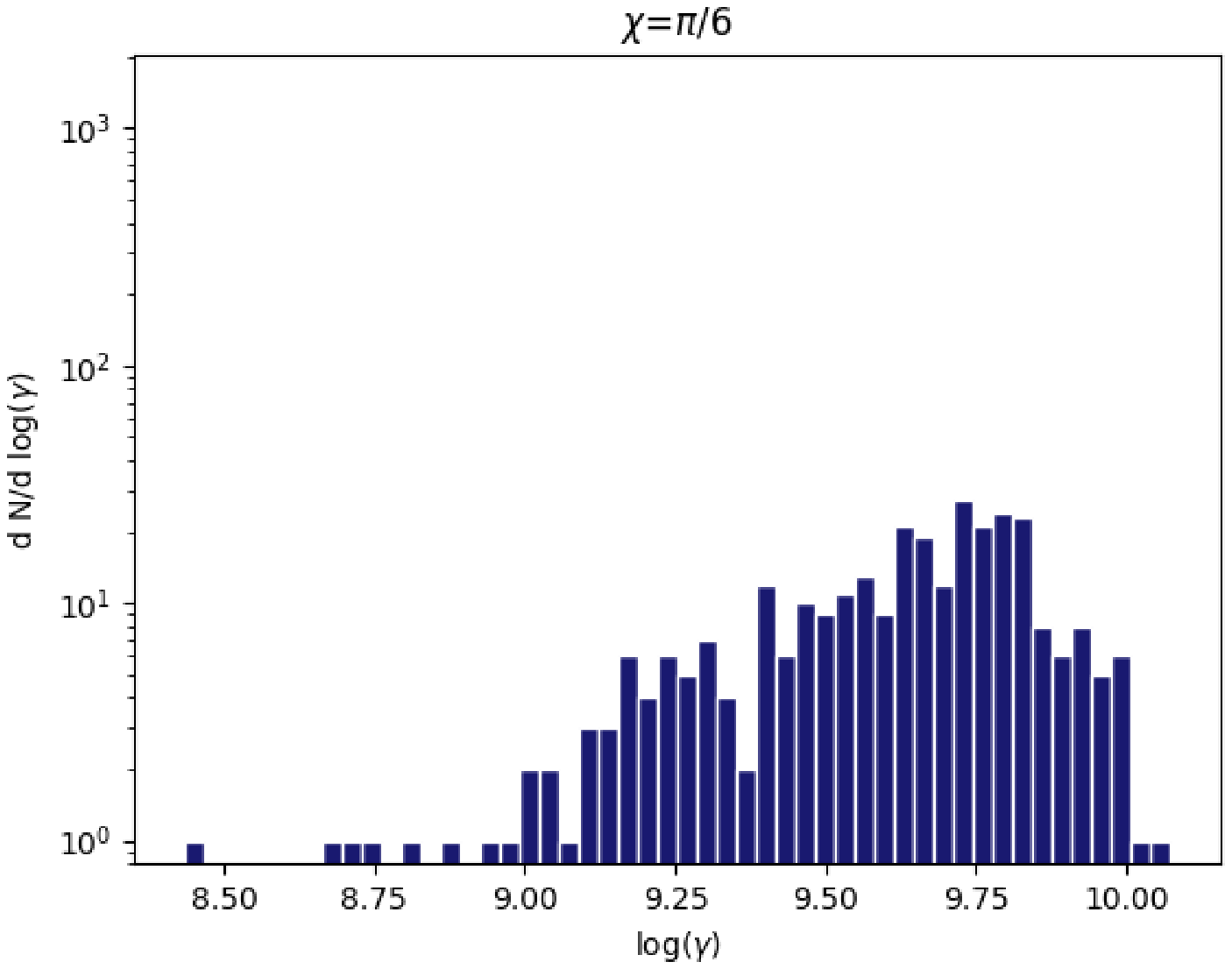}% Images in 100% size
%  \caption{Lorentz factor spectrum for ejected protons around a pulsar of inclination $\chi=\upi/6$}
\label{fig:Spectre_prot_chi1sur6_eject_rand}
\end{subfigure}
  \begin{subfigure}[t]{0.03\textwidth}
    \textbf{(d)}
  \end{subfigure}
\begin{subfigure}{.46\textwidth}
  \centering
  \includegraphics[scale=0.4]{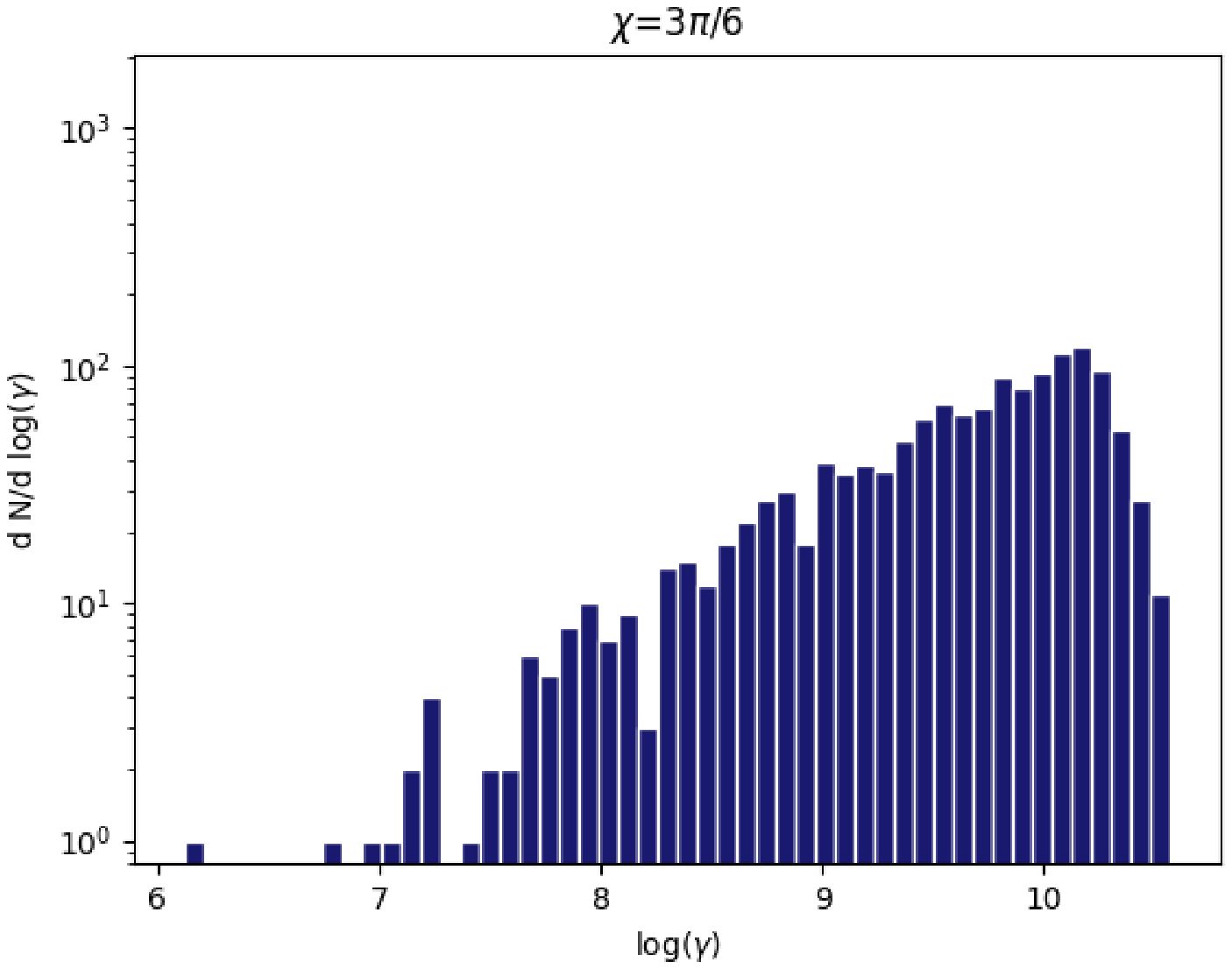}% Images in 100% size
%  \caption{Lorentz factor spectrum for ejected protons around a pulsar of inclination $\chi=\upi/2$}
\label{fig:Spectre_prot_chi3sur6_eject_rand}
\end{subfigure}
  \begin{subfigure}[t]{0.03\textwidth}
    \textbf{(e)}
  \end{subfigure}
\begin{subfigure}{.46\textwidth}
  \centering
  \includegraphics[scale=0.4]{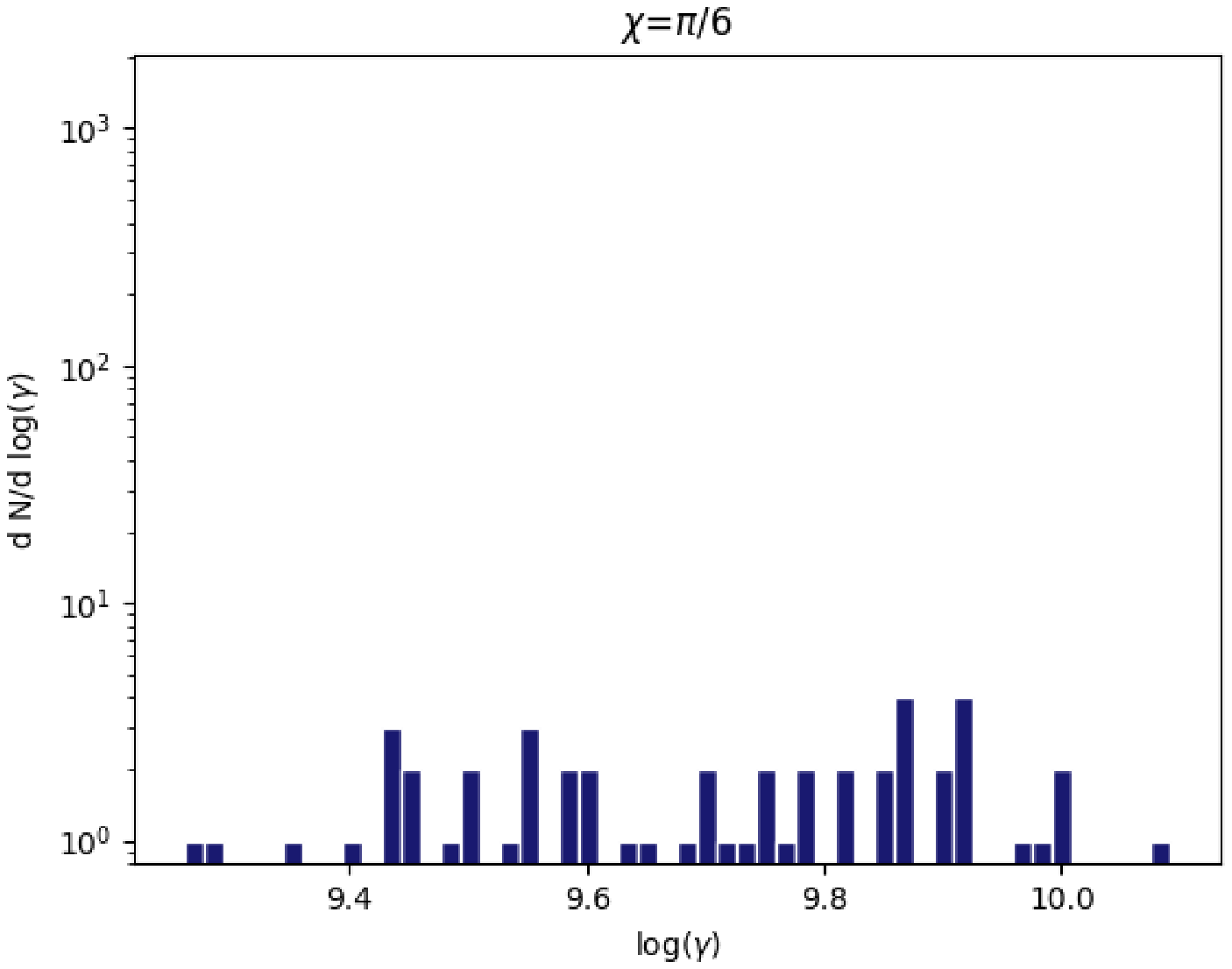}% Images in 100% size
%  \caption{Lorentz factor spectrum for trapped protons around a pulsar of inclination $\chi=\upi/6$}
\label{fig:Spectre_prot_chi1sur6_trap_rand}
\end{subfigure}
  \begin{subfigure}[t]{0.03\textwidth}
    \textbf{(f)}
  \end{subfigure}
\begin{subfigure}{.46\textwidth}
  \centering
  \includegraphics[scale=0.4]{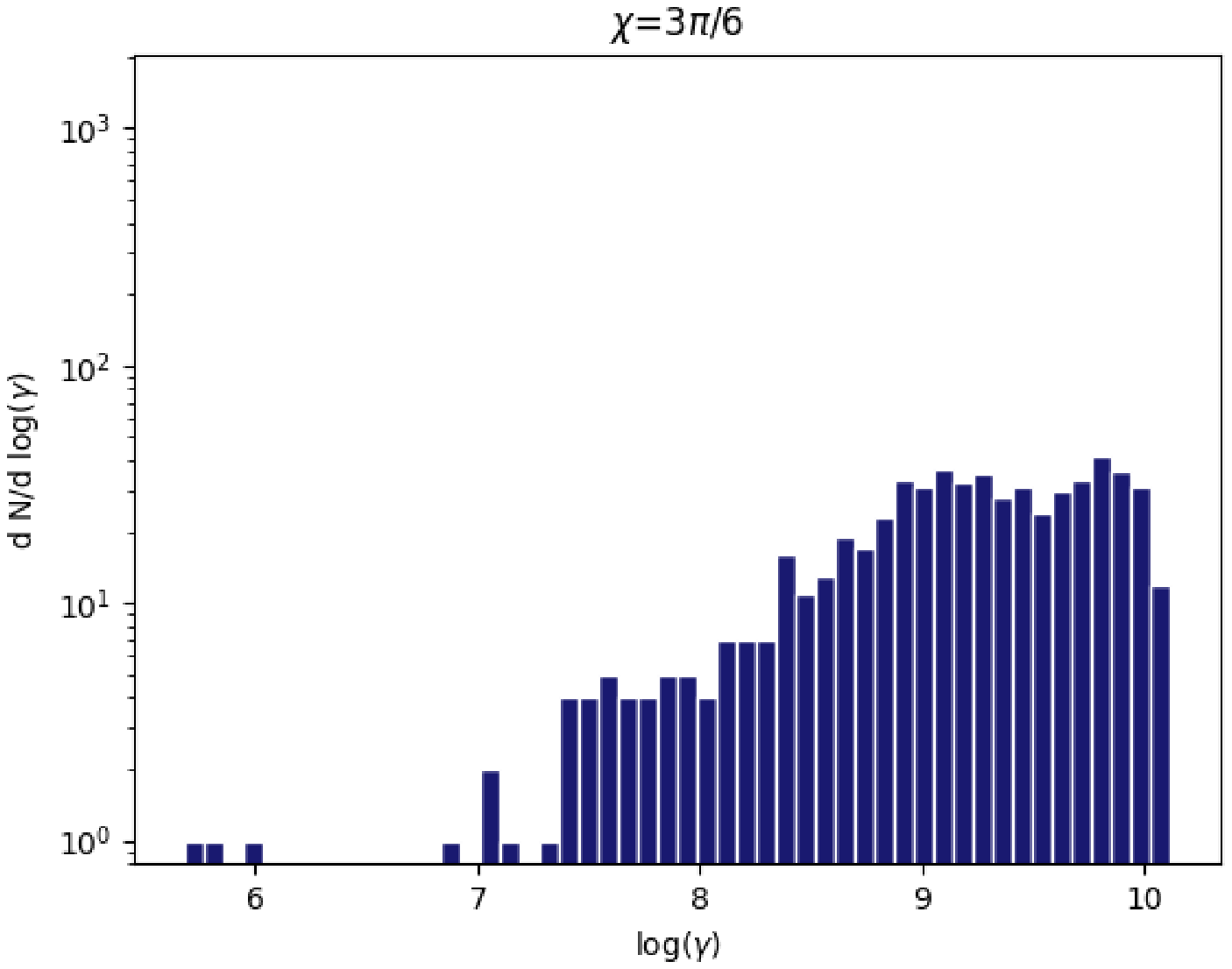}% Images in 100% size
%  \caption{Lorentz factor spectrum for trapped protons around a pulsar of inclination $\chi=\upi/2$}
\label{fig:Spectre_prot_chi3sur6_trap_rand}
\end{subfigure}
\caption{Protons distribution functions depending on their final state. (a) Crashed, $\chi=30^{\circ}$, (b) Crashed, $\chi=90^{\circ}$, (c) Ejected, $\chi=30^{\circ}$, (d) Ejected, $\chi=90^{\circ}$, (e) Trapped, $\chi=30^{\circ}$, (f) Trapped, $\chi=90^{\circ}$, for a random distribution of 2.048 particles.}
\label{fig:Spectre_prot_state_rand}
\end{figure}

\begin{figure}
  \begin{subfigure}[t]{0.03\textwidth}
    \textbf{(a)}
  \end{subfigure}
\begin{subfigure}{.46\textwidth}
  \centering
  \includegraphics[scale=0.4]{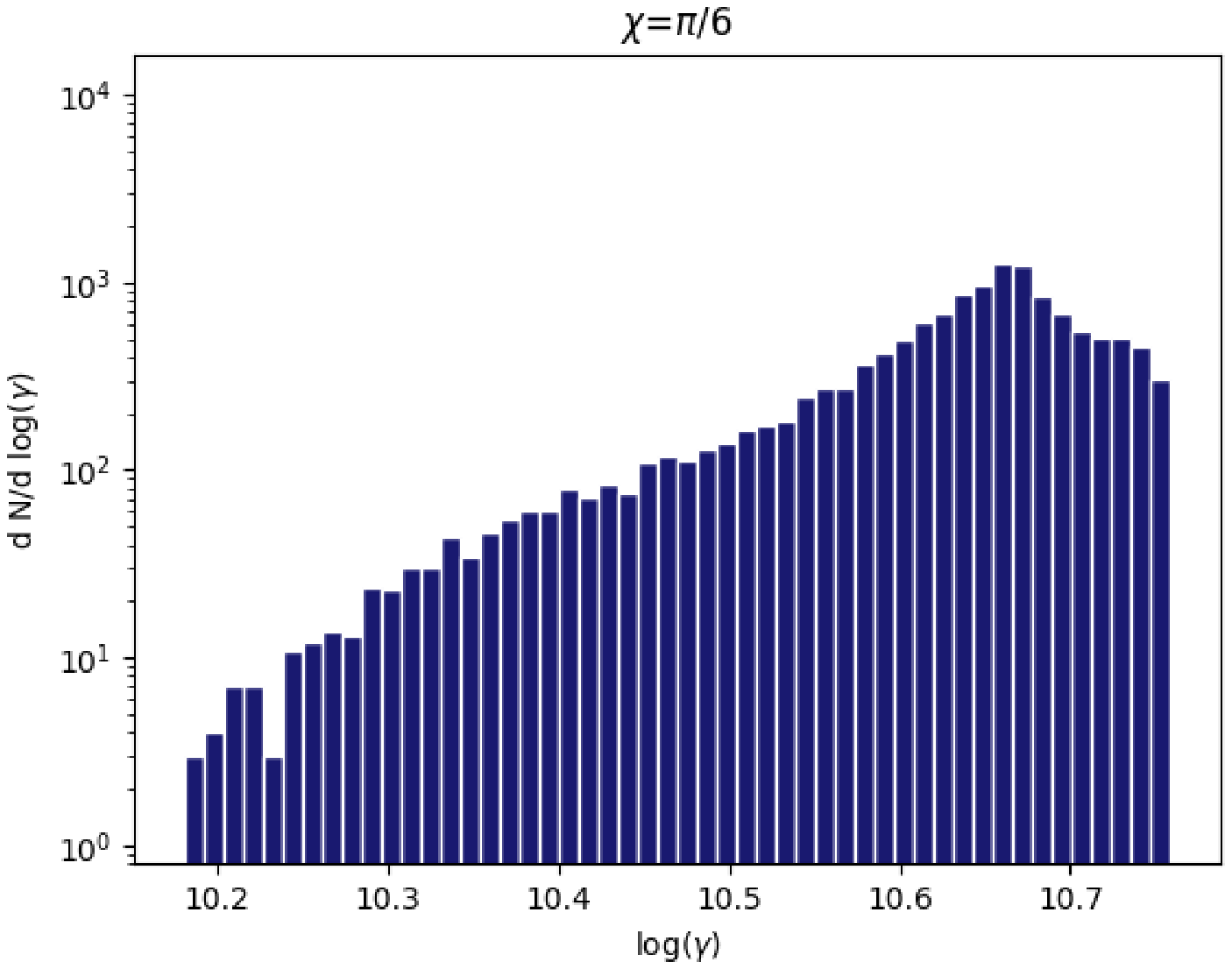}% Images in 100% size
%  \caption{Lorentz factor spectrum for crashed protons around a pulsar of inclination $\chi=\upi/6$}
\label{fig:Spectre_prot_chi1sur6_crash_rand_16384}
\end{subfigure}
  \begin{subfigure}[t]{0.03\textwidth}
    \textbf{(b)}
  \end{subfigure}
\begin{subfigure}{.46\textwidth}
  \centering
  \includegraphics[scale=0.4]{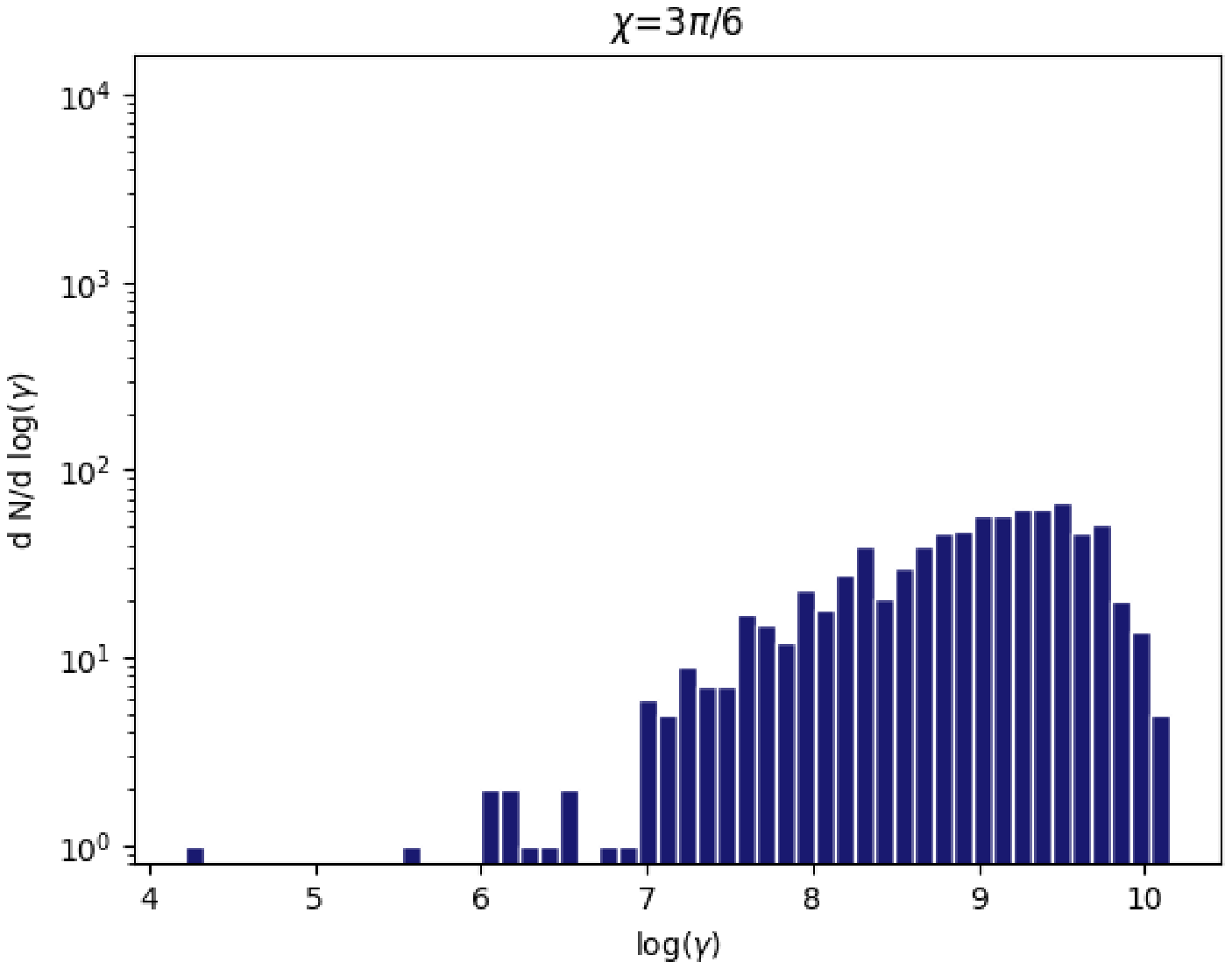}% Images in 100% size
%  \caption{Lorentz factor spectrum for crashed protons around a pulsar of inclination $\chi=\upi/2$}
\label{fig:Spectre_prot_chi3sur6_crash_rand_16384}
\end{subfigure}
  \begin{subfigure}[t]{0.03\textwidth}
    \textbf{(c)}
  \end{subfigure}
\begin{subfigure}{.46\textwidth}
  \centering
  \includegraphics[scale=0.4]{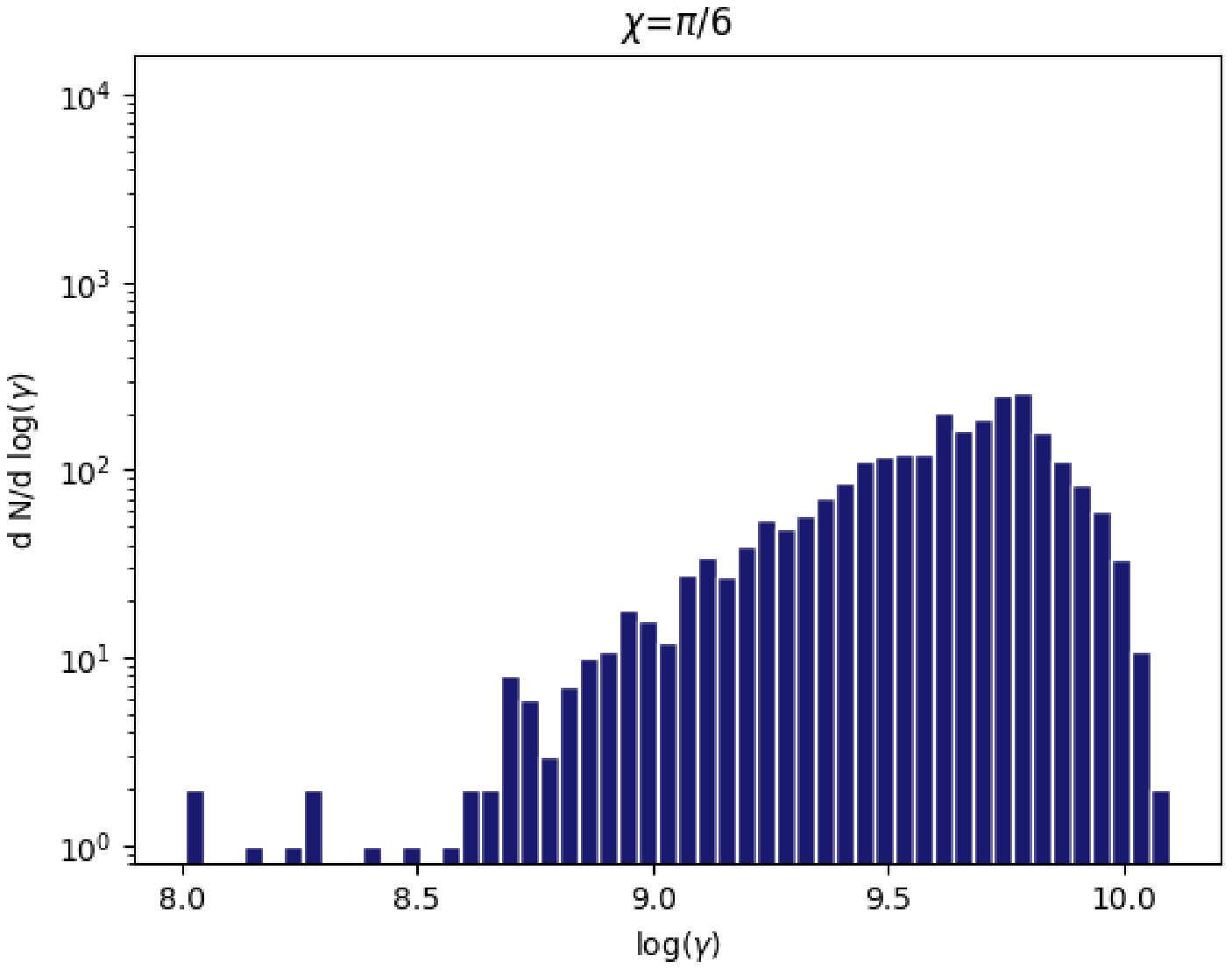}% Images in 100% size
%  \caption{Lorentz factor spectrum for ejected protons around a pulsar of inclination $\chi=\upi/6$}
\label{fig:Spectre_prot_chi1sur6_eject_rand_16384}
\end{subfigure}
  \begin{subfigure}[t]{0.03\textwidth}
    \textbf{(d)}
  \end{subfigure}
\begin{subfigure}{.46\textwidth}
  \centering
  \includegraphics[scale=0.4]{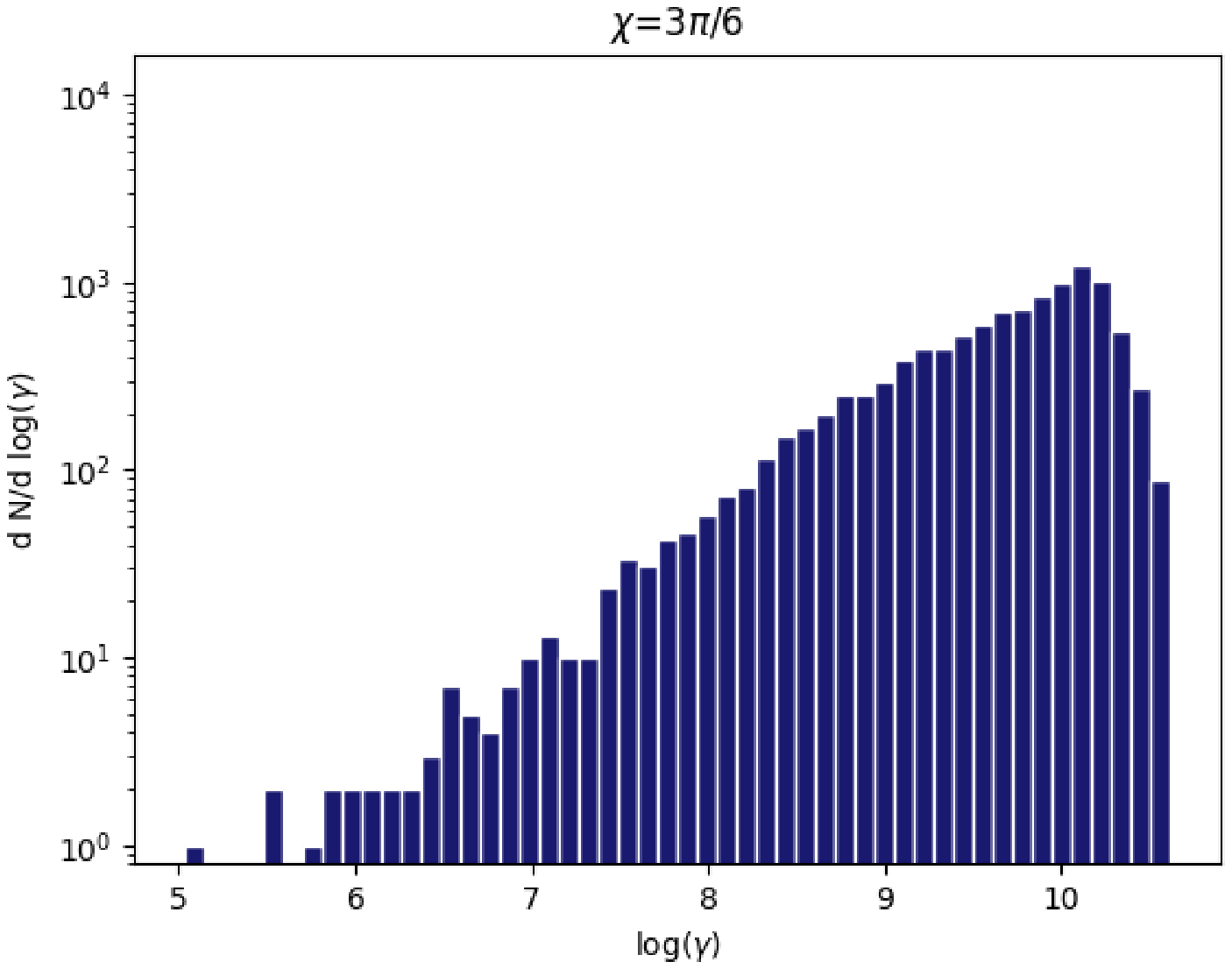}% Images in 100% size
%  \caption{Lorentz factor spectrum for ejected protons around a pulsar of inclination $\chi=\upi/2$}
\label{fig:Spectre_prot_chi3sur6_eject_rand_16384}
\end{subfigure}
  \begin{subfigure}[t]{0.03\textwidth}
    \textbf{(e)}
  \end{subfigure}
\begin{subfigure}{.46\textwidth}
  \centering
  \includegraphics[scale=0.4]{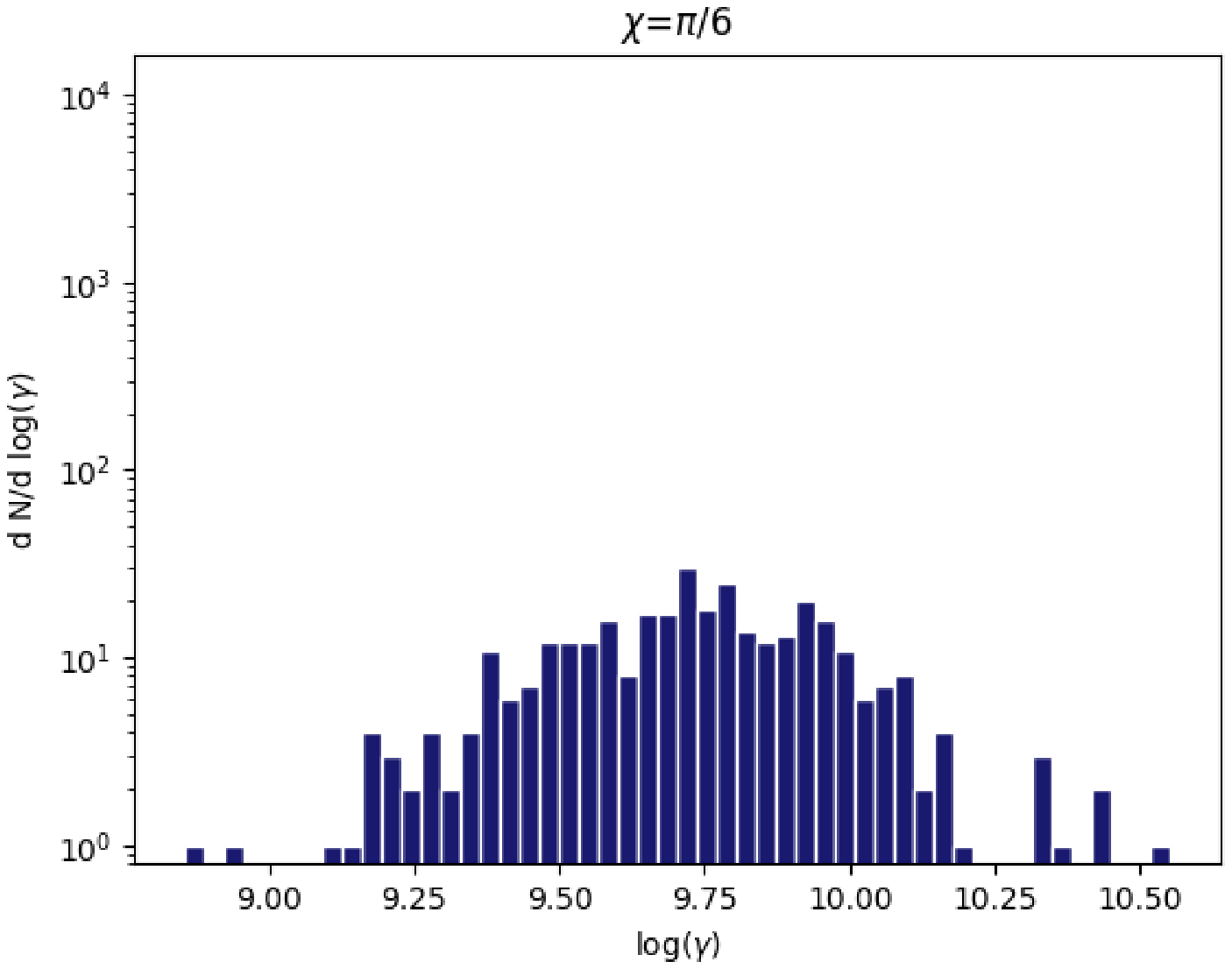}% Images in 100% size
%  \caption{Lorentz factor spectrum for trapped protons around a pulsar of inclination $\chi=\upi/6$}
\label{fig:Spectre_prot_chi1sur6_trap_rand_16384}
\end{subfigure}
  \begin{subfigure}[t]{0.03\textwidth}
    \textbf{(f)}
  \end{subfigure}
\begin{subfigure}{.46\textwidth}
  \centering
  \includegraphics[scale=0.4]{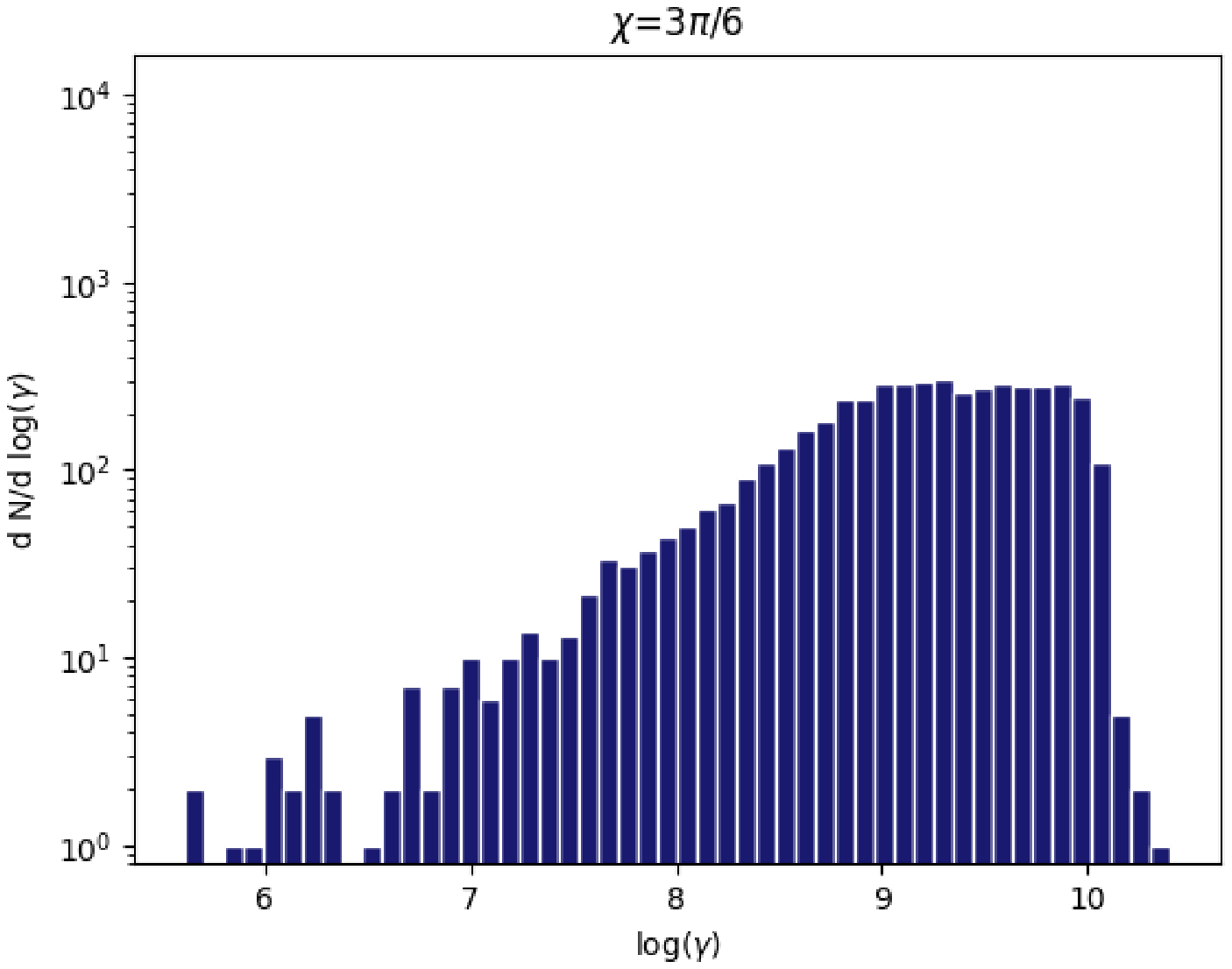}% Images in 100% size
%  \caption{Lorentz factor spectrum for trapped protons around a pulsar of inclination $\chi=\upi/2$}
\label{fig:Spectre_prot_chi3sur6_trap_rand_16384}
\end{subfigure}
\caption{Protons distribution functions depending on their final state. (a) Crashed, $\chi=30^{\circ}$, (b) Crashed, $\chi=90^{\circ}$, (c) Ejected, $\chi=30^{\circ}$, (d) Ejected, $\chi=90^{\circ}$, (e) Trapped, $\chi=30^{\circ}$, (f) Trapped, $\chi=90^{\circ}$, for a random distribution of 16.384 particles.}
\label{fig:Spectre_prot_state_rand_16384}
\end{figure}

In a similar way, comparing figure~\ref{fig:Spectre_prot_state_rand} to figure~\ref{fig:Spectre_prot_state} shows that randomly placed particles tend to fill some of the gaps while also smoothing the distribution.
And again, a higher number of particles tends to smooth the spectra, and the effect is even more noticeable on the spectra with low statistics, as shown in figure \ref{fig:Spectre_prot_state_rand_16384}.

\subsubsection{Final positions of the randomly placed particles}

\begin{figure}
  \begin{subfigure}[t]{0.03\textwidth}
    \textbf{(a)}
  \end{subfigure}
\begin{subfigure}{.46\textwidth}
  \centering
  \includegraphics[scale=0.4]{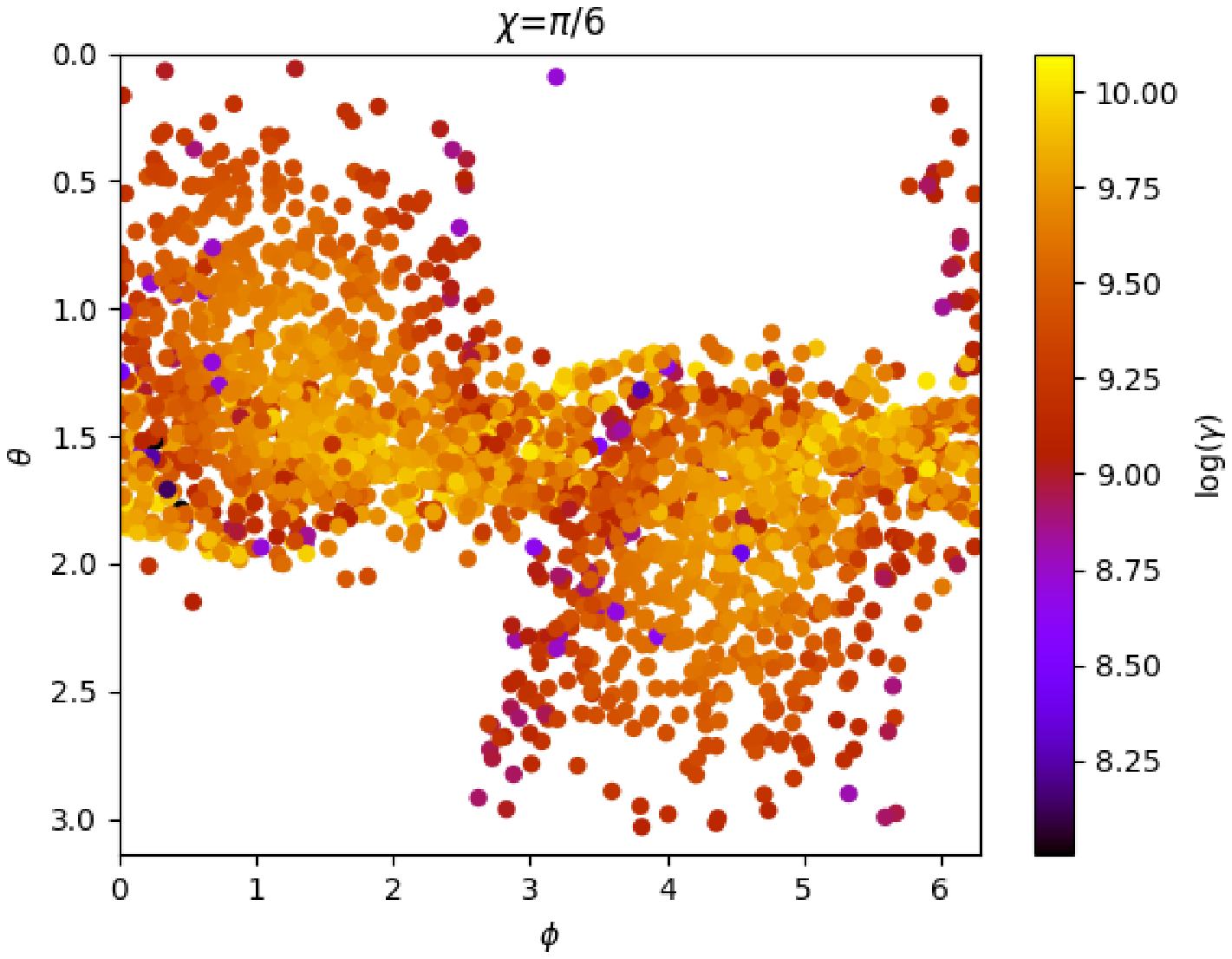}% Images in 100% size
  \caption{}
%  \caption{Lorentz factor spectrum for iron ions around a pulsar of inclination $\chi=\upi/6$}
\label{fig:Ejection_16384_chi1sur6_rand}
\end{subfigure}
  \begin{subfigure}[t]{0.03\textwidth}
    \textbf{(b)}
  \end{subfigure}
\begin{subfigure}{.46\textwidth}
  \centering
  \includegraphics[scale=0.4]{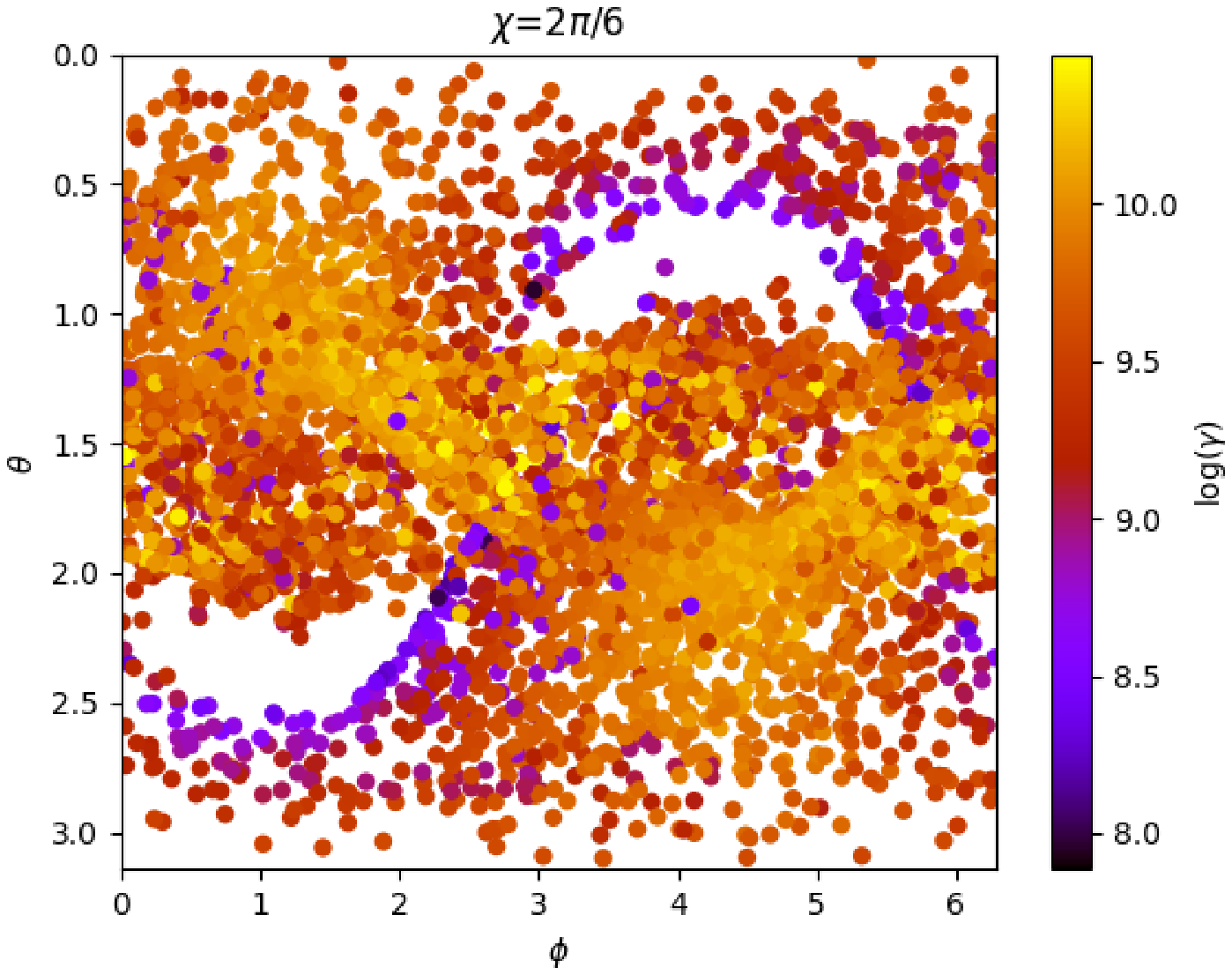}% Images in 100% size
  \caption{}
%  \caption{Lorentz factor spectrum for }
\label{fig:Ejection_16384_chi2sur6_rand}
\end{subfigure}
  \begin{subfigure}[t]{0.03\textwidth}
    \textbf{(c)}
  \end{subfigure}
\begin{subfigure}{.46\textwidth}
  \centering
  \includegraphics[scale=0.4]{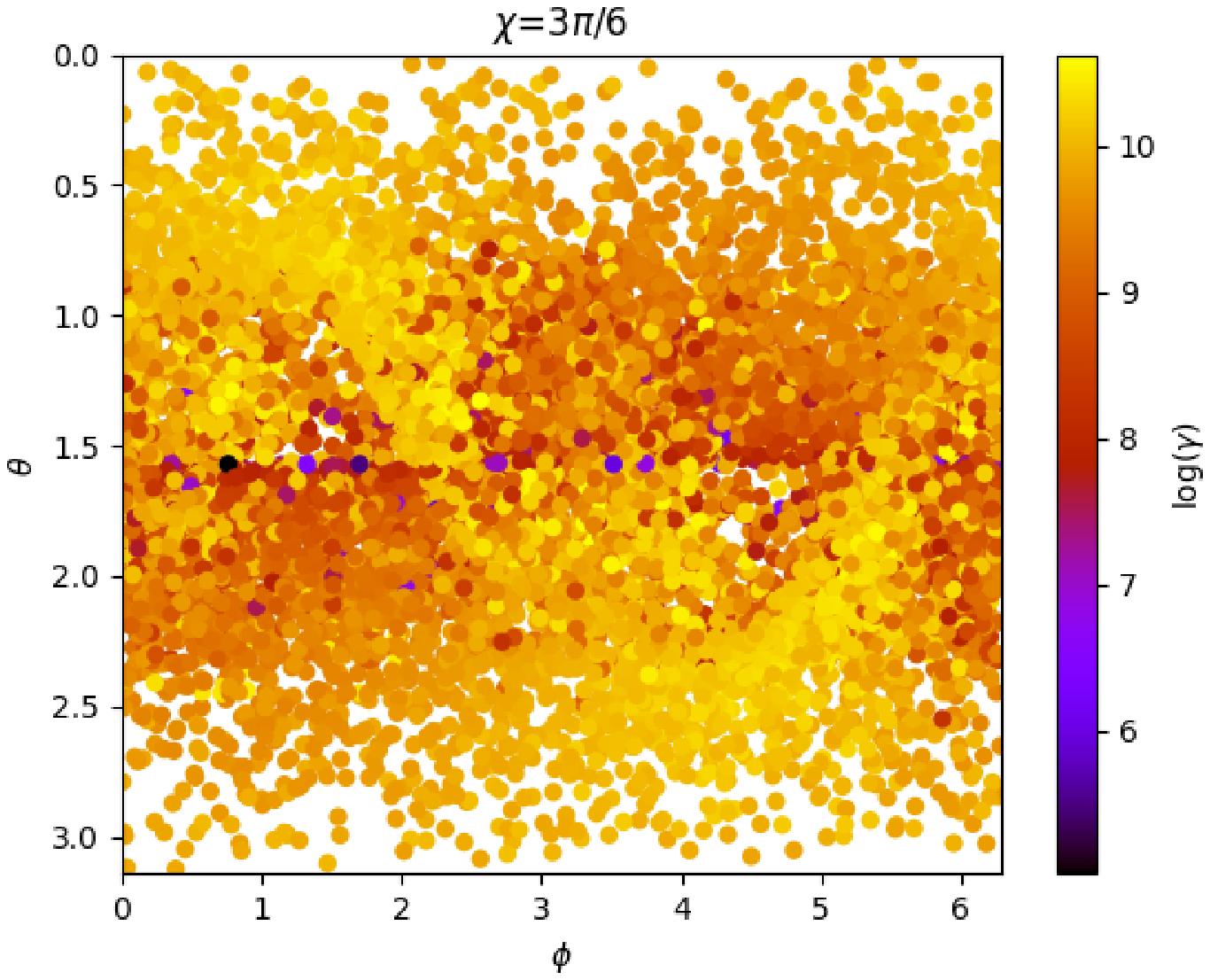}% Images in 100% size
  \caption{}
%  \caption{Lorentz factor spectrum for electrons around a pulsar of inclination $\chi=5\upi/6$}
\label{fig:Ejection_16384_chi3sur6_rand}
\end{subfigure}
  \begin{subfigure}[t]{0.03\textwidth}
    \textbf{(d)}
  \end{subfigure}
\begin{subfigure}{.46\textwidth}
  \centering
  \includegraphics[scale=0.4]{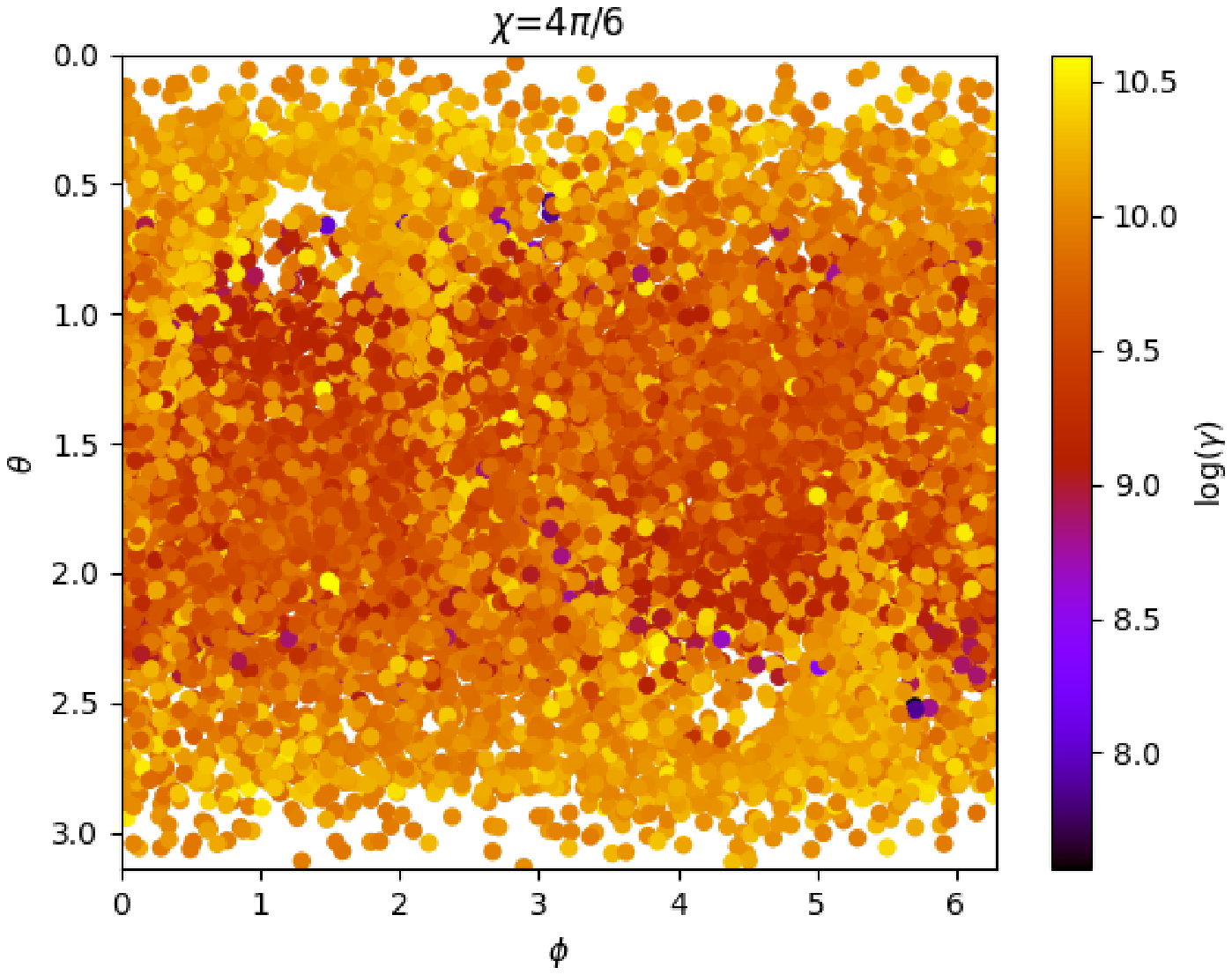}% Images in 100% size
  \caption{}
%  \caption{Lorentz factor spectrum for a fictive particle ($q=q_p$, $m=10^{6}m_p$) around a pulsar of inclination $\chi=\upi/6$}
\label{fig:Ejection_16384_chi4sur6_rand}
\end{subfigure}
\caption{Ejection maps for 16.384 protons following the random placement: (a) inclination $\chi=30^{\circ}$, (b) inclination $\chi=60^{\circ}$, (c) inclination $\chi=90^{\circ}$, (d) inclination $\chi=120^{\circ}$.}
\label{fig:Ejection_16384_rand}
\end{figure}

The simulations with 16.384 randomly placed particles still highlights over-densities and under-densities in the ejection maps. In addition, some structures where the particles have similar Lorentz factors still appear. In fact, figure~\ref{fig:Ejection_16384_rand} shows preferred directions for a given range of energy of particles.

\begin{figure}
  \begin{subfigure}[t]{0.03\textwidth}
    \textbf{(a)}
  \end{subfigure}
\begin{subfigure}{.46\textwidth}
  \centering
  \includegraphics[scale=0.4]{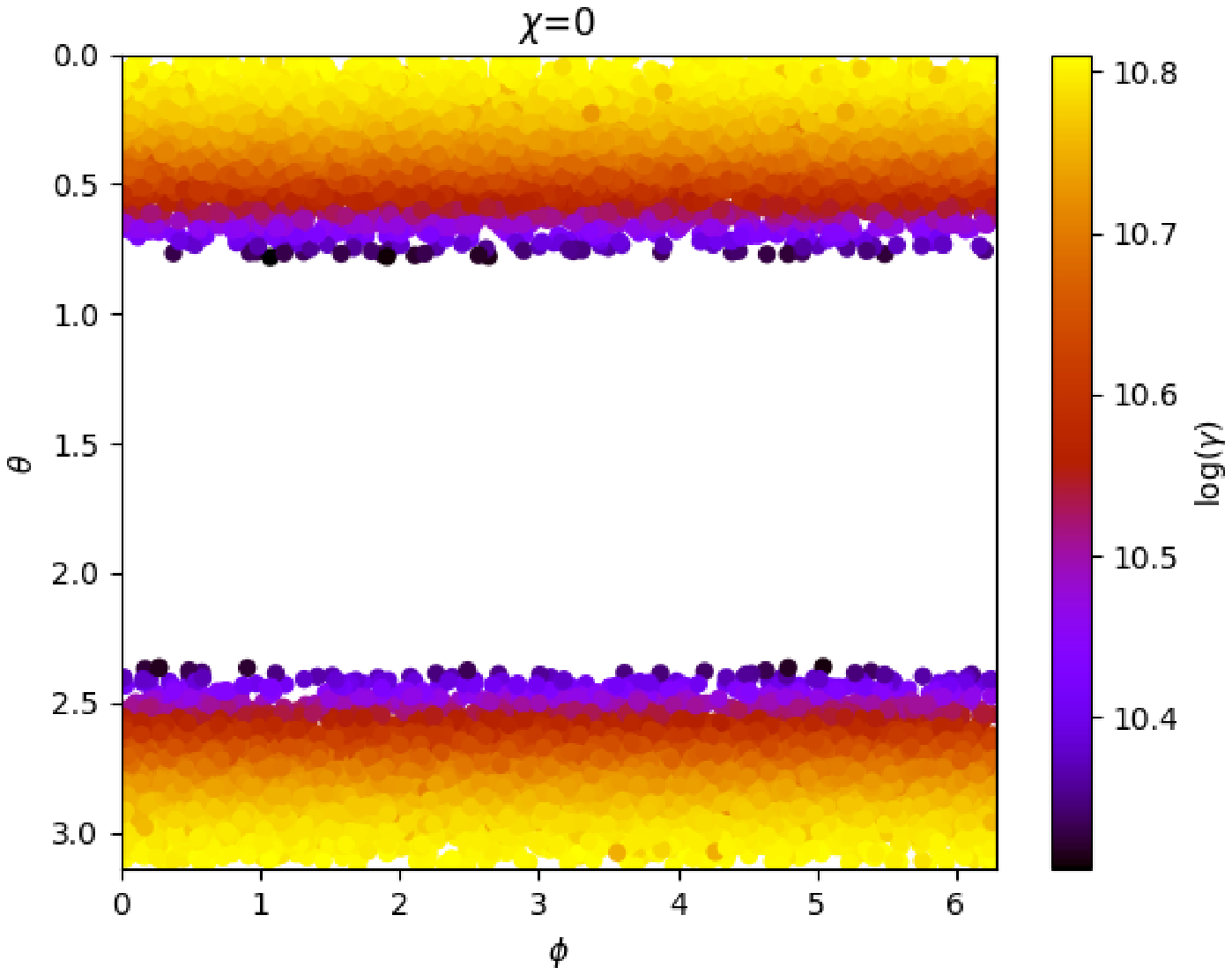}% Images in 100% size
  %\caption{$\chi=0\upi/6$}
\label{fig:Crash_prot_chi0sur6_16384_rand}
\end{subfigure}
  \begin{subfigure}[t]{0.03\textwidth}
    \textbf{(b)}
  \end{subfigure}
\begin{subfigure}{.46\textwidth}
  \centering
  \includegraphics[scale=0.4]{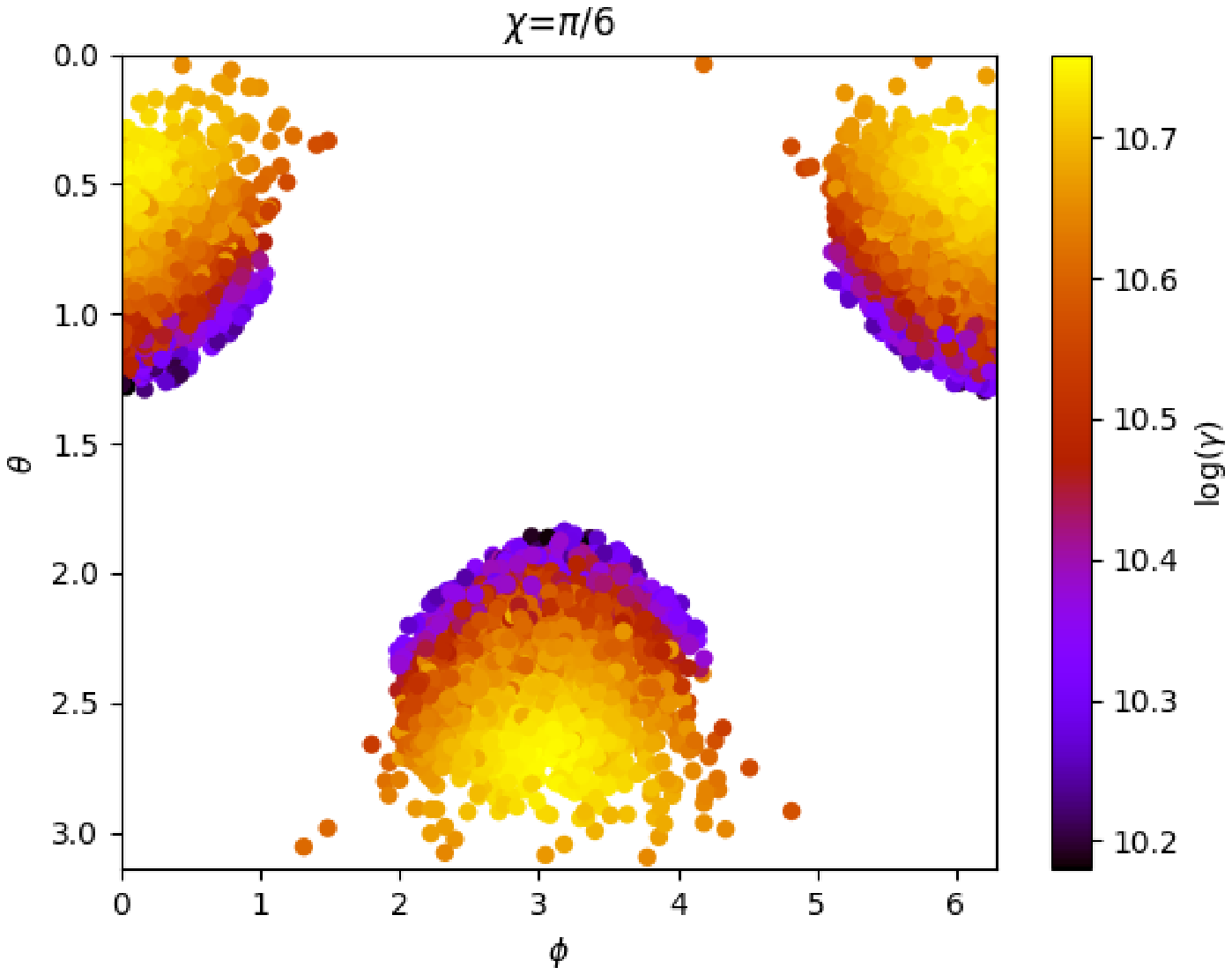}% Images in 100% size
  %\caption{$\chi=\upi/6$}
\label{fig:Crash_prot_chi1sur6_16384_rand}
\end{subfigure}

  \begin{subfigure}[t]{0.03\textwidth}
    \textbf{(c)}
  \end{subfigure}
\begin{subfigure}{.46\textwidth}
  \centering
  \includegraphics[scale=0.4]{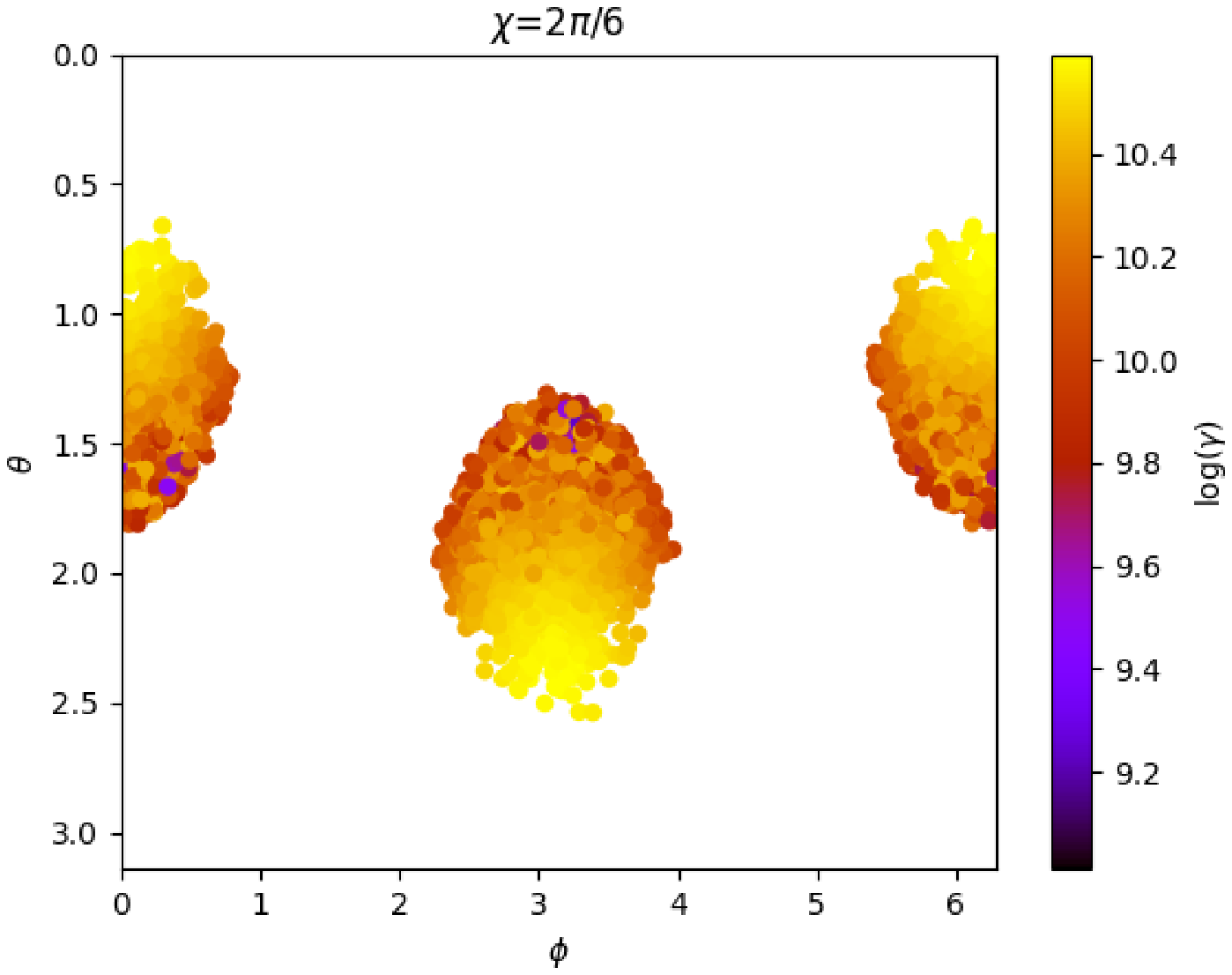}% Images in 100% size
  %\caption{$\chi=2\upi/6$}
\label{fig:Crash_prot_chi2sur6_16384_rand}
\end{subfigure}
  \begin{subfigure}[t]{0.03\textwidth}
    \textbf{(d)}
  \end{subfigure}
\begin{subfigure}{.46\textwidth}
  \centering
  \includegraphics[scale=0.4]{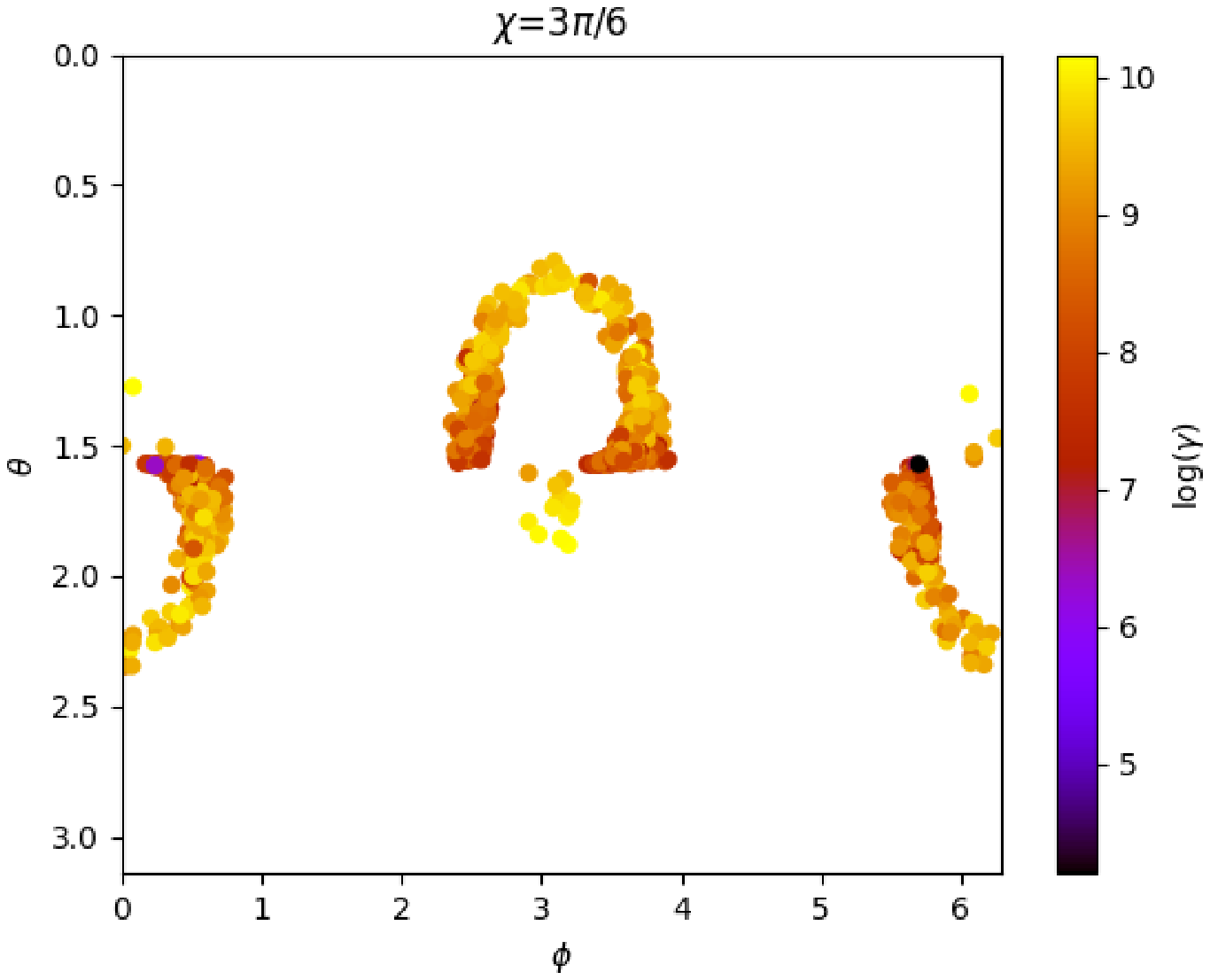}% Images in 100% size
  %\caption{$\chi=3\upi/6$}
\label{fig:Crash_prot_chi3sur6_16384_rand}
\end{subfigure}
  \caption{Impact maps of randomly placed protons on the neutron star surface for 16.384 particles, with $\chi=0^{\circ}$ (a), $\chi=30^{\circ}$ (b), $\chi=60^{\circ}$ (c) and $\chi=90^{\circ}$ (d).}
\label{fig:Crash_16384_rand}
\end{figure}

In figure~\ref{fig:Crash_16384_rand}, impact positions are similar as in figure~\ref{fig:Crash}, still with gradients of energy depending of the position of impact. Interestingly, the hotspots for $\chi=90^{\circ}$  are still crescent-shaped.

\subsection{Aligned and anti-aligned neutron stars}

\begin{figure}
  \begin{subfigure}[t]{0.03\textwidth}
    \textbf{(a)}
  \end{subfigure}
\begin{subfigure}{.46\textwidth}
  \centering
  \includegraphics[scale=0.4]{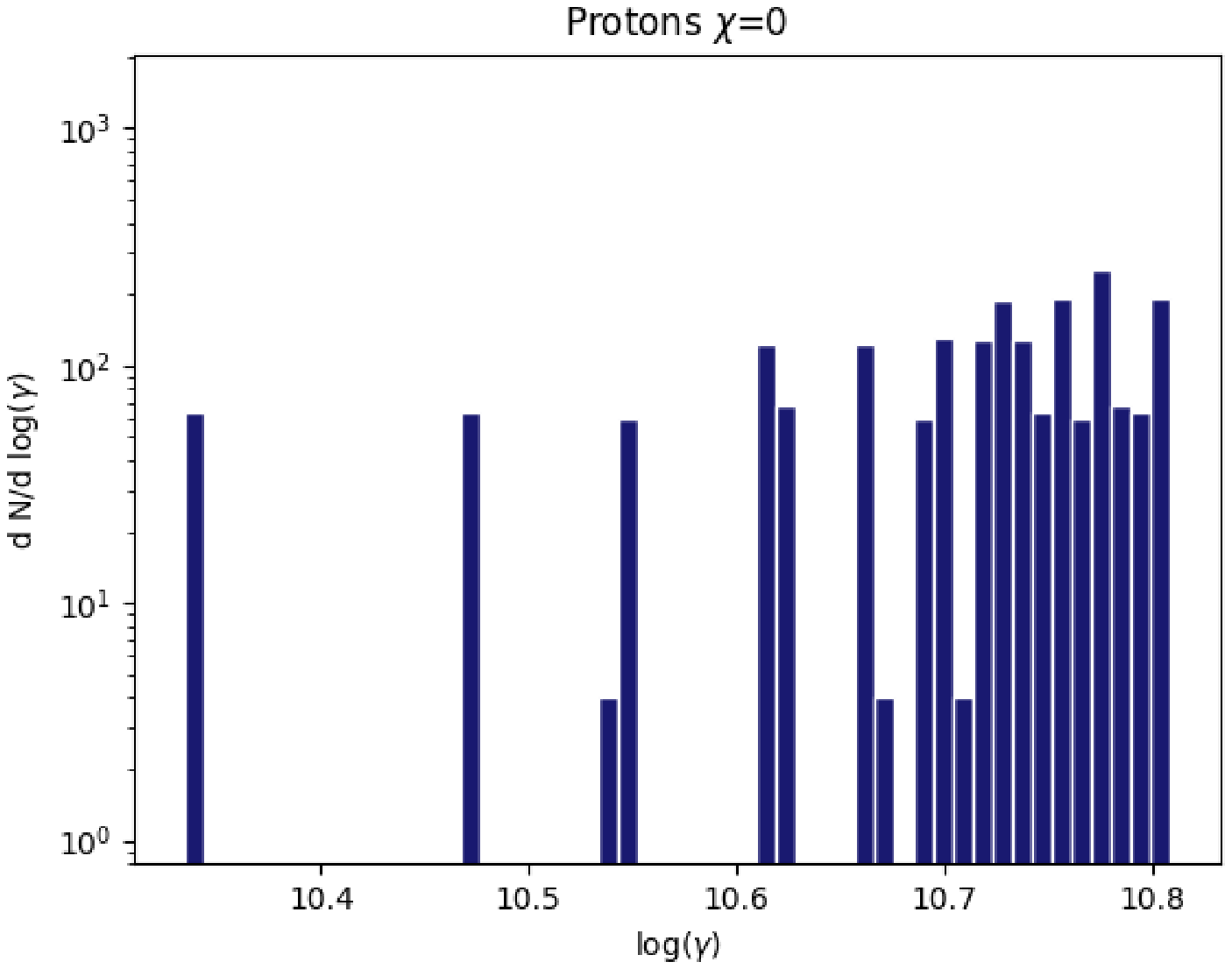}% Images in 100% size
%  \caption{Antiprotons}
%\label{fig:Eject_aprot_chi3sur6}
\end{subfigure}
  \begin{subfigure}[t]{0.03\textwidth}
    \textbf{(b)}
  \end{subfigure}
\begin{subfigure}{.46\textwidth}
  \centering
  \includegraphics[scale=0.4]{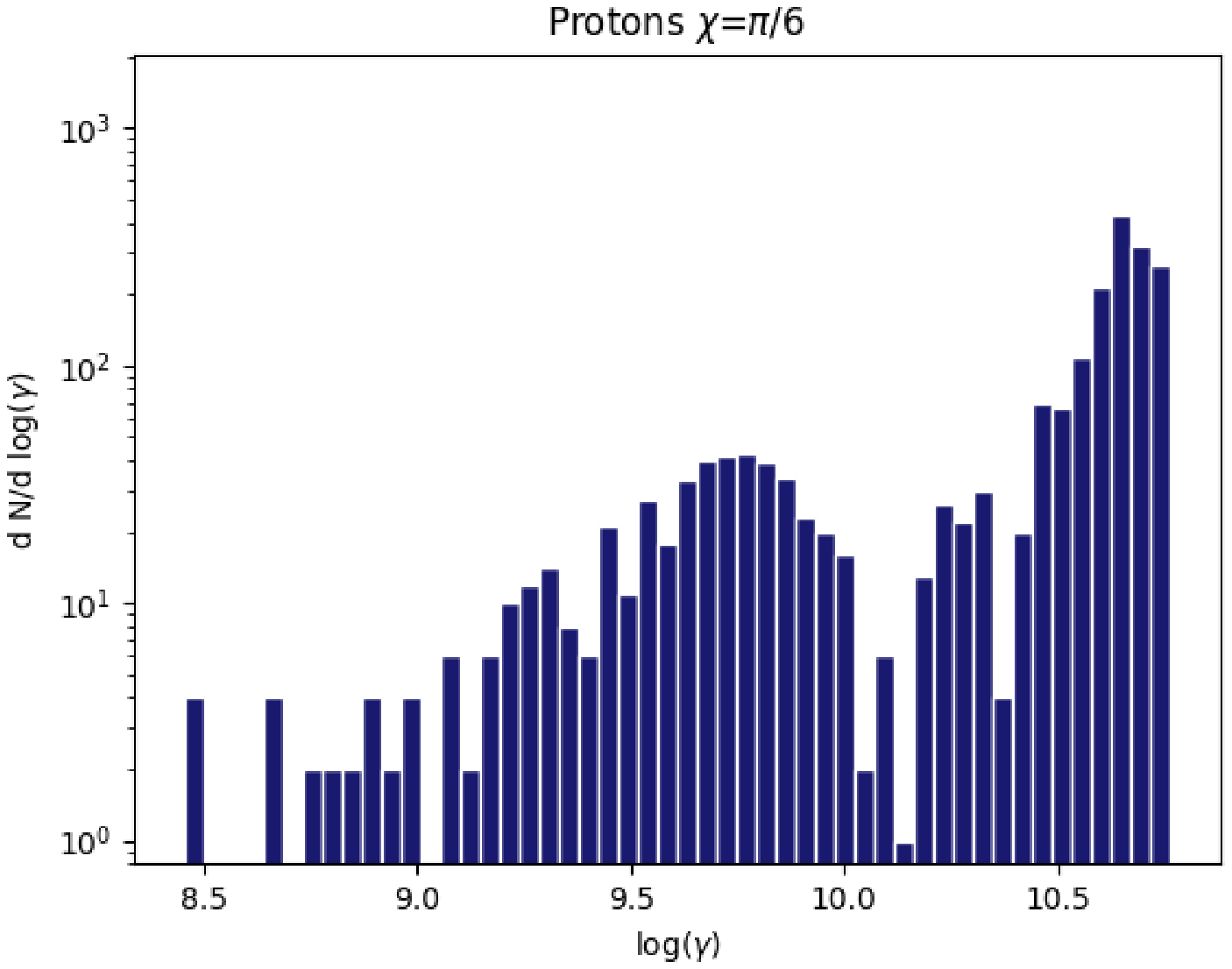}% Images in 100% size
%  \caption{Electrons}
%\label{fig:Eject_elec_chi3sur6}
\end{subfigure}
  \caption{Spectra of protons in the aligned (a) and anti-aligned (b) cases, 2.048 protons placed on rhe regular pattern.}
\label{fig:Spectre_ali_antiali}
\end{figure}

A neutron star is aligned (respectively anti-aligned) when $\chi=0^{\circ}$ (respectively $\chi=180^{\circ}$). In these cases, it is easier to interpret the trajectories of particles. For protons near an aligned neutron star, the poles of the star are negatively charged while the equator is positively charged. This leads the protons to be pushed away from the equator and pulled by the poles. However, even if the magnetic field is strong enough, particles still reach the poles by following the field lines, allowing them to crash easily, and since as particles get closer to the surface the electric field gets more intense, they reach an even higher Lorentz factor as proven by the left spectrum (a) of figure~\ref{fig:Spectre_ali_antiali}.

Inversely, protons around an anti-aligned neutron stars will mostly be trapped around it since they are pulled toward the equator which is now negatively charged, but they cannot reach it due to the too strong magnetic field. These protons will feel an electric drift similar to that in figure~\ref{fig:B_dom} but this drift forces them to rotate around the neutron star, leading to a high spread in Lorentz factors on the right spectrum (b) of figure~\ref{fig:Spectre_ali_antiali}. 

\begin{figure}
  \begin{subfigure}[t]{0.03\textwidth}
    \textbf{(a)}
  \end{subfigure}
\begin{subfigure}{.46\textwidth}
  \centering
  \includegraphics[scale=0.4]{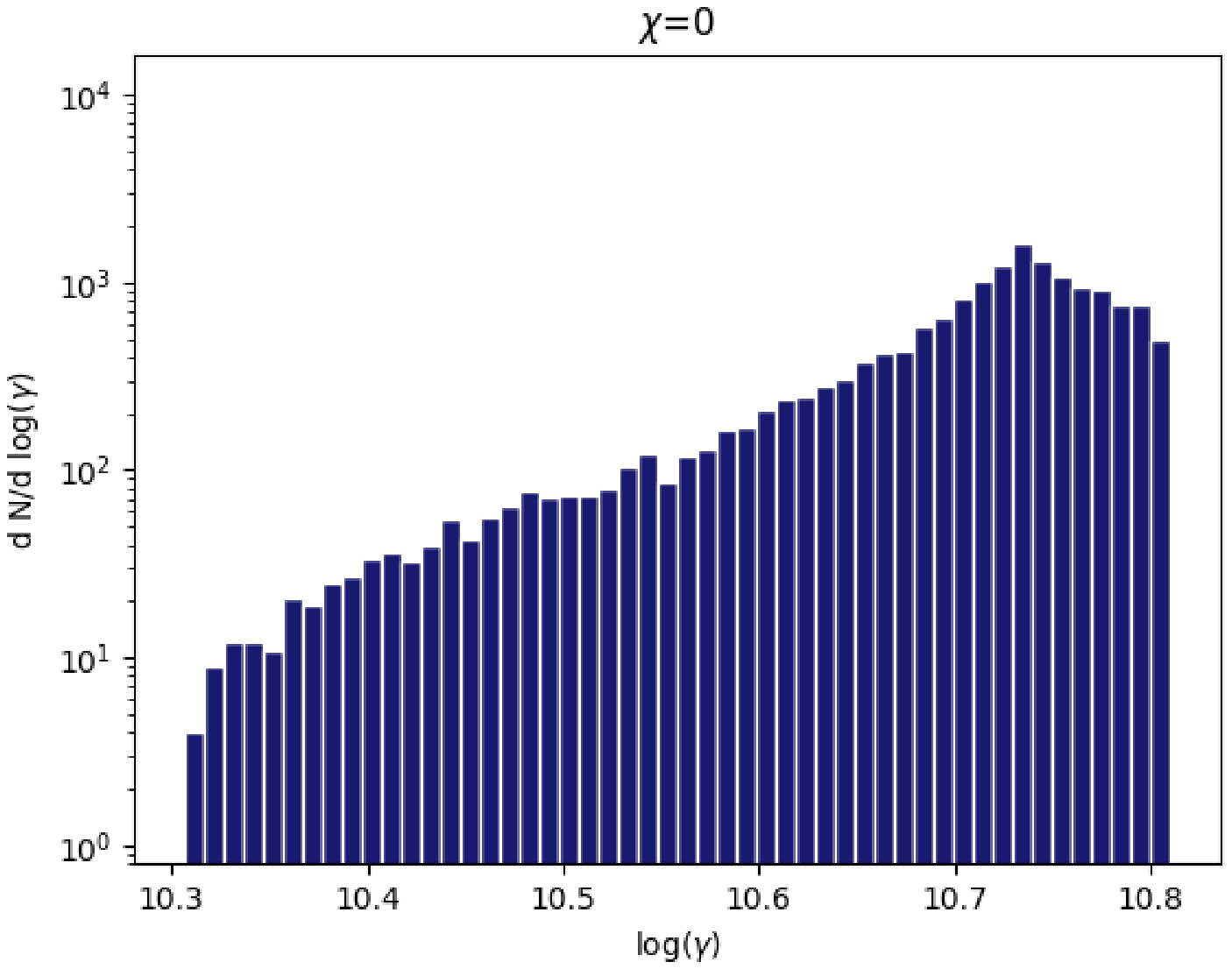}% Images in 100% size
%  \caption{Antiprotons}
%\label{fig:Eject_aprot_chi3sur6}
\end{subfigure}
  \begin{subfigure}[t]{0.03\textwidth}
    \textbf{(b)}
  \end{subfigure}
\begin{subfigure}{.46\textwidth}
  \centering
  \includegraphics[scale=0.4]{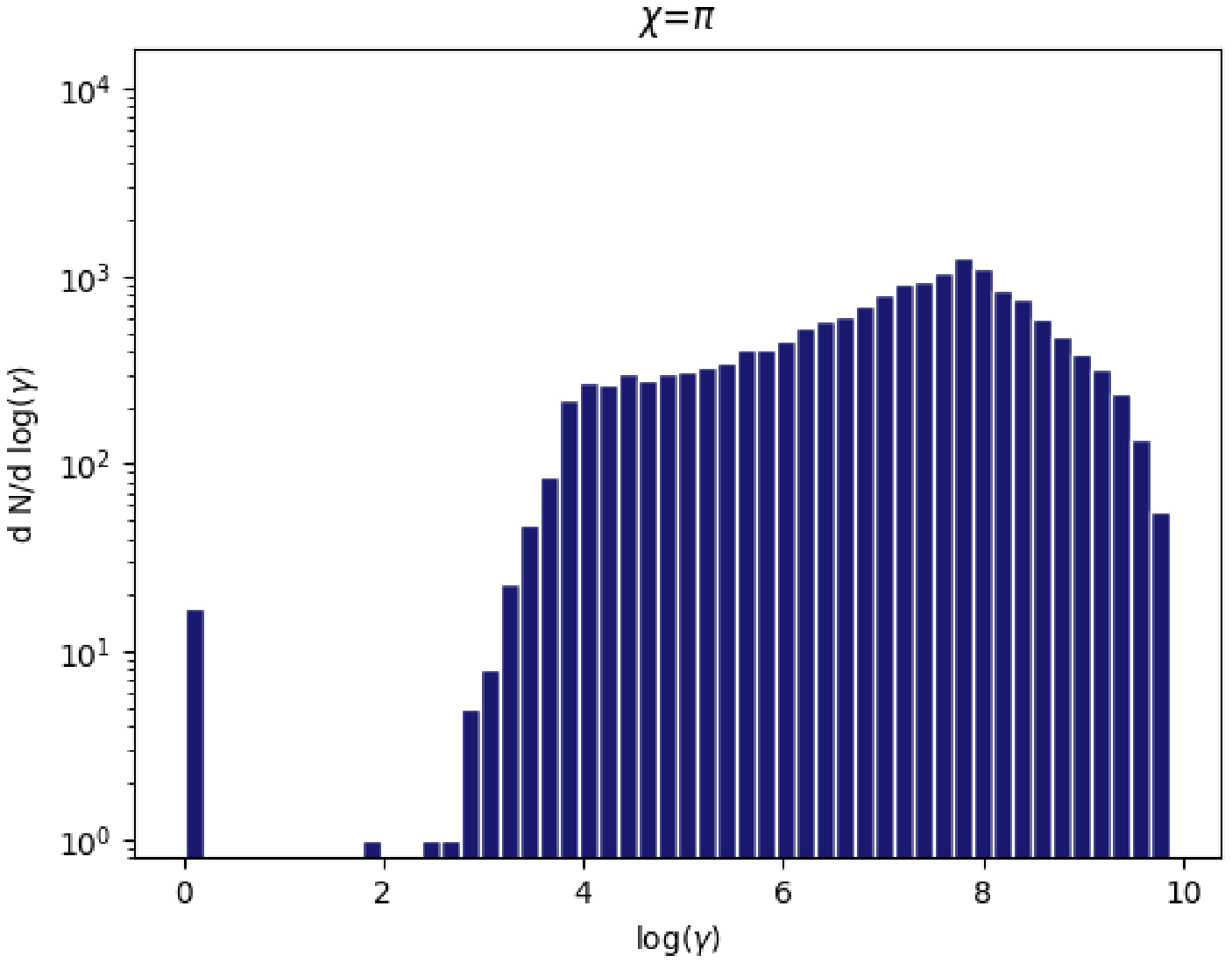}% Images in 100% size
%  \caption{Electrons}
%\label{fig:Eject_elec_chi3sur6}
\end{subfigure}
  \caption{Spectra of 16.384 randomly placed protons in the aligned (a) and anti-aligned (b) cases.}
\label{fig:Spectre_ali_antiali_16384}
\end{figure}

And with 16.384 randomly placed particles, the spectra appear to be more continuous and a peak not visible in figure~\ref{fig:Spectre_ali_antiali} appears, as shown in figure~\ref{fig:Spectre_ali_antiali_16384}, while keping the same maximum Lorentz factor.

\section{Conclusion}
\label{sec:Conclusion}

In this paper, we proved that our new algorithm to simulate particles acceleration in realistic ultra-strong neutron star electromagnetic fields is efficient and accurate. We put a constraint on the highest Lorentz factors that charged particles can reach thanks to our preliminary idealised model of a rotating magnet in vacuum.

Interestingly, the maximum Lorentz factors attainable by the particles scale linearly with their charge to mass ratio~$q/m$. The code also retrieved two expected symmetries: a charge-latitude symmetry and a central symmetry relative to the centre of the pulsar. The inclination of the pulsar has an influence on the particle distribution functions and on their final positions. Actually, positive charges hit the surface of the neutron star only for $\chi \leq 90^{\circ}$.

This preliminary work is encouraging but still not straightforwardly applicable per se to real neutron stars electrodynamics. Nevertheless, we will pursue our effort by studying the acceleration efficiency in a plasma filled magnetosphere, which is a more realistic case than the Deutsch vacuum field. Good starting points would be resistive magnetospheres \citep{li_resistive_2012} as well as radiative magnetospheres \citep{petri_radiative_2020, petri_electrodynamics_2020}. 

Because of the very high Lorentz factors, radiation reaction cannot be ignored as shown by \cite{laue_acceleration_1986}. We plan to add this damped motion in an upcoming work. Our long standing goal is to inject particle due to pair creation and to implement the pusher into a particle in cell (PIC) or Vlasov code to fully and self-consistently simulate a pulsar magnetosphere with ultra-strong fields, generating current sheets and acceleration gaps. We expect a decrease of the Lorentz factors by several orders of magnitudes compared to the present investigation thanks to the screening of the field and the energy lost via radiation.
%\revtwo{
%6) The authors mention their ambition to ultimately incorporate this particle pusher in a PIC approach. Some comments as to how efficient (compute-intensive computations are needed for doing the (inverse) rotations, (inverse) Lorentz boosts, ...) this pusher really is are in order: does the computational cost stay manageable?
%}
 Our long lasting goal is to obtain kinetic solution of realistic pulsar magnetospheres.
The feasibility of a PIC code implementing our analytical pusher has been shown by \cite{petri_relativistic_2020} who designed a 1D relativistic electromagnetic PIC code to simulate plasma oscillations, the relativistic two-stream instability and a strongly magnetized ultra-relativistic shock. We found there that the computational cost remains manageable, with only a small overhead compared for instance with the fully implicit scheme presented by \cite{petri_fully_2017-1} and \cite{vay_simulation_2008}. The additional cost to boost particles and rotate the coordinate axes are largely compensated by the ability to compute realistic trajectories in ultra strong fields.

\section*{Acknowledgement}
We are grateful to the referees for helpful comments and suggestions. This work has been supported by CEFIPRA grant IFC/F5904-B/2018. We also acknowledge the High Performance Computing center of the University of Strasbourg for supporting this work by providing scientific support and access to computing resources. Part of the computing resources were funded by the Equipex Equip@Meso project (Programme Investissements d'Avenir) and the CPER Alsacalcul/Big Data.

%\bibliographystyle{jpp}
%% Note the spaces between the initials
%
%\bibliography{jpp-instructions,/home/petri/zotero/Ma_bibliotheque}

\end{document}